\documentclass[acmsmall]{acmart}

\setcopyright{cc}
\setcctype{by}
\acmDOI{10.1145/3798202}
\acmYear{2026}
\acmJournal{PACMPL}
\acmVolume{10}
\acmNumber{OOPSLA1}
\acmArticle{94}
\acmMonth{4}
\received{2025-10-10}
\received[accepted]{2026-02-17}

\usepackage{algorithmic}
\usepackage{graphicx}
\usepackage{textcomp}
\usepackage{xcolor}
\usepackage{listings}
\usepackage{xspace}
\usepackage{amsmath,amsthm,verbatim,amscd}
\usepackage{caption}
\usepackage{subcaption}
\usepackage[ruled, vlined]{algorithm2e}
\usepackage{booktabs, multirow}
\usepackage{tikz}
\usepackage{pifont}
\usepackage{enumitem}
\usepackage{url}
\usepackage{wrapfig}
\usepackage[T1]{fontenc}
\usepackage{mdframed}
\usepackage[frozencache,cachedir=.]{minted}
\usepackage{etoolbox}
\usepackage{multicol}
\usepackage{booktabs}
\usepackage[skins]{tcolorbox}

\newtheorem{definition}{Definition}
\newcommand{\eg}{\textit{e.g.,}~}
\newcommand{\ie}{\textit{i.e.,}~}
\newcommand{\etal}{\textit{et al.}~}

\newcommand{\nameheading}{\textsc{CFM-SE}\xspace}
\newcommand{\name}{\textsc{cfm-se}\xspace}
\newcommand{\code}[1]{\texttt{#1}}

\newcommand{\methodK}{\textbf{K}\xspace}
\newcommand{\methodC}{\textbf{C}\xspace}
\newcommand{\methodSM}{\textbf{SM}\xspace}
\newcommand{\methodCSM}{\textbf{C-SM}\xspace}

\newcommand{\circled}[1]{\raisebox{.5pt}{\textcircled{\raisebox{-.9pt} {#1}}}} 
\def\BibTeX{{\rm B\kern-.05em{\sc i\kern-.025em b}\kern-.08em
    T\kern-.1667em\lower.7ex\hbox{E}\kern-.125emX}}
    
\newtoggle{bv}
\toggletrue{bv}

\begin{document}

\title{Taming the Hydra: Targeted Control-Flow Transformations for Dynamic Symbolic Execution}

\author{Charitha Saumya}
\authornote{Equal contribution.}
\authornote{Work done while at Purdue University.}
\orcid{0000-0001-9900-2804}
\affiliation{%
	\institution{Intel Corporation}
	\city{Santa Clara}
	\country{USA}
}
\email{charitha.saumya.gusthinna.waduge@intel.com}

\author{Muhammad Hassan}
\authornotemark[1]
\orcid{0009-0009-2773-7524}
\affiliation{%
	\institution{Virginia Tech}
	\city{Blacksburg}
	\country{USA}
}
\email{mhassan01@vt.edu}

\author{Rohan Gangaraju}
\authornotemark[2]
\orcid{0009-0002-7193-0144}
\affiliation{%
	\institution{University of Texas at Austin}
	\city{Austin}
	\country{USA}
}
\email{rgangar@utexas.edu}

\author{Milind Kulkarni}
\orcid{0000-0001-6827-345X}
\affiliation{%
	\institution{Purdue University}
	\city{West Lafayette}
	\country{USA}
}
\email{milind@purdue.edu}

\author{Kirshanthan Sundararajah}
\orcid{0000-0001-6384-062X}
\affiliation{%
	\institution{Virginia Tech}
	\city{Blacksburg}
	\country{USA}
}
\email{kirshanthans@vt.edu}

\ccsdesc[500]{Software and its engineering~Compilers}
\ccsdesc[500]{Software and its engineering~Software testing and debugging}

\keywords{Control-flow Transformation, Branch Elimination, Dynamic Symbolic Execution, Path Explosion}

\begin{abstract}
	Dynamic Symbolic Execution (DSE) suffers from the path explosion problem when the target program has many conditional branches.
	The classical approach for managing the path explosion problem is dynamic state merging. 
	Dynamic state merging combines similar symbolic program states to avoid the exponential growth in the number of states during DSE.
	However, state merging still requires solver invocations at each program branch, even when both paths of the branch are feasible.
	Moreover, the best path search strategy for DSE may not create the best state merging opportunities.
	Some drawbacks of state merging can be mitigated by compile-time state merging (\ie branch elimination by converting control-flow into data flow).
	In this paper, we propose a non-semantics-preserving but failure-preserving compiler transformation for removing expensive symbolic branches in a program to improve the scalability of DSE. 
	We have developed a framework for detecting spurious bugs that our transformation can insert.
	Finally, we show that our transformation can significantly improve the performance of DSE on various benchmark programs and help improve the performance of coverage and bug discovery of large real-world programs.
 \end{abstract}

\maketitle

\section{Introduction}
\label{sec:intro}

The \emph{Dynamic Symbolic Execution (DSE)} is a popular dynamic analysis technique used for software testing and verification~\cite{khurshid-se, klee}.
DSE executes a program using symbolic values instead of concrete ones as input.
All feasible paths in the program can be explored by DSE, with some variables declared as symbolic.
A program path is \emph{feasible} if at least one input exists that will exercise that path.
For each explored path, DSE computes a \emph{path condition} which is essentially the conjunction of branching conditions that are true along the path. 
When DSE reaches a branch where the branching condition is symbolic, it continues the execution on both directions of the branch (\emph{true} and \emph{false}) if they are feasible.
This path condition can be solved using a Satisfiability Modulo Theory (SMT) solver~\cite{open-smt, z3} to find a concrete input which exercises that path.
Because DSE explores all feasible paths in a program, it can be used to find bugs or prove the program works correctly for all possible inputs.

\begin{figure}[t]
\begin{minted}[frame=single, fontsize=\footnotesize, numbersep=1mm]{c}
while ((*hp != '\r') && (*hp != '\n')) {
  if (*hp) {
    if ((*hp >= '0') && (*hp <= '9'))
      hp++;
    else if (*hp == '.')
      hp++;
    else {
      ...
      return OSIP_SYNTAXERROR;
    }
  }
  else {
    ...
    return OSIP_SYNTAXERROR;
  }
}
\end{minted}
\vspace{-1em}
\caption{Symbolic branch within a loop in \code{libosip}.}
\label{lst:example}
\vspace{-1em}
\end{figure}

Unfortunately, DSE suffers from the \textit{path explosion problem}, wherein the number of paths grows exponentially with the number of symbolic branches in the program~\cite{cute, dart}.
Complex control-flow (\ie code with many branches) is the main contributing factor to path explosion. 
For example, consider this code snippet from GNU oSIP-4.0.0 (\ie \code{libosip})~\cite{gnu-osip} (shorten for brevity) with a symbolic branches inside a loop in Figure~\ref{lst:example}.
At every iteration of the loop, DSE needs to explore four potential paths through the program. 
If the loop runs for $N$ iterations, the total number of explored paths amounts to $4^N$.
Out of the various techniques~\cite{dse-survey, anand08, xie16, postconditioned-se, qi13, underconstrained-se} proposed to avoid path explosion problem, \emph{dynamic state merging}~\cite{efficient-state-merging} is considered as one of the foundational approaches.
In DSE engines like KLEE~\cite{klee}, each explored path is associated with a \emph{state}, which maintains the values of all symbolic variables, memory, stack, and registers at that point in the program.
Often, multiple paths in a program share very similar program states.
State merging exploits this observation by merging sufficiently similar states to reduce the number of paths that need to be explored.
Even though state merging can reduce the number of paths that need to be explored significantly, it still requires calling the SMT solver at symbolic branches.
At conditional branches where both the {\em true} and {\em false} paths are feasible, calling the SMT solver to check the path feasibility is an unnecessary overhead in cases where the two forked states get merged.
Therefore, even with dynamic state merging, DSE can still suffer from SMT solver overhead if the program has a lot of symbolic branches.

Prior work suggests that static program transformations can also be used to improve the performance of dynamic test generation~\cite{targeted-transformations,tfuzz}.
In this context of DSE, both semantics-preserving~\cite{perry17} and non-semantics-preserving program transformations~\cite{converse-17} have been proposed to improve its performance.
The root cause of path explosion is the symbolic branches in a program.
If the number of symbolic branches can be reduced, the number of paths that need to be explored will also be reduced.
For example, Collingbourne \etal~\cite{collingbourne11} used aggressive {\em phi-node folding}~\cite{phi-predication} to reduce the number of symbolic branches in image processing applications and showed that it could heavily improve the performance of DSE on programs operating over images.
Compiler optimizations like code hoisting/sinking~\cite{llvm-simplifycfg} or tail merging~\cite{gen-tail-merge-sas03} can also be used to reduce the number of symbolic branches in a program.
Tail merging can eliminate \code{if-then-else} branches if both paths contain identical sequences of operations.
This is done by merging instructions with identical operations and using \code{select} instructions for choosing operands for those merged instructions.
LLVM~\cite{llvm} optimizer contains control-flow graph simplification pass (\code{-simplifycfg}) that can eliminate branches by hoisting or sinking instructions out of \code{if-then-else} or \code {if-then} statements when the compiler can prove it is safe to do so.
In KLEE, the simplification pass is enabled by default to perform these branch elimination optimizations.

Recent developments in compiler transformations such as DARM~\cite{darm} and HyBF~\cite{hybf} have shown how to exploit code similarity within conditional branches to improve the performance and code size.
DARM employs a hierarchical sequence alignment technique to identify isomorphic control-flow regions that contains similar instruction sequences within them.
If two isomorphic regions are similar enough (according to a cost model~\cite{llvm-cost-model}), DARM merges them into a single region.
By changing the alignment models, DARM can be applied to different applications such improving performance for GPUs or reducing code size for CPUs.
DARM provides a flexible way to exploit code similarity at control-flow region level and, it is more general than traditional compiler optimizations such as code sinking/hoisting, tail merging~\cite{gen-tail-merge-sas03} or branch fusion~\cite{branch-fusion} that exploits code similarity only at the basic block level.
However, it is important to note that DARM is not designed to {\em eliminate branches} and increases number of branches and \code{select} instructions in the generated code if applied to conditional branches with non-identical instruction sequences.
This can hurt the performance of DSE as it can increase the number of symbolic branches and increase the complexity of the path constraints due to additional \code{select} instructions inserted.

In this paper, we propose \name, a targeted control-flow transformation designed to remove expensive symbolic branches from a program to improve the performance of DSE.
First, \name uses a static taint analysis to identify expensive symbolic branches to explore in DSE.
Then, it identifies the code similarity between the paths of conditional branches.
Next, \name inserts minimal {\em extra} instructions to \code{if-then} and \code{else} blocks (possibly by adding empty \code{else} blocks first for \code{if-then} statements) to make them look identical in terms of operation sequences.
Finally, \name merges the identical instruction sequence within the \code{if-then} and \code{else} blocks into a single basic block to eliminate the expensive symbolic branch.

As unconditional executions of particular instructions (\eg load/stores) are unsafe, \name transformation is not semantics-preserving. 
This can introduce new bugs that were not present in the original program.
However, \name transformation is {\em failure-preserving}. 
A failure-preserving transformation ensures that any bug in the original program is also present in the transformed one (\ie \name transformation does not erase a bug in the original program).
\name is failure-preserving because the additional instructions inserted into the program do not alter the original dataflow or program memory state.
Any crashing input resulting after the application of failure-preserving transformation can be checked against the original program to verify whether the crash is a true positive.
We use this property to develop a framework to detect false-positive bugs introduced by failure-preserving transformations in \name that are not semantic-preserving.

The main contributions of this paper are as follows:
\begin{itemize}
\item {\bf Targeted Control-Flow Transformation:} We propose \name, a {\em non-semantics-preserving}, but {\em failure-preserving} control-flow transformation designed to remove expensive symbolic branches of a program to improve the performance of DSE.
\item {\bf Detection of False-Positive Crashes:} A framework for detecting false-positives caused by \name transformation in the context of DSE.
\item {\bf Performance Enhancement of DSE:} Evaluation of \name, showing its ability to improve the performance of DSE and its downstream tasks such as {\em code coverage} and {\em bug discovery} on various benchmark programs and real-world applications.
\end{itemize}

The rest of the paper is organized as follows.
In Section~\ref{sec:motivating_example}, we illustrate an example where performing control-flow melding improves the performance of DSE.
We describe all parts of our approach in Section~\ref{sec:design} and present its evaluation in Section~\ref{cfmse:sec:eval}. 
Section~\ref{cfmse:sec:discuss} discusses the limitations of our approach.
We discuss related work in Section~\ref{cfmse:sec:relatedwork} and Section~\ref{sec:conclusion} concludes the paper.

\section{Motivating Example}
\label{sec:motivating_example}

\begin{figure}[ht]
\begin{minted}[linenos=true, frame=single, fontsize=\footnotesize, numbersep=1mm, xleftmargin=3mm]{c}
void to_upper(char *text) {
  for (int i = 0; i < SIZE; i++) {
    if ((text[i] >= 'a') & (text[i] <= 'z')) //symbolic branch
      text[i] = text[i] - 'a' + 'A';
  }
}
void to_upper_branchless(char *text) {
  for (int i = 0; i < SIZE; i++) {
    unsigned is_lower = (text[i] >= 'a') & (text[i] <= 'z');
    unsigned diff = (is_lower == 0) ? 0 : 'a' - 'A';
    text[i] = text[i] - diff;
  }
}
int main() {
  char text[SIZE];
  klee_make_symbolic(&text, sizeof(text), "text");
  to_upper(text);
  for (int i = 0; i < SIZE; i++){
    klee_assert(!((text[i] >= 'a') & (text[i] <= 'z')));
  }
  return 0;
}
\end{minted}
\caption{Motivating example}
\label{fig:motivating_example}
\end{figure}

Our motivating example is the \code{to\_upper} function that converts all elements of a \code{char} array to upper case, and it is shown in Figure~\ref{fig:motivating_example}. 
This figure also shows the driver in \code{main} method to symbolically execute \code{to\_upper} function with a char array of size \code{SIZE} as input.
After executing \code{to\_upper}, the driver also asserts that the output array contains no lowercase characters.
In this case, the scalability of this example is limited since the symbolic execution engine has to fork the execution at every iteration on the symbolic branching condition inside the loop (line 3). 
When this program is run with KLEE\footnote{KLEE-2.3+LLVM-14.0.0} using default settings, it explores 1024 program paths and invokes the SMT solver 21 times for input size 10. 
If constraint caching is disabled, the number of SMT solver invocations increases to 90.

The conditional branch inside the loop can be removed by {\em converting the control-flow into dataflow}.
The transformed function \code{to\_upper\_branchless} is also shown in Figure~\ref{fig:motivating_example}.
The idea here is to compute how each character value must be shifted to convert it to upper case.
If the character is lowercase, the shift value is \code{'a' - 'A'}; otherwise, it is zero.
We compute this value (\ie \code{diff}) conditioned (line 10) on the character being lower case (\ie \code{is\_lower}) (line 9).
Then, we apply the shift to the character value (line 11).
Note that the conditional assignment to \code{diff} is translated into a \code{select} instruction in LLVM-IR.
KLEE converts \code{select} instructions into \code{ite} expressions. 
Therefore, executing the loop does not require SMT solver invocations.
If we run the transformed version in KLEE, it explores only one program path and invokes the SMT solver only 11 times (20 with constraint caching disabled).
This example shows the utility of converting control-flow into dataflow in the context of DSE.
Loops with symbolic conditionals are a common source of scalability issues in DSE.
Targeted compiler transformations can be used to remove such bottlenecks.

Converting control-flow into dataflow is unsafe when computations with side-effects are present under the branch.
In \code{to\_upper} function, store to \code{text[i]} is an operation with a side-effect because it modifies the input array.
Therefore, the compiler cannot sink the store outside the conditional branch.
In transformed code, the store is executed unconditionally, but if the character is uppercase, the stored value is the same as the original value.
Even though this transformation is safe in this example, reasoning about its safety at compile time is not always trivial.

We argue that branch-eliminating transformations help us manage the scalability issues of DSE.
Although applying such transformations to the program can change the semantics, it may lead to better scalability of DSE. 
This can increase coverage of DSE within a limited amount of time and help identify bugs faster.
In the context of DSE, the dynamic state merging has similar goals as branch-eliminating transformations.
However, in some cases, branch-eliminating transformations can act as a better alternative to dynamic state merging since they perform lightweight compile-time state merging and offer portability due to being free of annotation (it requires the user to annotate the parts of program to indicate where the dynamic state merging must be performed). 

\begin{table}[t]
    \centering
    \caption{Comparison of the number of conditional branches and total instructions with and without \name.}
    \resizebox{0.95\textwidth}{!}{%
    \begin{tabular}{l||cc|cc|cc}
		\toprule
		\multirow{2}{*}{{\bf Benchmark}} & 
		\multicolumn{2}{c}{{\bf \#Conditional Branches}} &
		\multicolumn{2}{|c}{{\bf \#Total Instructions}} &
        \multicolumn{2}{|c}{{\bf \#Nesting Levels}} \\
		\cmidrule{2-7}
		& Baseline & Transformed & 
		Baseline & Transformed & 
        Branch Nesting & Control-flow Nesting \\
		\midrule
        toupper              & 2  & 1  & 75   & 73   & 1 & 2 \\
		bitonic sort         & 4  & 4  & 136  & 129  & 1 & 2 \\
		connected components & 14 & 12 & 287  & 273  & 2 & 5 \\
		prim                 & 12 & 7  & 316  & 309  & 1 & 3 \\ 
		merge sort           & 9  & 9  & 261  & 246  & 1 & 2 \\
		transitive closure   & 10 & 9  & 338  & 323  & 1 & 4 \\
		dilation             & 18 & 17 & 399  & 375  & 3 & 7 \\
		detect edges         & 62 & 53 & 1292 & 1261 & 2 & 4 \\
        floyd warshall       & 10 & 9  & 311  & 297  & 1 & 4 \\
		erosion              & 17 & 16 & 378  & 353  & 3 & 7 \\
		dijkstra             & 15 & 14 & 285  & 273  & 1 & 3 \\
		\bottomrule
    \end{tabular}%
    }
    \label{tab:select_instr_label}
    \vspace{-1em}
\end{table}

In the following sections, we describe a compiler transformation called \name that converts control-flow into dataflow to eliminate branches in a program to improve the scalability of DSE.
We further motivate the performance enhancement of DSE with \name by presenting the structural metrics for the small benchmark programs in Table~\ref{tab:select_instr_label} (the same set of benchmark programs used in Section~\ref{cfmse:subsec:accel_se}). For both the original and \name-transformed versions, the table shows the number of conditional branches and total instructions. 
For the original version, the table additionally shows branch nesting depth (considering only \texttt{if-then}/\texttt{if-then-else} constructs) and control-flow nesting depth (considering all control-flow constructs including branches, loops, and \texttt{switch} statements) for the deepest branch \name transformation is applied.
Here, it is noticeable that the number of conditional branches in \name-transformed version is less than or equal to the original one and number of total instructions (LLVM-IR) of the transformed version is always smaller than the original one.
Therefore, performance enhancement of DSE for these programs with \name transformation is expected.

In the next section, we describe \name, a non-semantics preserving transformation that can introduce new bugs to the program, in detail.
Then, we illustrate a system that allows us to filter out the false positives and verify if the bug discovered after \name transformation is indeed a real bug in the original program.

\section{Detailed Design}
\label{sec:design}
In this section, we describe the transformation used by \name (referred to as \name transformation) to statically merge paths to accelerate the DSE of that program. 
We illustrate the extra code insertion phase used to make the instruction sequences within \code{if-then-else} statements identical and the final code generation phase used to eliminate expensive branches for DSE to explore.
Also, we define and provide a proof sketch for the failure-preserving property of the transformation and describe a system for detecting new failures introduced by it.
Finally, we discuss the static analysis used to find the symbolic branches (\ie branch condition depends on a symbolic variable) in the program.

\subsection{\nameheading Transformation}
\label{subsec:transformation}

The \name transformation is based on Control-flow Melding (CFM)~\cite{darm}.
CFM is a compiler optimization that improves the performance of GPU programs by statically merging divergent program paths.
CFM aims to identify \code{if-then-else} branches with similar basic blocks (or isomorphic control-flow regions) and merge the common instructions within those blocks into convergent blocks.
If the operation sequence inside both sides of the branch is identical, then the branch can be eliminated.
However, this scenario is rare in real-world programs.
If the instruction alignment step finds non-identical operations, CFM moves the non-identical portions of the operations into new basic blocks and allows them to execute conditionally.
This process can increase the number of branches in the program making the program less desirable for DSE.

The key idea of \name is to insert extra instructions into the two sides of the conditional branch to make the sequences of operations identical.
Extra instruction is needed when the instruction alignment contains {\em unaligned} instructions.
For the following definitions, consider a program with an \code{if-then-else} branch with two basic blocks $B_t$ and $B_f$ (\ie diamond-shaped control-flow) and $I_t$ and $I_f$ are the instruction sequences of those basic blocks, respectively.

\begin{definition}
	\textbf{Compatible Pair:} Let $(a, b)$ be a pair of instructions. It is a compatible pair, if and if only the operation of $a$ and $b$ are identical (\eg operation of $a$ and $b$ are \code{iadd}). It is not necessary for instructions $a$ and $b$ to have identical operands for the pair $(a, b)$ to be compatible.
\end{definition}

\begin{definition}
  \textbf{Instruction Alignment:} Let $I_t = \{i_1^t \prec \ldots \prec i_n^t\}$ and $I_f = \{i_1^f \prec \ldots \prec i_m^f\}$ be the ordered sequence of instructions in basic blocks $B_t$ and $B_f$ respectively.
  An instruction alignment is an ordered sequence of pairs $A = \{(a_1, b_1) \prec \ldots \prec (a_k, b_k)\}$ such that $a_i \in I_t \cup \{\psi\}$, $b_i \in I_f \cup \{\psi\}$, $k = \max(n, m)$ and, $\forall i \in [1, k], (a_i, b_i) \neq (\psi, \psi) $. 
  If $a_i \neq \psi$ and $b_i \neq \psi$, then $(a_i, b_i)$ is a compatible pair for merging. 
  Here, $\psi$ denotes the absence of an instruction (\ie empty slot).
\end{definition}

\begin{definition}
  \textbf{Unaligned Instruction:} Let $(a_i, b_i) \in A$ be a pair in an instruction alignment $A$ such that $a_i = \psi$ or $b_i = \psi$. 
  Let $i'$ be the valid instruction in the pair $(a_i, b_i)$, then $i'$ is called an unaligned instruction.
\end{definition}

\begin{definition}
  \textbf{Complete Alignment:} An instruction alignment $A'$ is called complete if it does not contain any unaligned instructions.
\end{definition}

If the instruction alignment for $B_t$ and $B_f$ is {\em complete}, we can fully merge $B_t$ and $B_f$ into a single basic block, eliminating the conditional branch.
The first step of \name is to transform the alignment $A$ into a complete alignment $A'$.

\begin{definition}
  \textbf{Extra Instruction:} Let $I_t$ contains an unaligned instruction $i'$. 
  An instruction $i''$ is inserted into $I_f$ such that $i'$ and $i''$ are compatible and form a pair in the new alignment $A'$, then $i''$ is known as the extra instruction.
\end{definition}

Inserting extra instructions is necessary to complete the alignment, allowing us to eliminate the conditional branch. 
But, inserting extra instructions can change the semantics of the program and can introduce new bugs.
However, \name transformation ensures the following conditions to minimize the number of new bugs the transformation introduces.

\begin{enumerate}
  \item Any original instruction in the program does not use a value produced by any extra instruction.
  \item Instructions considered by the transformation can only be a side-effect-free Arithmetic Logic Unit (ALU) instructions or a memory read/write instructions.
  \item An extra memory operation is allowed to read from any memory location, but it is not allowed to overwrite any memory location with a value other than the value immediately read from that location.
\end{enumerate}

The condition \circled{1} ensures that all the original instructions in the program do not use any values produced by a extra instruction.
The value flow in the original program is still preserved after inserting a extra instruction.
The condition \circled{2} states that extra code insertion is not supported for all types of instructions (\ie all opcodes).
For example, if the unaligned instruction is a call instruction, we cannot insert an extra call instruction because the function call can have side effects.
In \name, we only insert extra instructions for ALU and memory read/write operations.
Supported ALU operations include arithmetic, logical, comparison, bitwise, and conversion (\ie casting) operations~\cite{llvm-language-ref}.
The condition \circled{3} allows us to execute load/store operations unconditionally.

\paragraph{{\bf Select minimization:}}
In the code generation process of CFM, extra select operations are inserted if the operands of the two merged instructions do not match.
This process can increase the number of instructions in the program, and extra select operations can make the dataflow more complex.
In DSE, select operations translate to \code{ite} expressions resulting in more complex constraints for the solver to solve.
Therefore, minimizing the number of select instructions generated by \name transformation is essential.
Select operations can be minimized if both sides of the conditional branch have similar def-use chains.
More precisely, let $i_t = op(o_t^1, o_t^2)$ and $i_f = op(o_f^1, o_f^2)$ be two aligned instructions in the alignment $A$. 
Merging $i_t$ and $i_f$ does not require additional select operations if $o_t^1 = o_f^1$ and $o_t^2 = o_f^2$ or $(o_t^1, o_f^1)$ and $(o_t^2, o_f^2)$ are also produced by aligned instructions in the $A$. 

\paragraph{{\bf Setting operands for extra ALU instructions:}}
There is some flexibility in setting operands for extra ALU instructions.
We can set all the operands of the extra instruction to some safe, constant value (\eg $0$) depending on the semantics of the instruction.
On the other hand, we can set operands such that def-use chains are preserved and selects are minimized.
In \name, we do a mix of both.
We preserve the def-use chains and minimize select operations if the extra instructions cannot result in new bugs (such as overflow, underflow, division by zero, or undefined behavior). 
The operand setting process is explained with an example at the end of this section.

\paragraph{{\bf Challenges in merging memory operations:}}
DSE engines such as KLEE experience significant performance overhead if the program contains memory accesses to symbolic addresses~\cite{perry17}.
Merging memory operations can result in symbolic memory accesses even if the two aligned memory operations access concrete addresses individually.
For example, consider two load-aligned instructions $i_t = load(a)$ and $i_f = load(b)$ with a {\em symbolic} branching condition $c$.
If we merge them into a single load, the resulting load will be $i = load(select(c, a, b))$.
Even if $a$ and $b$ are concrete addresses, the resulting load will be symbolic because the address is a function of the branching condition.

\paragraph{{\bf Setting operands for extra load/store instructions:}}
We follow the following criteria when aligning memory operations to avoid creating more symbolic memory accesses.
\begin{enumerate}
  \item If we can statically determine two aligned memory operations accessing the same address, we merge them into a single memory operation. 
  The merged instruction cannot have a symbolic address if the addresses are identical.
  \item If two aligned memory operations access different concrete memory locations.
  We convert them into unaligned memory operations. 
  This essentially linearizes the two memory operations but avoids creating symbolic memory addresses.
  \item If at least one of the two aligned memory operations accesses a symbolic address, we do not apply \name to that branch.  
\end{enumerate}

In Case \circled{2}, linearizing guarded load/store instructions is unsafe and can lead to new program bugs because the guarding condition might protect the memory operation against out-of-bounds access.
Therefore, \name transformation can result in new bugs.
Using the original program, we can identify if the \name transformation introduces a memory out-of-bounds access bug. 
We describe an automated way to do this in Section~\ref{cfmse:subsec:driver}.
If the memory location is valid, then \name transformation does not change the semantics of the program.
This is because no other instruction uses the result of an extra load.
For an extra store, the stored value is the same as the value read from the same location immediately before the store.
This can be achieved by inserting an extra load right before the extra store.

\begin{figure*}[ht]
  \begin{subfigure}[b]{0.24\textwidth}
  \begin{minted}[linenos, fontsize=\footnotesize, numbersep=1mm]{c}
  // ...
  if ((text[i] >= 'a') 
    & (text[i] <= 'z')) {
    t1 = text[i] - 'a';
    t2 = t1 + 'A';
    t3 = text[i];
    text[i] = t2;
  } else {
  }
  \end{minted}
  \caption{}
  \label{fig:codegen_step1}
  \end{subfigure}
  \hfill
  \begin{subfigure}[b]{0.24\textwidth}
  \begin{minted}[linenos, fontsize=\footnotesize,  numbersep=1mm]{c}
  // ...
  if ((text[i] >= 'a') 
    & (text[i] <= 'z')) {
    t1 = text[i];
    t2 = t1 - 'a';
    t3 = t2 + 'A';
    t4 = text[i];
    text[i] = t3;
  } else {
    t5 = text[i];
    t6 = t5 - 0;
    t7 = t6 + 0;
    t8 = text[i];
    text[i] = t8;
  }
  \end{minted}
  \caption{}
  \label{fig:codegen_step2}
  \end{subfigure}
  \hfill
  \begin{subfigure}[b]{0.4\textwidth}
  \begin{minted}[linenos, fontsize=\footnotesize,  numbersep=1mm]{c}
  // ...
  unsigned is_lower = 
    (text[i] >= 'a') & (text[i] <= 'z');
  t1_t5 =  text[i];
  // select
  s1    = is_lower == 0 ? 0 : 'a'; 
  t2_t6 = t1_t5 - s1;
  s2    = is_lower == 0 ? 0 : 'A'; 
  // select
  t3_t7 = t2_t6 + s2;
  t4_t8 = text[i];
  // select
  s3    = is_lower == 0 ? t4_t8 : t3_t7; 
  text[i] = s3;
  \end{minted}
  \caption{}
  \label{fig:codegen_step3}
  \end{subfigure}
\caption{\name transformation example}
\label{fig:codegen}
\end{figure*}

\paragraph{\textbf{Example:}} 
Now we explain how \name transformation works in action using our running example (Figure~\ref{fig:motivating_example}).
Figure~\ref{fig:codegen} shows how \code{to\_upper} function is transformed at each stage.
Figure~\ref{fig:codegen_step1} shows \code{to\_upper} function with an empty \code{else} section inserted. 
This is an extra canonicalization step of \name that converts \code{if-then} to \code{if-then-else} form, which allows it to merge \code{if-then} branches.
Also, instructions are shown on separate lines (Lines 4-7) for better readability.
Figure~\ref{fig:codegen_step2} shows the code after extra code insertion. 
Here, the \code{else} path is empty; therefore, all the instructions are unaligned.
The \code{else} path contains inserted extra instructions.
For example, the load operation in Line 4 is repeated in Line 10 after the extra code insertion.
\name also tries to preserve def-use chains and minimize select operations required for merging.
For example, instruction \code{t7} at Line 12 uses instruction \code{t6} at Line 11.
This is similar to instruction \code{t3} using \code{t2} as its first operand.
\code{t7} uses \code{0} as its second operand to avoid any overflow/underflow bugs.
This example also demonstrates how store instructions are handled during extra code insertion.
On \code{if} path, there is a store (Line 8) of value \code{t3} to \code{text[i]}.
On \code{else} path, the same store is performed (Line 14), but the stored value is \code{text[i]} (\ie \code{t8}).
This requires loading \code{text[i]} before the store (Line 13). 
Because of this, we insert a redundant load to the \code{if} path (Line 7) to complete the alignment.
Figure~\ref{fig:codegen_step3} shows the code after the merging step.
Extra select instructions (shown as \code{C} ternary operator) are inserted to select operands if input operands do not match.
In this example, executing all the memory operations unconditionally is safe. 
Therefore, the transformation is semantics-preserving.
The transformed program is much faster to symbolically execute in KLEE than the original one (Section~\ref{sec:motivating_example}).

\begin{algorithm}[t]
  \SetAlgoLined
  \SetKw{Continue}{continue}
  \SetKw{False}{false}
  \SetKw{True}{true}
  \SetKw{Is}{is}
  \KwIn{Original Program $P$, False Positive Locations $L$}
  \KwOut{Transformed Program $P'$}
  \caption{\name Transformation Algorithm}
  \label{alg:cfmse}

   \For{F $\in$ functions(P)}{
   	Changed $\gets$ \False\;
   	\Repeat{Changed \Is \False}{
   		ValidBranches $\gets$ collectValidBranches(F, L)\;
	   	\For{BI $\in$ ValidBranches}{
	   		BL, BR $\gets$ getDiamondBlocks(BI)\;
	   		A $\gets$ computeAlignment(BL, BR)\;
	   		\If{canCompleteAlignment(A, BL, BR) \Is \False}{
	   			\Continue
	   		}
	   		BL', BR' $\gets$ insertExtraInsts(A, BL, BR)\;
	   		mergeBlocks(BL', BR')\;
	   		Changed $\gets$ \True\;
	   		
	   		\If{Changed \Is \True}{
	   			simplify(F)\;
	   		}
	   	}
   	}
   }

\end{algorithm}

\paragraph{\textbf{Putting all together:}} 
We describe the overall \name transformation for a program in Algorithm~\ref{alg:cfmse}.
\name transformation iterates through all the functions in a program.
For each function, it collects all the valid branches for applying \name transformation.
In this step, it filters out branches that can lead to new bugs if they are previously identified as false positive locations (Section~\ref{cfmse:subsec:driver}).
Then for each valid branch, it computes the instruction alignment for the two sides of the branch.
If the alignment can be made complete, it inserts extra instructions to complete the alignment.
Finally, it merges the two blocks and simplify the function removing the branch.
We repeatedly perform these steps until there are no changes to the function. 

\begin{definition}
	{\bf Failure Preservation Property:} Let the program $P$ transformed into program $P'$. If all failures observerable in $P$ are also observed in $P'$ then, the transformation $P \rightarrow P'$ is failure-preserving.
\end{definition}

\paragraph{\textbf{Properties of \name Transformation:}}
Consider a single application of \name transformation to a branch in program $P$. 
Let program $P^d$ is obtained after inserting extra instructions to $P$, and $P'$ is the final result of the \name transformation after merging the instructions.
Overall, the transformation is $P \rightarrow P^d \rightarrow P'$.
A failure in $P^d$ can occur at an original instruction or an extra instruction.
If a failure occurs at an original instruction, it is also present at $P$.
Any failure in $P^d$, whether at an original or extra instruction, is present in $P'$ since it was obtained by merging aligned instructions in $P^d$.
Any value produced by an extra instruction in $P^d$ is not used by original instructions come from $P$.
Also, values of the extra instructions cannot flow outside the merged branch because the $\phi$-nodes filter them out at the end.
Any extra store operation inserted into $P^d$ does not mutate the memory space because the extra store operation writes the value that has been immediately read from that memory location.
Therefore, any additional program bug introduced in $P \rightarrow P^d$ must be realized within the branch merged by \name. 
These properties of \name transformation allow us to detect and filter out false positive bugs introduced by \name.

\paragraph{\textbf{Proof Sketch for Preservation of Failure:}}
We make the case that failures in the original program $P$ are preserved after the \name transformation by arguing that every state of the original program is part of the corresponding states in the program $P^d$ we get after insertion of extra instructions.
This proof sketch is specific to the nature of LLVM IR.
Nevertheless, that is not a fundamental limitation.
We model the state of a program represented in LLVM IR as a map of registers and memory locations to the values they hold (\ie state map). 
The number of registers can be infinite since the LLVM IR is in Static Single Assignment (SSA) form.
First, consider the most straightforward program where we can apply \name transformation.
Let $P$ be an \code{if-then-else} region with diamond control-flow with basic blocks $B_t$ and $B_f$ and instruction sequences $I_t$ and $I_f$, respectively. 
As described previously, we aligned the instructions of $P$ in the program $P^d$ and added appropriate extra instructions to achieve complete alignment.
Let $B^d_t$ and $B^d_f$ be the basic blocks and $I^d_t$ and $I^d_f$ be the corresponding instruction sequences in the program $P^d$ (Note that $I^d_t$ and $I^d_f$ are aligned perfectly).
Based on the branch condition, we will execute instructions on either of the basic blocks.
Without the loss of generality, when executing the instructions in $I^d_t$, we will encounter two types of instructions: original and extra.
An original instruction is the one present in $I_t$. 
If this is an ALU instruction $i = op(o^1_t, o^2_t)$, the operands $o^1_t$ and $o^2_t$ are either constants or registers contain values not produced by a prior extra instruction and $i$ is a new register (\ie SSA).
Executing this instruction causes a state change, which adds the new register $i$ to the state map.
On the other hand, if this ALU instruction is extra, $i$ will never be read by any original instructions.
If we assume $I_t$ contains {\em only} ALU instructions, we can make the following observations:
\begin{itemize}
	\item Any state occurs while executing $I^d_t$, the registers that hold the values produced by extra instructions are {\em separable} (\ie if these registers are removed from a state map, we will end up with a state map when executing $I_t$).
	\item All state transitions when executing $I_t$ will also be present when executing $I^d_t$.
	\item There will be additional state transitions due to extra instructions when executing $I^d_t$.
	\item These additional state transitions will never change any register that is part of the dataflow of the original instructions.
\end{itemize}
Therefore, in this simple case, we can reason that a failure during state transition in the original program $P$ will also be there in $P^d$.
We can extend this reasoning to memory operations when they appear in $I^d_t$.
All the loads and stores in $I_t$ also appear in $I^d_t$. 
An extra load instruction can read from any memory address, which does not change the memory part of the state map.
However, the value read is in a register and is never accessed by original instructions.
An extra store instruction always stores the value that has been immediately read from the memory location.
Hence, the extra store alone does not cause a state transition.
We can conclude that the same observations before hold in the presence of memory instructions as well.
In summary, the effect of extra instructions on the state is {\em separable}, and the original state transitions are preserved in $P^d$.
Therefore, a failure that occurs during a state transition in the original program is preserved in the program with extra instructions.
We further argue that since $P^d$ has complete alignment and merging instructions with select statement is semantics preserving, the application of \name transformation to a single branch is {\em failure-preserving}.
The \name transformation is monotonic in terms of instructions and state transitions (while the state being {\em separable}), and we can use this fact to argue that {\em repeated} application of \name transformation preserves the failures in the original program.

\subsection{False Positive Detection}
\label{cfmse:subsec:driver}


Assume the program we are testing using DSE is $P$. 
Let $P'$ be the program after \name transformation.
Let $\alpha_{crash}$ be a program input that causes a crash in $P'$. 
We have a false positive failure if executing the same input on $P$ does not cause a crash.
As discussed in Section~\ref{subsec:transformation}, \name can introduce new failures in the program, and it is essential to detect them.
These additional failures inserted by \name transformation must be realized within the code region where the transformation is applied.
In other words, the extra instructions inserted by \name are not used by any original instructions. 
Therefore, their values cannot flow outside the merged branch, and a false positive bug added by \name must materialize at an instruction produced by the transformation.
\begin{wrapfigure}{r}{0.4\textwidth}
  \vspace{-1em}
  \begin{center}
  	\includegraphics[width=0.37\textwidth]{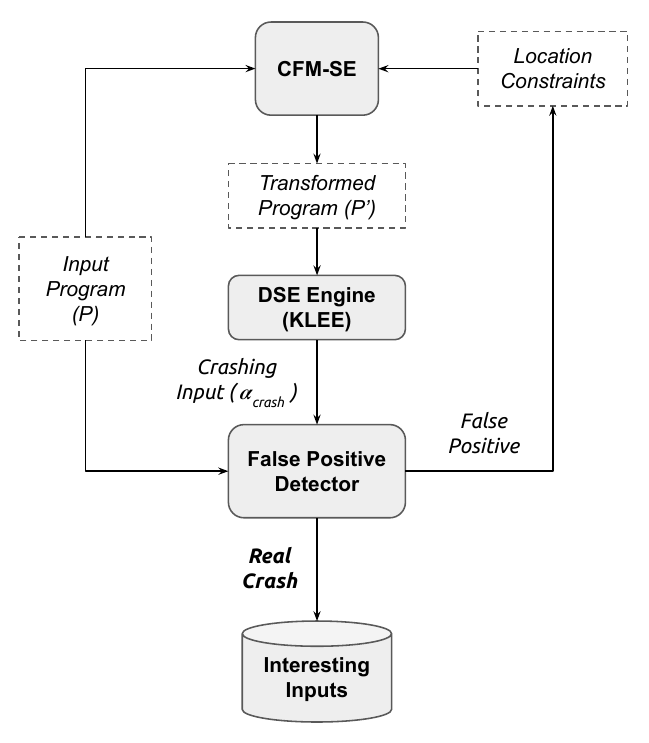}
  \end{center}
  \vspace{-1.5em}
  \caption{Driver for detecting false positives.}
  \label{fig:driver}
  \vspace{-1em}
\end{wrapfigure}
An example would be a memory out-of-bounds access caused by unconditionally executing memory instructions. 
This failure will be realized at the program location of this memory access.
We can find this program location and avoid applying \name to that location and re-execute the program symbolically.
This will prevent that specific false positive failure from occurring again.
The false positive detection and re-execution driver is shown in Figure~\ref{fig:driver}.
The driver starts by symbolically executing the \name transformed program $P'$ in KLEE.
If a crashing input ($\alpha_{crash}$) is detected during symbolic execution, execution is stopped, and the driver checks if $\alpha_{crash}$ is valid.
This can be done by re-executing the original $P$ with $\alpha_{crash}$. 
The $\alpha_{crash}$ is collected as an interesting input if it is a real crash.
Otherwise, the driver obtains the program location where the crash occurred and updates location constraints for the \name transformation (\ie transformation is not applied to that location). 
Program $P$ is recompiled with updated location information, and the driver loop continues.
\newpage

\subsection{Symbolic Variable Analysis}
\label{sec:symvar_analysis}

\begin{figure}[ht]
\begin{minted}[linenos, frame=single, fontsize=\footnotesize, numbersep=1mm, xleftmargin=3mm]{c}
int foo(int x, int y) {
  if (x > y) return x;
  if (y > 0) return y;
  return x + y;
}
int main() {
  int a;
  klee_make_symbolic(&a, sizeof(a), "a");
  ...
  int p = foo(a, 10);
  ...
  int q = foo(-5, a);
  ...
}
\end{minted}
  \caption{Symbolic variable analysis example}
  \label{fig:symvar_example}
\end{figure}

 
In DSE, if the outcome of a conditional branch depends on a symbolic variable, it is called a {\em symbolic branch}.
If both outcomes of a symbolic branch ({\em true} and {\em false}) are feasible, the DSE has to fork the execution and explore both outcomes of the branch.
To minimize the number of paths explored by DSE, we need to apply the \name transformation only at expensive symbolic branches for DSE to explore.
This requires identifying symbolic branches in the program at compile time.
Our symbolic variable analysis is based on {\em Divergence Analysis}~\cite{llvm-div-analysis}.
Divergence analysis is a dataflow analysis used to identify divergent variables in GPGPU programs (or programs that use Single Instruction Multiple Data (SIMD) instructions).
A program variable is marked as {\em divergent} if its value is different across different threads (or SIMD lanes) of a GPGPU kernel.
Divergence analysis tracks data and control dependences across registers to identify divergent variables. 
This analysis cannot be adapted directly to identify symbolic variables because it is intra-procedural.
But symbolic values flow across function boundaries. 
Next, we describe how we address this limitation in designing an inter-procedural symbolic variable analysis.

The first step of symbolic variable analysis is identifying symbolic sources. 
Symbolic sources refer to variables explicitly marked as symbolic by the user.  
For example, in KLEE~\cite{klee}, a variable can be marked as symbolic using the call to function \code{klee\_make\_symbolic}. 
Any user arguments to the program are also considered symbolic sources (\ie the arguments to the \code{main} function).
Similar to divergence analysis, if we use an intra-procedural dataflow analysis after marking symbolic sources, it will only mark a subset of truly symbolic instructions in the program.
This is because the symbolic property of a variable is not propagated to callees from a call site with symbolic arguments.

Consider the example program in Figure~\ref{fig:symvar_example}.
An intra-procedural symbolic variable analysis will identify the variable \code{a} in the function \code{main} as a symbolic source first and mark the call sites at Lines 10 and 12 as symbolic due to their data dependence on \code{a}.
However, none of the instructions in the function \code{foo} will be marked as symbolic because function foo does not have any explicit symbolic sources, and symbolic variables are not propagated to callees at their call sites.
We address this limitation by updating the symbolic sources of the callees at each call site and re-processing the callee if the symbolic sources change.

First, we mark the symbolic sources for all the functions in a program and insert each function into a work list.
Then, we process each function in the work list and propagate the symbolic property to other instructions within the function based on data- or sync-dependences (\ie similar to divergence analysis).
Once the function is processed, we check each call site. 
If the call site is marked as symbolic, at least one function argument is symbolic. 
If an argument is not marked symbolic, we mark it symbolic and insert the function into the work list for re-processing.
This analysis is inter-procedural but context-insensitive. 
A context-sensitive analysis is unnecessary because we only want to apply the \name transformation at symbolic branches at compile-time.
In the example program (Figure~\ref{fig:symvar_example}), processing the function main initially will mark call sites at Lines 10 and 12 as symbolic. 
Then, re-processing of function \code{foo} will mark all operations in that function as symbolic (Lines 2, 4 because of input argument \code{x} and Line 3 because of argument \code{y}).
Note that for functions with a variable number of arguments ({\em variadic functions})~\cite{variadic-func}, we mark all accesses to the variable arguments as symbolic if at least one of them is found to be symbolic at a call site.

\section{Evaluation}
\label{cfmse:sec:eval}

First, we discuss the implementation and the experimental setup to evaluate our approach.
Then, we discuss the three major research questions that were answered by the results of our evaluation.

\paragraph{{\bf Experimental setup.}} 
We implemented the \name as an LLVM-IR (LLVM-14.0.0) transformation pass .
\name transformation also keeps track of the source line number information of the merged instructions. 
For example, if two instructions $i_1$ and $i_2$ with source line numbers $l_1$ and $l_2$ are merged, the merged instruction $i_m$ will get attached additional debug information that contains the source line numbers $l_1$ and $l_2$. 
This helps us to compute the line coverage of the transformed program.
We refer to this as \emph{merged line coverage}.
We ran all our experiments on a system with a Intel(R) Xeon(R) Gold 6430 (64 cores) processor and 256 GB of memory running Ubuntu 20.04.1 LTS.
We used KLEE DSE engine (KLEE-2.3) that can choose to run \name pass as a pre-processing step (\ie expose \name pass to KLEE) to the symbolic execution. 
Note that KLEE usually runs a selected set of LLVM passes (\eg \code{mem2reg}) on the program to make it favorable for symbolically execute its IR instructions and \name transformation is exposed to this pipeline.

\paragraph{{\bf Objectives of our evaluation.}}
We evaluated the utility of \name transformation by designing our experiments to answer the following questions:
\begin{itemize}
    \item \textbf{[RQ1] Performance of DSE:} How effective is the branch elimination of \name transformation in reducing the number of solver calls and mitigating the path explosion problem?
    \item \textbf{[RQ2] Performance of Coverage:} Can a program transformed with \name achieve higher line coverage within a given time budget?
    \item \textbf{[RQ3] Performance of Bug Discovery:} Can a program be transformed with \name to achieve faster bug discovery?
\end{itemize}
\begin{table}[t]
  \centering
  \caption{Description of the benchmarks for RQ1}
  \vspace{-1em}
  \begin{tabular}{p{0.2\linewidth} p{0.7\linewidth}}
  \toprule
  \multicolumn{1}{c}{\textbf{Benchmark}}                          & \multicolumn{1}{c}{\textbf{Description}} \\ \midrule
  toupper & Converts lowercase to uppercase in a fixed length \code{char} array.                                       \\
  \cmidrule{1-2}
  bitonic sort  & Use bitonic mergesort to sort an \code{int} array~\cite{bitonic}.                                          \\
  \cmidrule{1-2}
  connected \newline components & Computes the number of connected components using Shiloach-Vishkin algorithm~\cite{shiloach82} in a graph represented as an adjacency matrix.                                          \\
  \cmidrule{1-2}
  prim  & Finds a minimum spanning tree for a weighted undirected graph.                                     \\
  \cmidrule{1-2}
  merge sort  & Recursive top-down merge sort for sorting an \code{int} array.                                         \\
  \cmidrule{1-2}
  transitive \newline closure & Computes the reachability from vertex $i$ to vertex $j$ for all vertex pairs $(i,j)$ in a directed graph.                                         \\
  \cmidrule{1-2}
  dilation & Applies morphological dilation to a binary image over a $3 \times 3$ neighborhood~\cite{Phillips-2008}.                                          \\
  \cmidrule{1-2}
  detect edges  & Applies the Sobel-Feldman operator~\cite{sobel1968} to an input image to for edge detection.                                          \\
  \cmidrule{1-2}
  floyd \newline warshall  & Finds the shortest paths between all pairs of vertices in a weighted directed graph~\cite{CLRS}.                                    \\
  \cmidrule{1-2}
  erosion  & Applies morphological erosion to a binary image over a $3 \times 3$ neighborhood~\cite{Phillips-2008}.                                            \\
  \cmidrule{1-2}
  dijkstra  & Finds the shortest distance between two nodes in a weighted graph.
                \\ \bottomrule
  \end{tabular}

  \label{tab:bench_desc}
  \vspace{-1em}
\end{table}
For RQ1, we used a benchmark suite of 11 programs consisting of well-known graph algorithms, sorting algorithms, image processing algorithms, and our motivating example (\ie \code{toupper}) from Section~\ref{sec:motivating_example}.
The functionality of each benchmark and the nature of their inputs are described in Table~\ref{tab:bench_desc}.
We selected these benchmarks because they are small enough so that KLEE can enumerate all feasible execution paths for sufficiently smaller input sizes within a reasonable time. 
This allows us to evaluate how effectively different techniques can mitigate the path explosion problem compared to vanilla KLEE.
\newpage

\subsection{Performance of DSE (RQ1)}
\label{cfmse:subsec:accel_se}
\begin{table*}[]
	\caption{KLEE symbolic execution statistics for the approaches \methodK, \methodSM, \methodC, and \methodCSM.
		(OOT = Out of time)}
	\resizebox{\textwidth}{!}{
		\begin{tabular}{@{}llllllllllllllllll@{}}
			\toprule
			\multicolumn{1}{c}{\multirow{2}{*}{\textbf{Benchmark}}}                         & \multicolumn{1}{c}{\multirow{2}{*}{\textbf{\begin{tabular}[c]{@{}c@{}}Input\\ Size\end{tabular}}}} & \multicolumn{4}{c}{\textbf{Time(s)}} & \multicolumn{4}{c}{\textbf{Number of Queries}} & \multicolumn{4}{c}{\textbf{Average Query Size}} & \multicolumn{4}{c}{\textbf{Explored Paths}}                                                                                                                                                                                                                         \\
			\cmidrule(l){3-6} \cmidrule(l){7-10} \cmidrule(l){11-14} \cmidrule(l){15-18}
			\multicolumn{1}{c}{}                                                            & \multicolumn{1}{c}{}                                                                               & \textbf{K}                           & \textbf{SM}                                    & \textbf{C}                                      & \textbf{C-SM}                               & \textbf{K}       & \textbf{SM}      & \textbf{C}       & \textbf{C-SM}    & \textbf{K} & \textbf{SM}      & \textbf{C} & \textbf{C-SM}    & \textbf{K}       & \textbf{SM}      & \textbf{C}       & \textbf{C-SM}    \\ \midrule
			\multirow{3}{*}{toupper}                                                        & 10                                                                                                 & 0.2                                  & 0.13                                           & 0.00                                            & 0.00                                        & 11               & 11               & 0                & 0                & 11         & 11               & 0          & 0                & 1.02$\times10^3$ & 11               & 1                & 1                \\
			                                                                                & 50                                                                                                 & OOT                                  & 0.69                                           & 0.00                                            & 0.00                                        & 27               & 51               & 0                & 0                & 11         & 11               & 0          & 0                & 1.05$\times10^7$ & 51               & 1                & 1                \\
			                                                                                & 100                                                                                                & OOT                                  & 1.51                                           & 0.00                                            & 0.00                                        & 26               & 101              & 0                & 0                & 11         & 11               & 0          & 0                & 1.03$\times10^7$ & 101              & 1                & 1                \\ \midrule

			\multirow{3}{*}{bitonic sort}                                                   & 4                                                                                                  & 0.58                                 & 0.12                                           & 0.00                                            & 0.00                                        & 37               & 7                & 0                & 0                & 103        & 145              & 0          & 0                & 28               & 7                & 1                & 1                \\
			                                                                                & 8                                                                                                  & OOT                                  & 1.81                                           & 0.00                                            & 0.00                                        & 9.68$\times10^4$ & 25               & 0                & 0                & 406        & 1.23$\times10^3$ & 0          & 0                & 5.90$\times10^4$ & 25               & 1                & 1                \\
			                                                                                & 16                                                                                                 & OOT                                  & 34.03                                          & 0.00                                            & 0.00                                        & 9.01$\times10^4$ & 81               & 0                & 0                & 400        & 9.34$\times10^3$ & 0          & 0                & 1.09$\times10^5$ & 81               & 1                & 1                \\ \midrule

			\multirow{3}{*}{\begin{tabular}[c]{@{}l@{}}connected\\ components\end{tabular}} & 3                                                                                                  & 0.11                                 & 0.17                                           & 0.07                                            & 0.07                                        & 10               & 44               & 12               & 12               & 4          & 145              & 151        & 151              & 512              & 29               & 3                & 3                \\
			                                                                                & 4                                                                                                  & 77.94                                & 12.39                                          & 0.12                                            & 0.11                                        & 17               & 118              & 20               & 20               & 4          & 8.96$\times10^3$ & 434        & 434              & 6.55$\times10^4$ & 71               & 4                & 4                \\
			                                                                                & 5                                                                                                  & OOT                                  & OOT                                            & 0.30                                            & 0.29                                        & 26               & 122              & 30               & 30               & 4          & 1.51$\times10^6$ & 955        & 955              & 1.78$\times10^7$ & 96               & 5                & 5                \\ \midrule

			\multirow{3}{*}{prim}                                                           & 4                                                                                                  & 20.35                                & 1.34                                           & 0.08                                            & 0.09                                        & 386              & 69               & 28               & 28               & 217        & 899              & 118        & 118              & 5.34$\times10^4$ & 37               & 1                & 1                \\
			                                                                                & 5                                                                                                  & OOT                                  & 6.74                                           & 0.16                                            & 0.15                                        & 6.84$\times10^3$ & 121              & 45               & 45               & 316        & 2.60$\times10^3$ & 228        & 228              & 3.33$\times10^6$ & 69               & 1                & 1                \\
			                                                                                & 6                                                                                                  & OOT                                  & 30.09                                          & 1.76                                            & 1.67                                        & 9.07$\times10^3$ & 187              & 66               & 66               & 288        & 6.04$\times10^3$ & 370        & 370              & 3.30$\times10^6$ & 111              & 1                & 1                \\ \midrule

			\multirow{3}{*}{merge sort}                                                     & 5                                                                                                  & 1.94                                 & 9.96                                           & 1.85                                            & 10.08                                       & 120              & 568              & 120              & 568              & 146        & 1.12$\times10^3$ & 146        & 1.12$\times10^3$ & 120              & 119              & 120              & 119              \\
			                                                                                & 10                                                                                                 & OOT                                  & OOT                                            & OOT                                             & OOT                                         & 1.01$\times10^5$ & 8.16$\times10^4$ & 1.03$\times10^5$ & 8.18$\times10^4$ & 385        & 2.46$\times10^3$ & 382        & 2.47$\times10^3$ & 1.09$\times10^5$ & 2.14$\times10^4$ & 1.04$\times10^5$ & 2.15$\times10^4$ \\
			                                                                                & 15                                                                                                 & OOT                                  & OOT                                            & OOT                                             & OOT                                         & 4.40$\times10^4$ & 9.91$\times10^4$ & 4.40$\times10^4$ & 9.88$\times10^4$ & 353        & 1.22$\times10^3$ & 355        & 1.22$\times10^3$ & 1.71$\times10^6$ & 4.36$\times10^4$ & 1.63$\times10^6$ & 4.40$\times10^4$ \\ \midrule

			\multirow{3}{*}{\begin{tabular}[c]{@{}l@{}}transitive \\ closure\end{tabular}}  & 3                                                                                                  & 3.47                                 & 0.39                                           & 0.00                                            & 0.00                                        & 772              & 27               & 0                & 0                & 164        & 913              & 0          & 0                & 49               & 24               & 1                & 1                \\
			                                                                                & 4                                                                                                  & 598.89                               & 1.33                                           & 0.00                                            & 0.00                                        & 7.51$\times10^4$ & 64               & 0                & 0                & 309        & 4.24$\times10^3$ & 0          & 0                & 2.04$\times10^3$ & 60               & 1                & 1                \\
			                                                                                & 5                                                                                                  & OOT                                  & 4.42                                           & 0.00                                            & 0.00                                        & 1.32$\times10^5$ & 125              & 0                & 0                & 386        & 1.36$\times10^4$ & 0          & 0                & 1.15$\times10^5$ & 120              & 1                & 1                \\ \midrule

			\multirow{3}{*}{dilation}                                                       & 4                                                                                                  & 0.47                                 & 0.39                                           & 0.27                                            & 0.31                                        & 42               & 28               & 28               & 24               & 40         & 59               & 14         & 15               & 81               & 9                & 16               & 5                \\
			                                                                                & 5                                                                                                  & 6.62                                 & 0.82                                           & 0.66                                            & 0.56                                        & 240              & 57               & 52               & 43               & 130        & 129              & 15         & 15               & 1.15$\times10^4$ & 19               & 512              & 10               \\
			                                                                                & 6                                                                                                  & OOT                                  & 1.57                                           & 36.06                                           & 0.89                                        & 1.01$\times10^3$ & 98               & 84               & 68               & 161        & 221              & 15         & 16               & 3.78$\times10^6$ & 33               & 6.55$\times10^4$ & 17               \\ \midrule

			\multirow{3}{*}{detect edges}                                                   & 3                                                                                                  & OOT                                  & 6.08                                           & 0.01                                            & 0.01                                        & 26               & 12               & 2                & 2                & 323        & 561              & 8          & 8                & 23               & 19               & 2                & 2                \\
			                                                                                & 4                                                                                                  & OOT                                  & 24.63                                          & 0.01                                            & 0.01                                        & 18               & 39               & 2                & 2                & 281        & 2.82$\times10^3$ & 8          & 8                & 16               & 50               & 2                & 2                \\
			                                                                                & 5                                                                                                  & OOT                                  & 56.03                                          & 0.01                                            & 0.01                                        & 20               & 84               & 2                & 2                & 293        & 6.66$\times10^3$ & 8          & 8                & 17               & 99               & 2                & 2                \\ \midrule

			\multirow{3}{*}{\begin{tabular}[c]{@{}l@{}}floyd \\ warshall\end{tabular}}      & 3                                                                                                  & OOT                                  & 0.60                                           & 0.00                                            & 0.00                                        & 1.98$\times10^3$ & 27               & 0                & 0                & 329        & 789              & 0          & 0                & 1.62$\times10^3$ & 23               & 1                & 1                \\
			                                                                                & 4                                                                                                  & OOT                                  & 3.41                                           & 0.00                                            & 0.00                                        & 6.75$\times10^4$ & 64               & 0                & 0                & 497        & 3.92$\times10^3$ & 0          & 0                & 5.34$\times10^4$ & 58               & 1                & 1                \\
			                                                                                & 5                                                                                                  & OOT                                  & 16.24                                          & 0.00                                            & 0.00                                        & 6.56$\times10^4$ & 125              & 0                & 0                & 505        & 1.32$\times10^4$ & 0          & 0                & 6.55$\times10^4$ & 117              & 1                & 1                \\ \midrule

			\multirow{3}{*}{erosion}                                                        & 4                                                                                                  & 2.01                                 & 0.44                                           & 0.23                                            & 0.27                                        & 130              & 32               & 28               & 24               & 255        & 172              & 15         & 15               & 59               & 9                & 16               & 5                \\
			                                                                                & 5                                                                                                  & 187.19                               & 1.05                                           & 0.50                                            & 0.46                                        & 7.35$\times10^3$ & 61               & 52               & 43               & 598        & 401              & 15         & 16               & 3.52$\times10^3$ & 19               & 512              & 10               \\
			                                                                                & 6                                                                                                  & OOT                                  & 2.35                                           & 38.61                                           & 0.76                                        & 8.54$\times10^4$ & 100              & 84               & 68               & 861        & 885              & 15         & 16               & 9.30$\times10^4$ & 33               & 6.55$\times10^4$ & 17               \\ \midrule

			\multirow{3}{*}{dijkstra}                                                       & 4                                                                                                  & 11.87                                & 3.06                                           & 0.61                                            & 0.62                                        & 774              & 95               & 46               & 46               & 281        & 754              & 178        & 178              & 368              & 33               & 1                & 1                \\
			                                                                                & 5                                                                                                  & OOT                                  & 13.85                                          & 2.17                                            & 2.07                                        & 9.81$\times10^4$ & 167              & 73               & 73               & 545        & 2244             & 313        & 313              & 5.43$\times10^4$ & 67               & 1                & 1                \\
			                                                                                & 6                                                                                                  & OOT                                  & 74.51                                          & 6.51                                            & 6.5                                         & 1.01$\times10^5$ & 260              & 106              & 106              & 430        & 5483             & 482        & 482              & 1.31$\times10^5$ & 113              & 1                & 1                \\ \bottomrule
		\end{tabular}}
	\label{tab:results_without_verify}

\end{table*}
We execute the 11 programs with symbolic inputs of different sizes for RQ1.
For comparison, we consider the following approaches:
\begin{itemize}
    \item \textbf{Vanilla KLEE (K):} KLEE with default optimization settings.
    \item \textbf{KLEE with State Merging (SM) :} KLEE with state merging enabled using its built-in state merging mechanism~\cite{klee}. 
    \item \textbf{KLEE with \nameheading (C):} KLEE with \name transformation enabled.
    \item \textbf{KLEE with \nameheading and State Merging (C-SM):} KLEE with \name transformation and state merging enabled. 
\end{itemize}
Because \name (\textbf{C}) is a compile-time technique, its applicability is limited compared to dynamic state merging (\textbf{SM}).
For some programs, applying both techniques is more beneficial (\textbf{C-SM}). 
\textbf{C-SM} has the benefits of both \name and state merging (\ie It can eliminate branches at compile-time and merge states at runtime).
We use STP solver~\cite{stp} as the solver backend in KLEE. For each KLEE execution, we use a time budget of 1 hour (\code{--max-time=3600s}) and a memory budget of 50 GBs (\code{--max-memory=51200}). 
We also used the option \code{--only-output-states-covering-new} which makes KLEE only output the states that cover new code paths.
For each run, we collect the time KLEE takes to explore all possible program paths (or timeout), the number of solver calls, the average solver query size, and the number of explored program paths.
To reduce the noise in time measurements, we repeated each experiment 5 times and reported the median.
We measured the standard deviation of these timings and, they are negligible compared to the run time of DSE.

Table~\ref{tab:results_without_verify} shows the results for RQ1.
For 8 out of the 11 benchmarks considered, \methodC outperforms both \methodK and \methodSM regarding the time taken to explore all feasible paths.
The gains come from the reduction in solver calls and the number of program paths DSE has to explore.
For these 8 benchmarks, \methodCSM behaves similarly to \methodC because most states are merged at compile time.
The only case where \methodC runs out of time is for the merge sort benchmark where \name does not apply due to memory operations with symbolic addresses.
For this benchmark, applying state merging makes the performance worse because it significantly increases the number of solver calls due to the symbolic memory addresses.
The other two exceptions are dilation and erosion benchmarks.
These two benchmarks contain symbolic branches that cannot be eliminated by \name (\eg \code{if-then} containing a loop inside).
Therefore, state merging has more opportunities to merge states, resulting in better performance than \methodC.
However, combining state merging with \name (\methodCSM) results in the best performance for these two benchmarks.
This shows that compile-time branch elimination and run-time state merging can complement each other, and combining them can help better manage path explosion.
Dynamic state merging can merge any random pair of states at runtime and \name (\ie static state merging) merges paths in a regular manner.
Therefore, performing dynamic state merging on the symbolic execution of a \name-transformed program help reduce random state merges and resulting in control the complexity of the solver queries.

The reason for the reduction in time spent on DSE after applying \name is that the fewer number of queries reaching the SMT solver and the number of paths explored.
\methodC explores significantly fewer paths compared to both \methodK and \methodSM except for the dilation and erosion benchmarks where \methodSM explores fewer paths.
The average query size is sensitive to the constraint caching mechanism in KLEE. 
The average query size becomes lower when there are more constraint cache hits because the need to construct newer queries is reduced.
We observe that \methodSM results in a higher average query size than \methodC. 
Dynamic state merging can merge any random pair of states at runtime, resulting in larger complex queries that are unlikely to be cached.
On the other hand, \methodC compile-time state merging is highly regular and can result in smaller queries that are more likely to be cached.

\iftoggle{bv}{
\paragraph{{\bf Accelerating bounded verification.}} 

So far, we have focused on mitigating the path explosion problem while performing DSE on programs.
If the {\em post conditions} of the program are known and easily expressible, we can use DSE to validate the correctness of these programs for a bounded input by appending them to the program as {\em assertions}.
In many cases, these conditions are non-trivial and computationally expensive (\eg graph algorithms).
For the following benchmark programs, these conditions are easily expressible.
\begin{itemize}
	\item {\em toupper:} Output string consists of only uppercase.
	\item {\em bitonic sort:} All output array values are in ascending order.
	\item {\em erosion:} Every output pixel value is less than or equal to the corresponding input pixel value.
	\item {\em dilation:} Every output pixel value is greater than or equal to the corresponding input pixel value.
\end{itemize}

We did bounded verification for these four programs using DSE.
In general, results suggest the \methodC is very effective in this task.
The observation here is that \name aggressively merges and transforms to have a single path for toupper and bitonic sort.
In toupper, \methodC and \methodCSM are the best-performing approaches, having close to $2 \times$  reduction in both runtime and number of solver calls compared to \methodSM.
In bitonic sort, the performance of \methodC and \methodSM have comparable performance even though \methodSM has significantly more solver calls (8 vs. 32).
The bitonic sort contains memory reads/writes with loop-carried dependences, and this results in complex path constraints after branch elimination.
For dilation and erosion, the best approach is \methodCSM because these benchmarks have opportunities for both static (\ie branch elimination) and dynamic state merging.
This shows the utility of transformations like \name in improving the performance of DSE.

}{}

\begin{figure}[t]
  \centering
  \begin{subfigure}[b]{0.32\linewidth}
    \includegraphics[width=\linewidth]{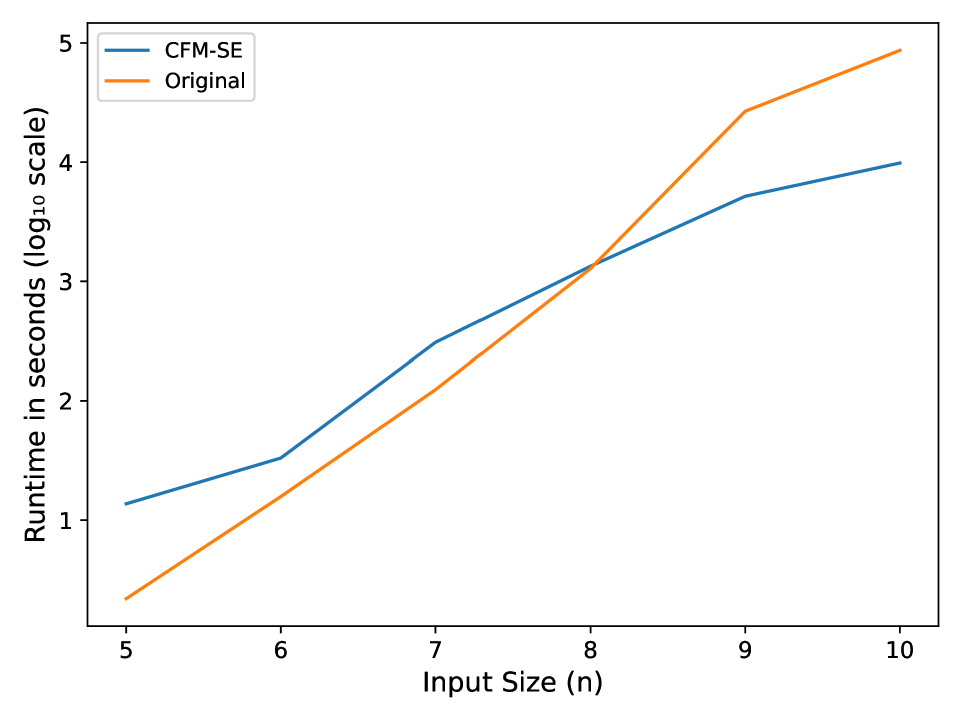}
    \caption{Runtime vs. input size.}
    \label{fig:ms-time}
  \end{subfigure}\hfill
  \begin{subfigure}[b]{0.32\linewidth}
    \includegraphics[width=\linewidth]{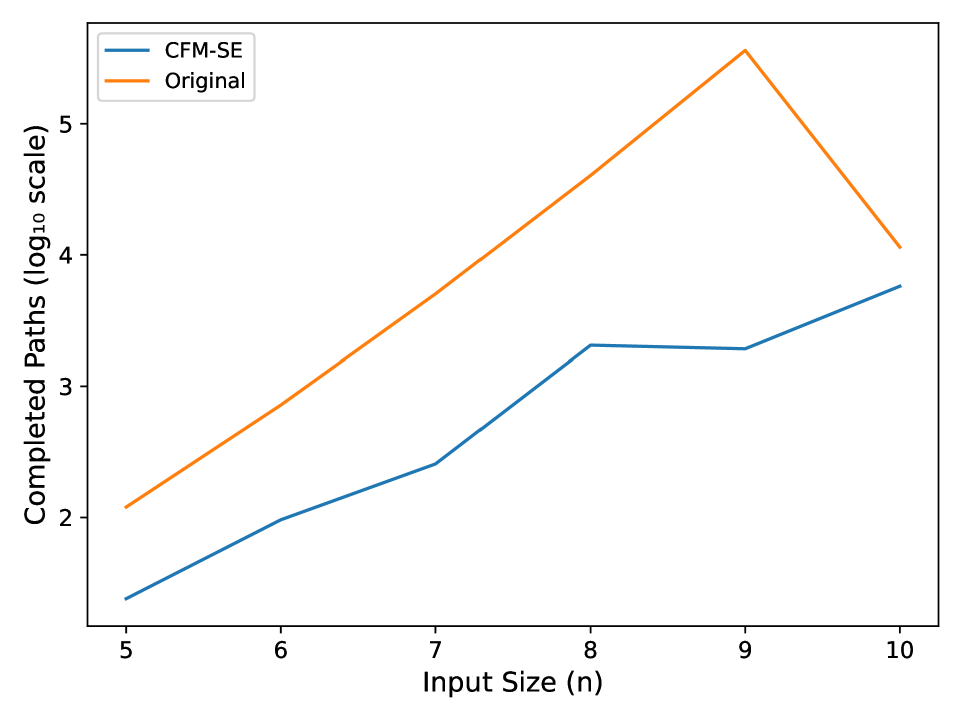}
    \caption{\#Completed paths vs. input size.}
    \label{fig:ms-completed-paths}
  \end{subfigure}\hfill
  \begin{subfigure}[b]{0.32\linewidth}
    \includegraphics[width=\linewidth]{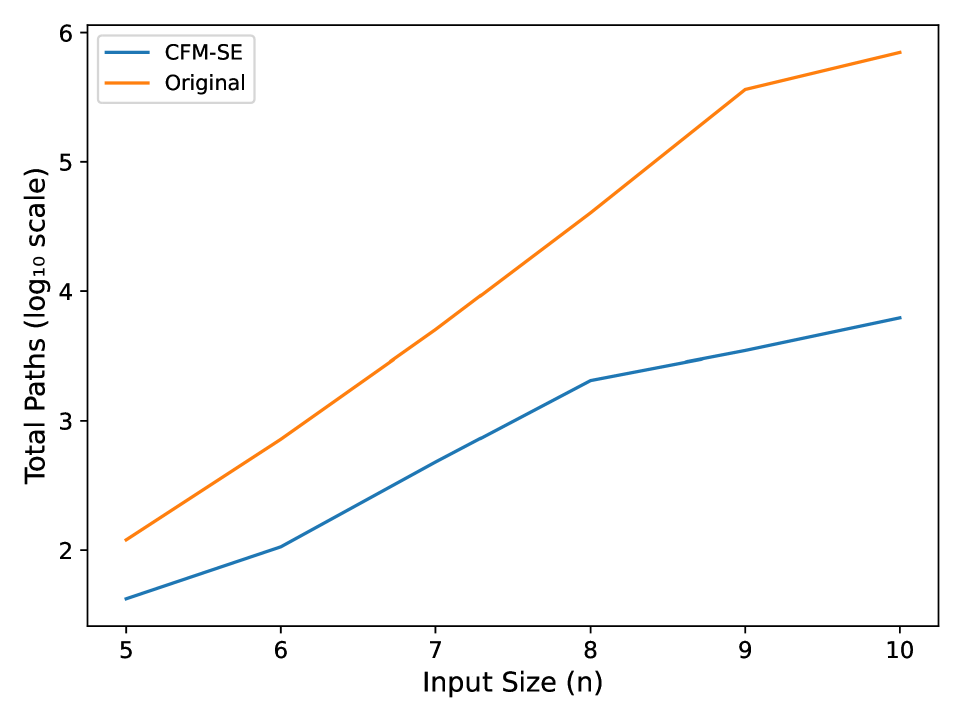}
    \caption{Total \#Paths vs. input size.}
    \label{fig:ms-total-paths}
  \end{subfigure}
  \caption{Merge sort performance across three metrics.}
  \label{fig:mergesort-three}
\end{figure}

\paragraph{{\bf Performance scaling of DSE}}
We proceed to explore how \name-transformed program scales compared to the original program in KLEE by evaluating merge sort for varying input sizes. 
The choice of merge sort is justified by the presence of {\em loop-carried dependence} in the program. 
The loop-carried dependences increase the size and complexity of the solver queries at each iteration of the loop. 
For smaller input size $n$, \name-transformed program is slower compared the original one, but as $n$ increases, the gap keeps shrinking. 
Figure ~\ref{fig:ms-time} clearly shows the steeper growth pattern for original program, as opposed to \name-transformed one. 
The intuition behind this is straightforward, once we consider how \name affects the solver queries. 
For the original program, we have multiple smaller solver queries as compared to fewer larger queries for \name-transformed one. 
At a smaller $n$, multiple smaller queries are easily managed by KLEE, thus it performs better.
However, as we increase the input size, the number of smaller queries start to explode, requiring a higher memory usage and a longer time to solve all these queries. 
As we approach a large input size, even though the total paths increase (Figure ~\ref{fig:ms-total-paths}), the paths that are actually completed begin to decrease for the original program since the solver is unable to manage the high workload.
For \name-transformed program, even though the per solver query time is significantly higher in comparison, the number of these queries is relatively low. 
This difference becomes apparent as we scale the input size of the program. 
Here, we can draw an analogy to performance scaling of parallel programs.
The parallel overhead is high for a small input size while at a large input size, the performance enhancement due to parallelism becomes observerable.
Similarly, \name-transformed starts to show performance (\ie shorter time to complete DSE) as we increase the input size.

\subsection{Performance of Coverage (RQ2)}
\label{cfmse:subsec:coverage}

Eliminating branches can help DSE to reach deeper parts of the programs faster.
We consider 9 real-world programs to assess how well \name transformation helps KLEE in achieving this goal: \code{GNU oSIP-4.0.0} (\ie \code{libosip})~\cite{gnu-osip}, \code{GNU libtasn1-2.11}~\cite{gnu-libtasn1}, \code{chcon}, \code{chown}, \code{mkdir}, and \code {mkfifo} utilities from coreutils-6.11~\cite{gnu-coreutils}, \code{json.h}~\cite{json-h}, \code{protobuf}, and \code{utf-8}~\cite{utf-8}.
\code{libosip} is a library for Voice Over IP (VoIP) applications. 
\code{libtasn1} is a library for encoding data objects 
in a machine-neutral fashion according to ASN.1 specification.
\code{chcon} utility in \code{coreutils} is used to change the SELinux security of a file~\cite{chcon}.
\code{chown} utility is used to change the file ownership and group~\cite{chown}.
\code{mkdir} utility is used to create new directories~\cite{mkdir}.
\code{mkfifo} utility is used to make new named pipes~\cite{mkfifo}.
\code{json.h} is a single header file C library for parsing \code{JSON} files~\cite{json-schema}.
\code{protobuf} is a C library that implements the {\em Google Protocol Buffer} data serialization format~\cite{googleprotobuf}.
\code{utf8} is a C library for working with UTF-8 encoded strings~\cite{utf-8}. 
We use these benchmark programs as they are moderately large, written in C, and they have been used in similar KLEE-based studies related to DSE~\cite{chopper,klee}.
For \code{libosip} and \code{libtasn1}, we use similar setup as in Chopper experiments~\cite{chopper}. 
This includes manually written test driver programs that initialize the library interfaces and invoke the library functions with symbolic inputs.
We followed the setup described in KLEE \code{coreutils} experiments~\cite{coreutils-experiment} for \code{chcon}, \code{chown}, \code{mkdir}, and \code{mkfifo}.
For \code{json.h}, we used a test harness that invokes the \code{json\_parse} function with a symbolic string of 20 characters.
We used a program that unpacks a message contained in a symbolic buffer of size 10 bytes for \code{protobuf}. 
For \code {utf-8}, we used two test harnesses, one that invokes the \code{utf8ncasecmp} with two symbolic strings of 15 characters and the other with two validated symbolic strings (\ie properly prepared strings avoid certain parsing errors according to the intended use of the library) of 20 characters.

\begin{figure*}[t]
    \centering
    \begin{subfigure}{0.22\textwidth}
        \includegraphics[width=0.9\linewidth]{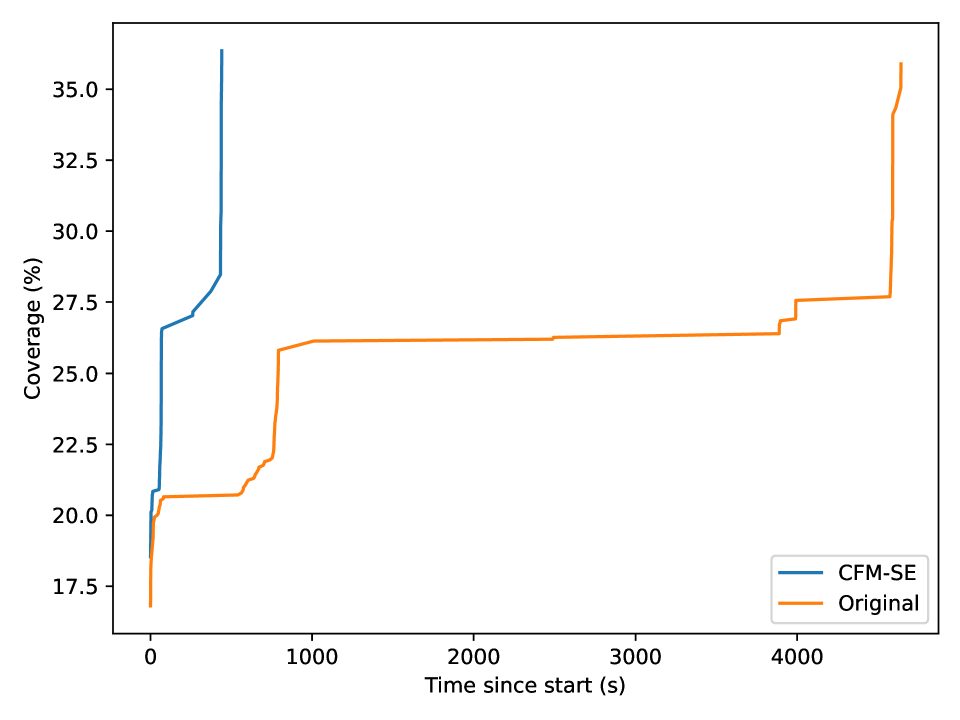}
        \caption{\code{libosip}}
        \label{fig:osip}
    \end{subfigure}
    \begin{subfigure}{0.22\textwidth}
        \includegraphics[width=0.9\linewidth]{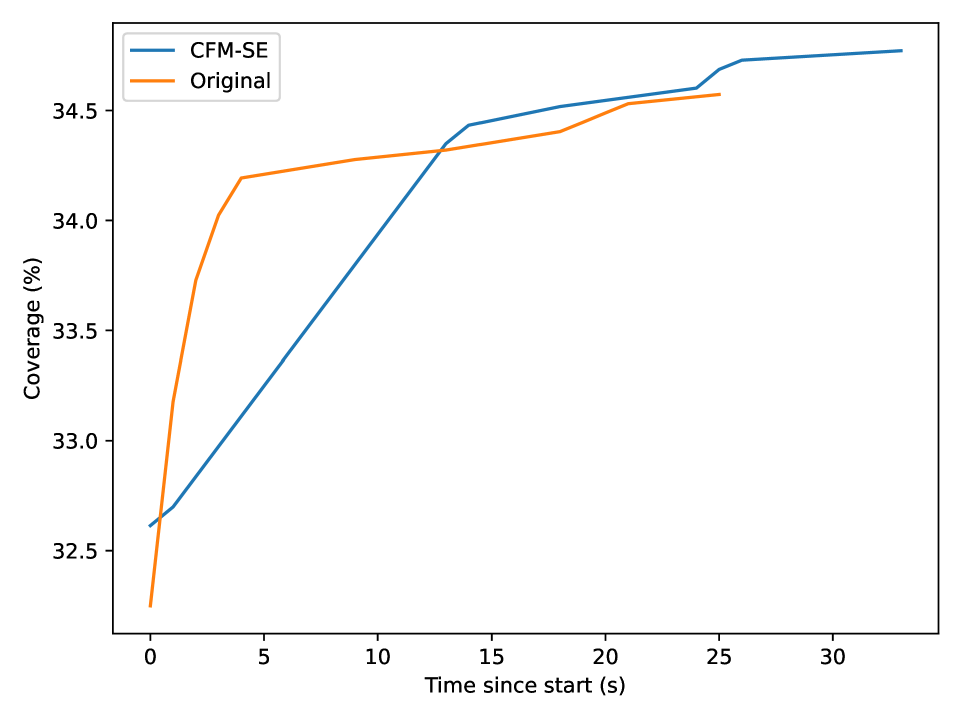}
        \caption{\code{libtasn1}}
        \label{fig:libtasn1}
    \end{subfigure}
    \begin{subfigure}{0.22\textwidth}
        \includegraphics[width=0.9\linewidth]{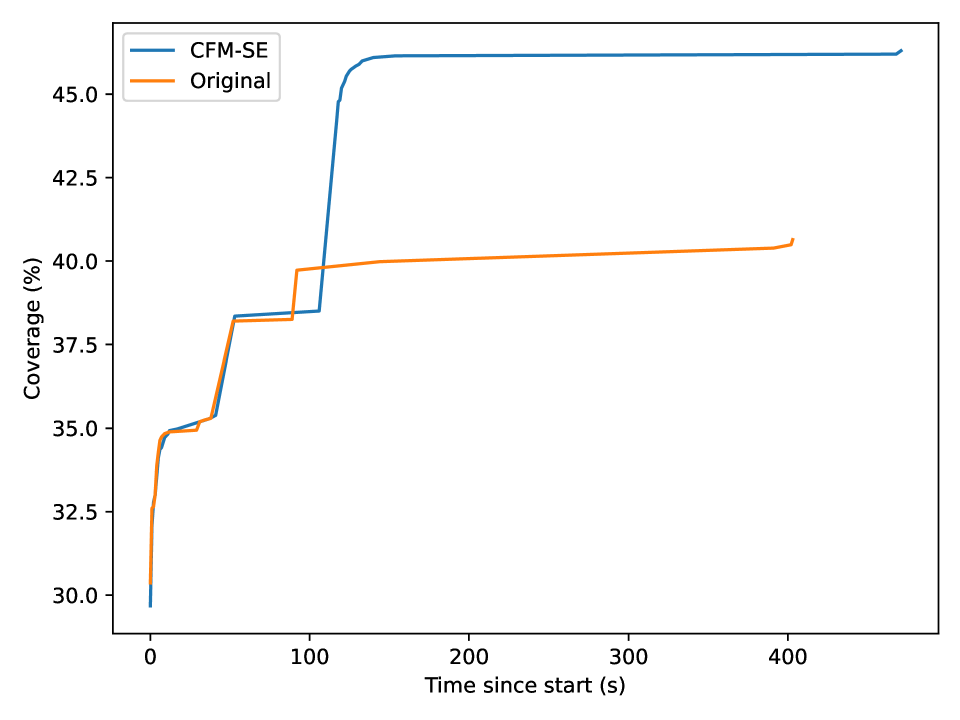}
        \caption{\code{chcon}}
        \label{fig:chcon}
    \end{subfigure}
    \begin{subfigure}{0.22\textwidth}
        \includegraphics[width=0.9\linewidth]{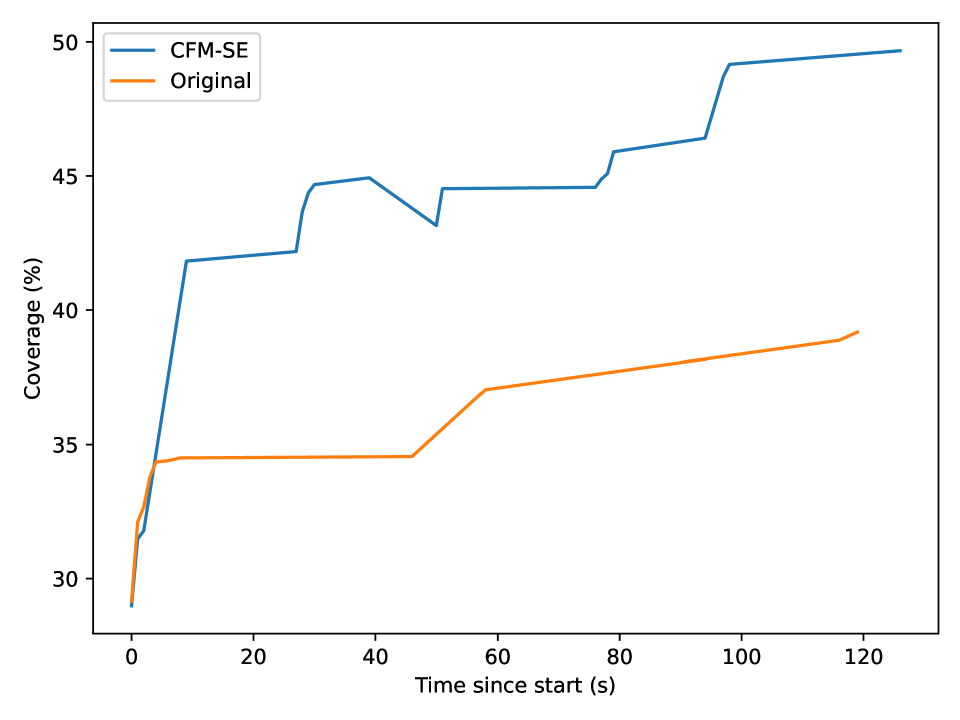}
        \caption{\code{chown}}
        \label{fig:chown}
    \end{subfigure}
    \begin{subfigure}{0.22\textwidth}
        \includegraphics[width=0.9\linewidth]{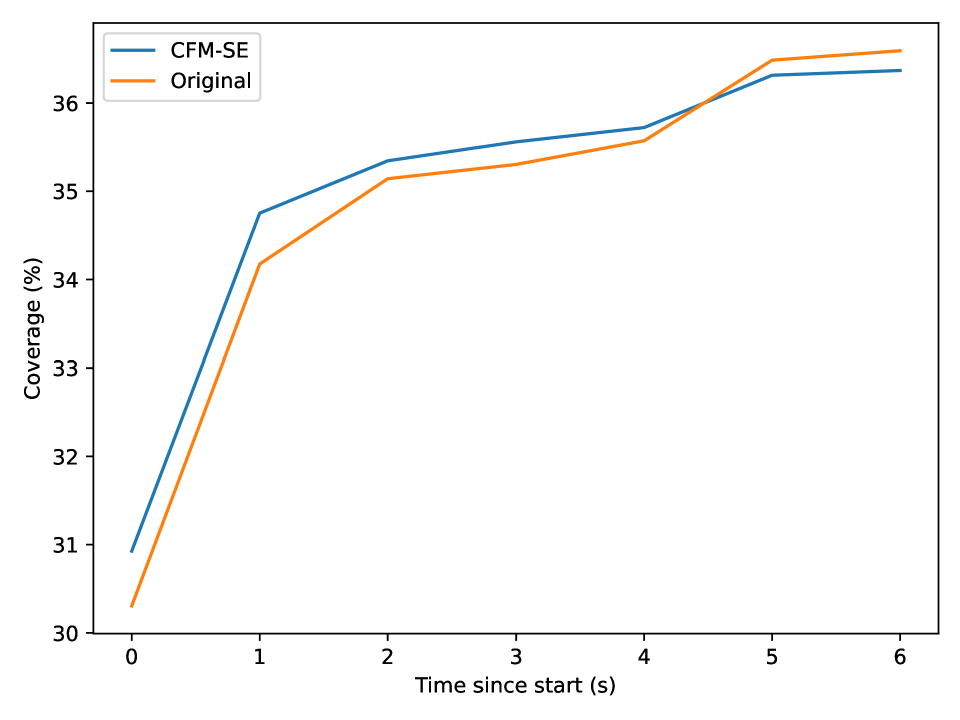}
        \caption{\code{mkdir}}
        \label{fig:mkdir}
    \end{subfigure}
    \begin{subfigure}{0.22\textwidth}
        \includegraphics[width=0.9\linewidth]{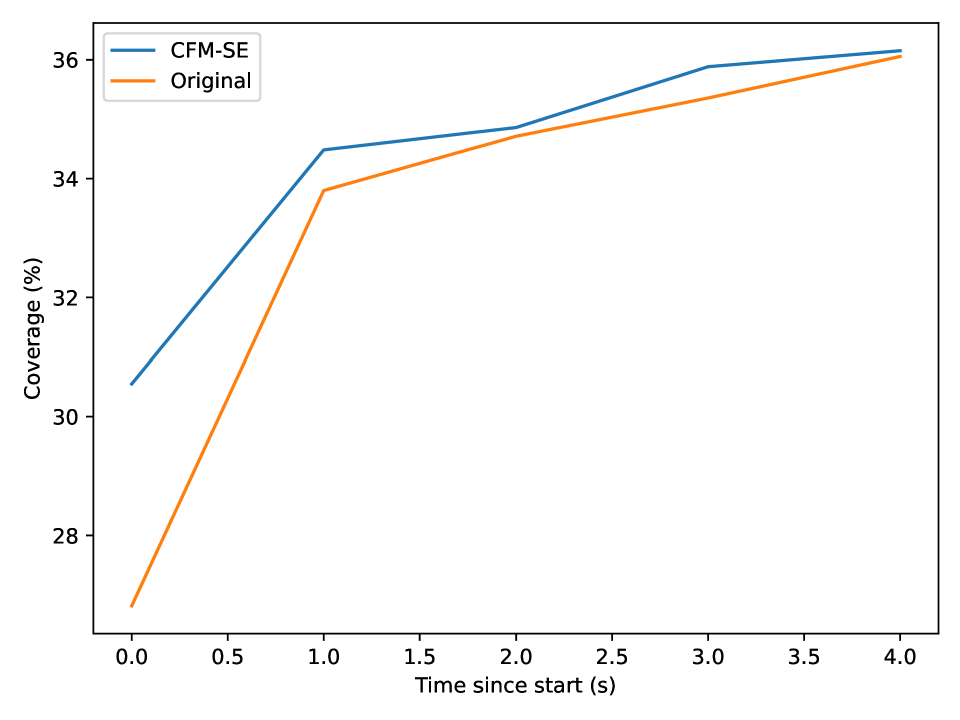}
        \caption{\code{mkfifo}}
        \label{fig:mkfifo}
    \end{subfigure}
    \begin{subfigure}{0.22\textwidth}
        \includegraphics[width=0.9\linewidth]{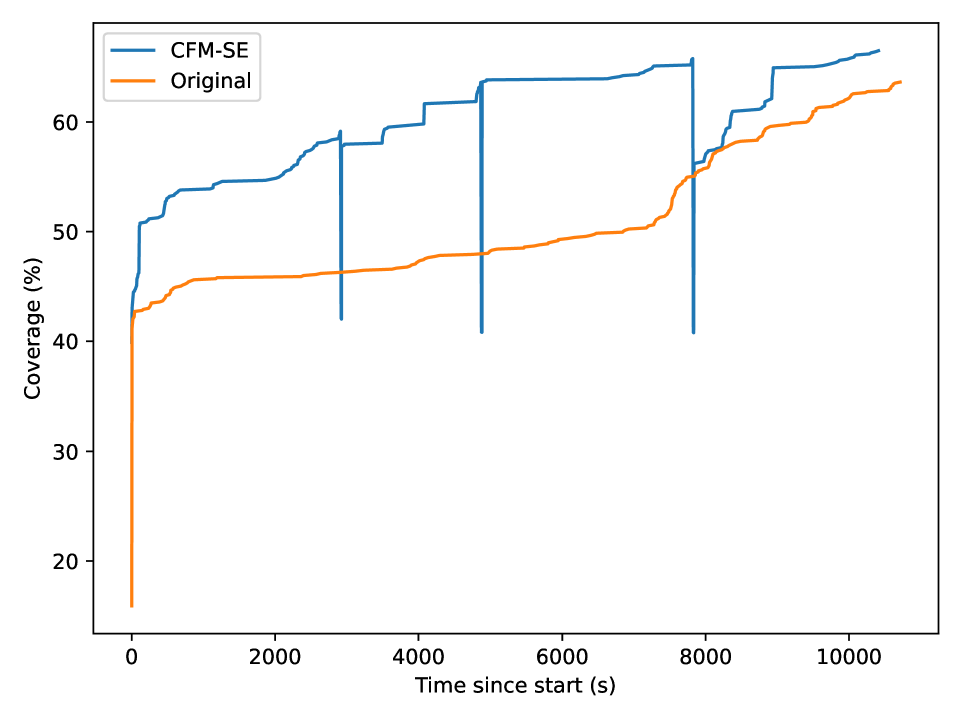}
        \caption{\code{json.h}}
        \label{fig:json}
    \end{subfigure}
    \begin{subfigure}{0.22\textwidth}
        \includegraphics[width=0.9\linewidth]{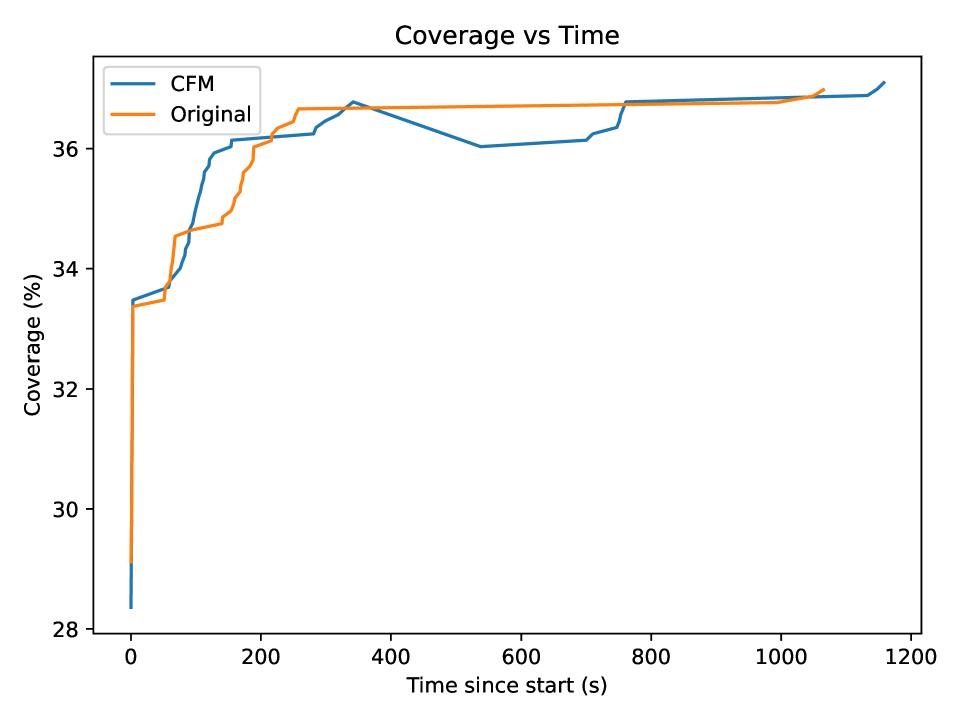}
        \caption{\code{protobuf}}
        \label{fig:protobuf}
    \end{subfigure}
    \begin{subfigure}{0.22\textwidth}
        \includegraphics[width=0.9\linewidth]{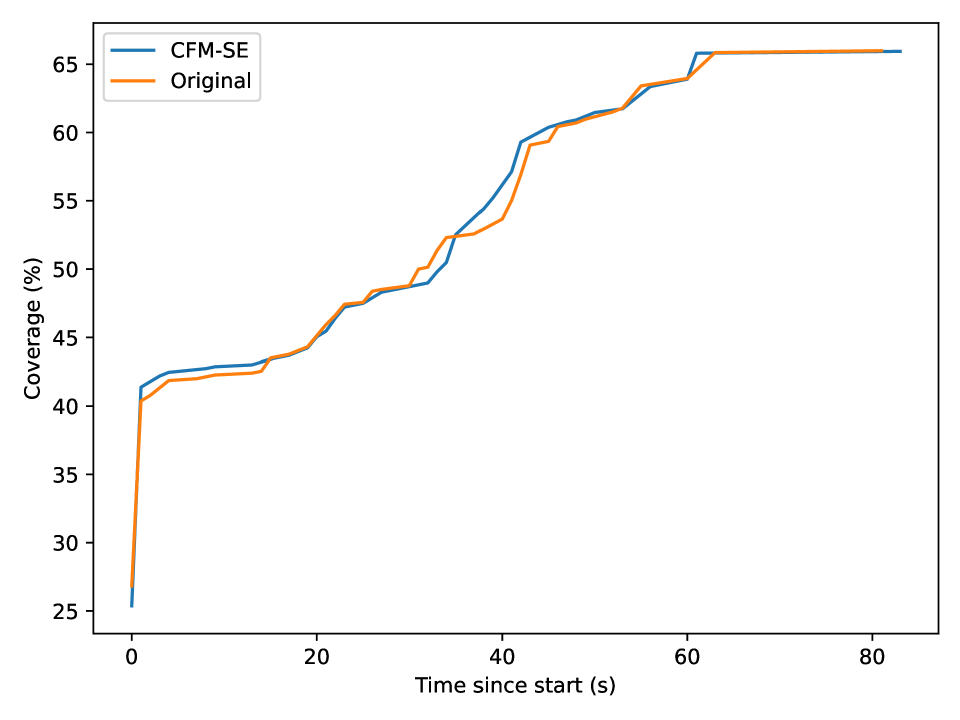}
        \caption{\code{utf8-valid}}
        \label{fig:utf8-val}
    \end{subfigure}
    \begin{subfigure}{0.22\textwidth}
        \includegraphics[width=0.9\linewidth]{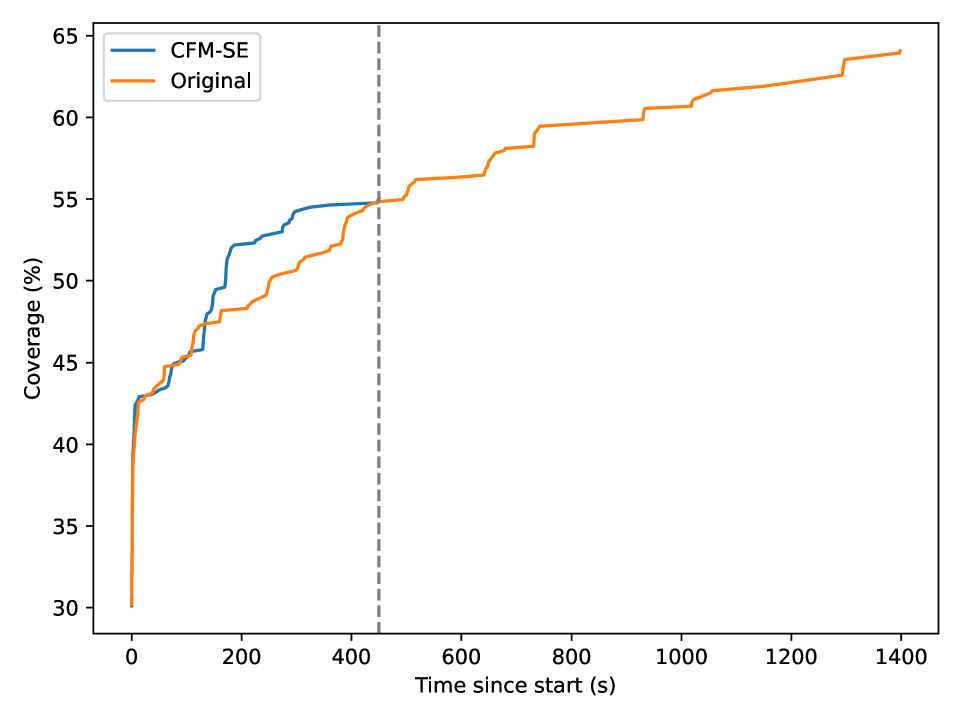}
        \caption{\code{utf8-nonvalid}}
        \label{fig:utf8-unval}
    \end{subfigure}
    \caption{Source line coverage vs time}
    \label{fig:coverage}
    \vspace{-1em}
\end{figure*}  

For each benchmark program considered, we run both KLEE with and without \name transformation using the driver program described in Section~\ref{cfmse:subsec:driver}.  
For \code{libosip}, \code{libtasn1}, and \code{json.h}, we set time limit of three hours. 
We set the time limit of an hour for \code{chcon} and \code{protobuf}. 
We used the default search strategy for KLEE path exploration.
Figure~\ref{fig:coverage} shows the source line coverage against time for each of the benchmark programs.
Line coverage is the percentage of distinct source lines that have been explored so far out of the total distinct source lines covered by all the LLVM-IR instructions in the compiled program.
For \code{libosip}, the maximum line coverage that can be achieved by both \name and KLEE is around 36\%. 
This is because the test driver only focuses on testing certain interfaces of the library, and some functions are not invoked at all.
Interestingly, \name can cover the maximum number of lines within less than 500 seconds, while KLEE takes close to an hour to reach the same coverage. 
Since \code{libosip} contains several loops that do not have early exits and contain symbolic branches, KLEE cannot explore other paths outside the loop without finishing the whole loop execution. 
Aggressively eliminating branches inside these loops allows KLEE to finish the execution of the loops faster and explore more paths outside them.
In this benchmark, we did not find any false positive bugs that are introduced by \name.
In other words, \name transformation is safe for the lines covered in this benchmark.
The fact that \name can achieve more coverage faster also means that the path queries generated by \name are not significantly more expensive than the ones generated by KLEE in total. 

Figure~\ref{fig:libtasn1} shows the source line coverage for \code{libtasn1} benchmark.
Here, both KLEE with and without and \name transformation crash due to a \code{malloc} call with symbolic size. 
KLEE with original program reaches this location faster than with \name transformation (27s vs 36s).
The \name encounters 3 false positive bugs (out of bound memory bounds due to unconditional execution of loads) before reaching the \code{malloc} call.
When a false positive is encountered, the driver program relaunches KLEE with additional location constraints on \name transformation so that the transformation does not apply to the false positive location again.
The initial slow rise in coverage for \name in Figure~\ref{fig:libtasn1} is caused by the fresh re-execution of the program by the driver loop. 

\code{chcon} benchmark also exhibits similar behavior as \code{libosip} benchmark (Figure~\ref{fig:chcon}).
\name transformation does not introduce any spurious bugs in this benchmark program and achieves more coverage faster than the original version.
The maximum source line coverage achieved in this benchmark is $\sim$47\% for the \name-transformed version and $\sim$41\% for the original version.
\name-transformed version achieves its maximum coverage in 471s, while the original version takes around 403s.
In \code{chown} benchmark, \name-transformed program achieves more coverage faster than the original version (Figure~\ref{fig:chown}). 
The maximum coverage achieved in this benchmark is $\sim$51\%. 
We observe that \name introduces 3 false positive bugs in this benchmark. 
Although the \code{mkdir} (Figure~\ref{fig:mkdir}) and \code{mkfifo} (Figure~\ref{fig:mkfifo}) benchmark programs do not show significantly higher coverage with \name, they are slightly higher and quicker than the original versions.

The \code{json.h} benchmark program, with \name transformation, achieves more coverage faster than the original version (Figure~\ref{fig:json}). 
The maximum coverage achieved in this benchmark is around 65\%.
We observe that \name transformation introduces 6 false positive bugs in this benchmark program. 
All the false positives are caused by \name transformation eliminating branches that are guarding against out-of-bound memory accesses.
The sudden drops in coverage are caused by the relaunch of KLEE to avoid false positives.
The re-execution with additional location constraints on \name quickly gains coverage because of the usage of test seeds from previous executions.
Figure~\ref{fig:protobuf} shows the source line coverage for \code{protobuf} benchmark program.
This benchmark program with \name transformation achieves slightly higher coverage than the original version, with both plateauing around 37\%. 
However, both the original version and \name-transformed one crash due to a \code{malloc} call with symbolic size.
Interestingly, \name-transformed one reaches this location faster than the original version (1738s vs. 2490s). 
Figure~\ref{fig:utf8-val} shows the source line coverage for \code{utf-8} benchmark program. 
Both of them eventually reach comparable final coverage.
Finally, Figure~\ref{fig:utf8-unval}, shows the same coverage, but both \name-transformed program and original one run into an error due to unintended use of the library~\cite{utf8-bug}. 
However, \name-transformed version reaches this error at the 500s mark, whereas the original one takes around 1450s.
In summary, this study shows that \name transformation can help in achieving code coverage faster in real-world programs compared to vanilla KLEE, at the same time helping us discover bugs faster, which will be further illustrated in the next section.

\subsection{Performance of Bug Discovery (RQ3)}
\label{cfmse:subsec:bug}

We evaluate how we can effectively use the \name transformation to accelerate bug detection. 
In this section, we compare the time it takes for KLEE to find the bug in \name-transformed program against in the original one. 
For this experiment, we select five small to moderate size open-source libraries, \code{tiny-regex-c}, \code{utf-8}, \code{json.h}, \code{libosip}, and \code{libyaml}. 
For each program, we manually inject a bug deep inside the program to serve as the target for DSE. 
For \code{json.h}, \code{utf-8}, and \code{libyaml} we modify the programs to introduce an assertion-based failure, whereas for \code{tiny-regex-c}, we modify an array access to introduce an invalid memory access. 
Lastly, for \code{libosip}, we use an older version with a minor crash-inducing bug.  

\begin{figure}[ht]
  \centering
  \begin{subfigure}[b]{0.4\textwidth}
    \centering
    \begin{minted}[fontsize=\scriptsize, framesep=2mm, breaklines=true]{c}
    if (json_get_value_size(state, 0)) {
        /* value parsing failed! */
        return 1; 
    }
    \end{minted}
    \caption{Original code segment}
    \label{fig:original_assertion}
  \end{subfigure}%
  \hfill
  \begin{subfigure}[b]{0.6\textwidth}
    \centering
    \begin{minted}[fontsize=\scriptsize, framesep=2mm, breaklines=true]{c}
    if (json_get_value_size(state, 0)) {
        /* value parsing failed! */
        _assert_fail("0", "json.c", 0, "json_get_array_size"); // bug
        return 1; 
    }
    \end{minted}
    \caption{Modified code segment with assertion-based bug}
    \label{fig:modified_assertion}
  \end{subfigure}
  \caption{An example of assertion-based bug in \code{json.h}.}
  \label{fig:assertion_bug}
  \vspace{-1em}
\end{figure}

\begin{figure}[ht]
  \centering
  \begin{subfigure}[b]{0.45\textwidth}
    \centering
    \begin{minted}[fontsize=\scriptsize, framesep=2mm, breaklines=true]{c}
    if (pattern[i+1] == 0) {
        /* incomplete pattern, missing non-zero char after '\\' */
        return 0;
    }
    ccl_buf[ccl_bufidx++] = pattern[i++];
    \end{minted}
    \caption{Original code segment}
    \label{fig:original_memory}
  \end{subfigure}%
  \hfill
  \begin{subfigure}[b]{0.45\textwidth}
    \centering
    \begin{minted}[fontsize=\scriptsize, framesep=2mm, breaklines=true]{c}
    if (pattern[i+1] == 0) {
        /* incomplete pattern, missing non-zero char after '\\' */
        return 0;
    }
    ccl_buf[ccl_bufidx++] = pattern[++i + 2]; // bug
    \end{minted}
    \caption{Modified code segment for memory-based bug}
    \label{fig:modified_memory}
  \end{subfigure}
  \caption{An example of a memory-based bug in \code{tiny-regex-c}.}
  \label{fig:memory_bug}
  \vspace{-1em}
\end{figure}

\begin{figure}[t]
  \centering
  \begin{subfigure}[b]{0.30\linewidth}
    \includegraphics[width=\linewidth]{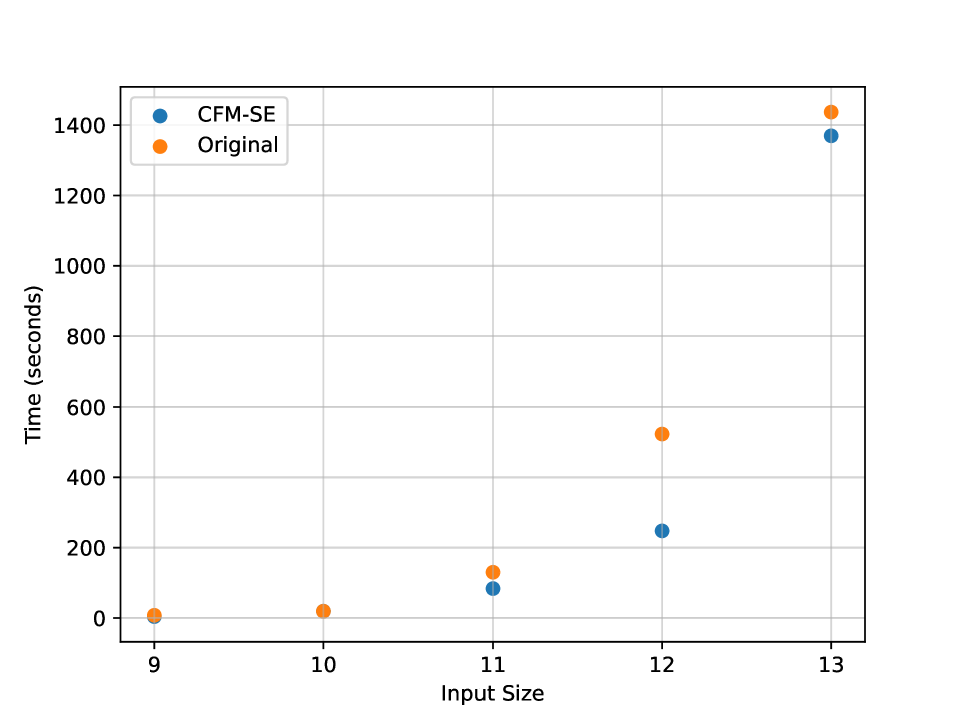}
    \caption{\code{tiny-regex-c}}
    \label{fig:subfig_regex}
  \end{subfigure}
  \begin{subfigure}[b]{0.30\linewidth}
    \includegraphics[width=\linewidth]{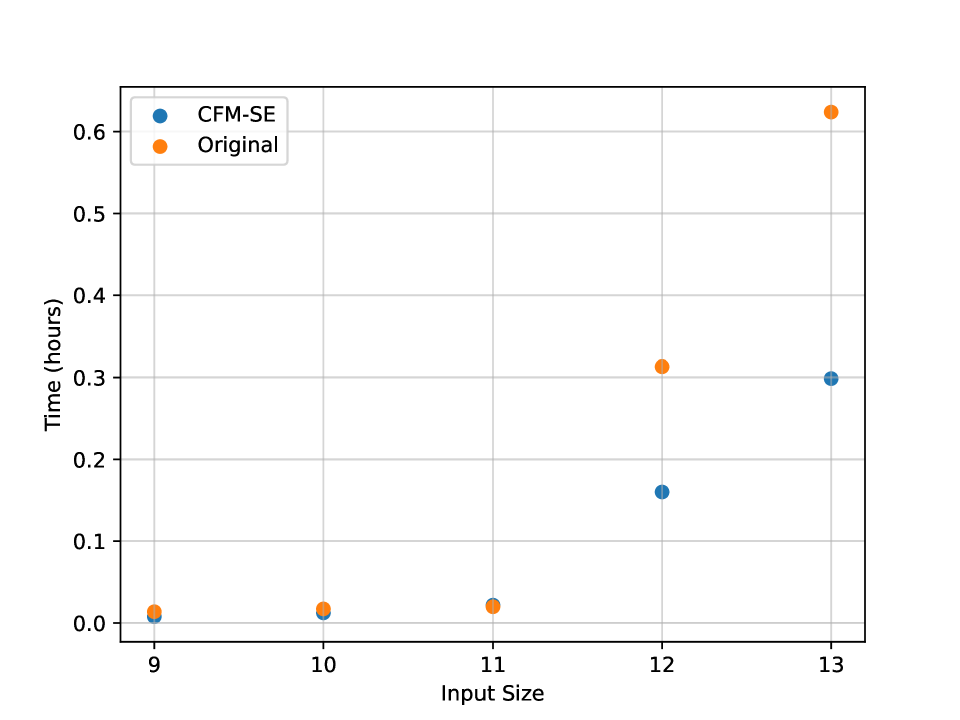}
    \caption{\code{utf-8}}
    \label{fig:subfig_utf}
  \end{subfigure}
  \begin{subfigure}[b]{0.30\linewidth}
    \includegraphics[width=\linewidth]{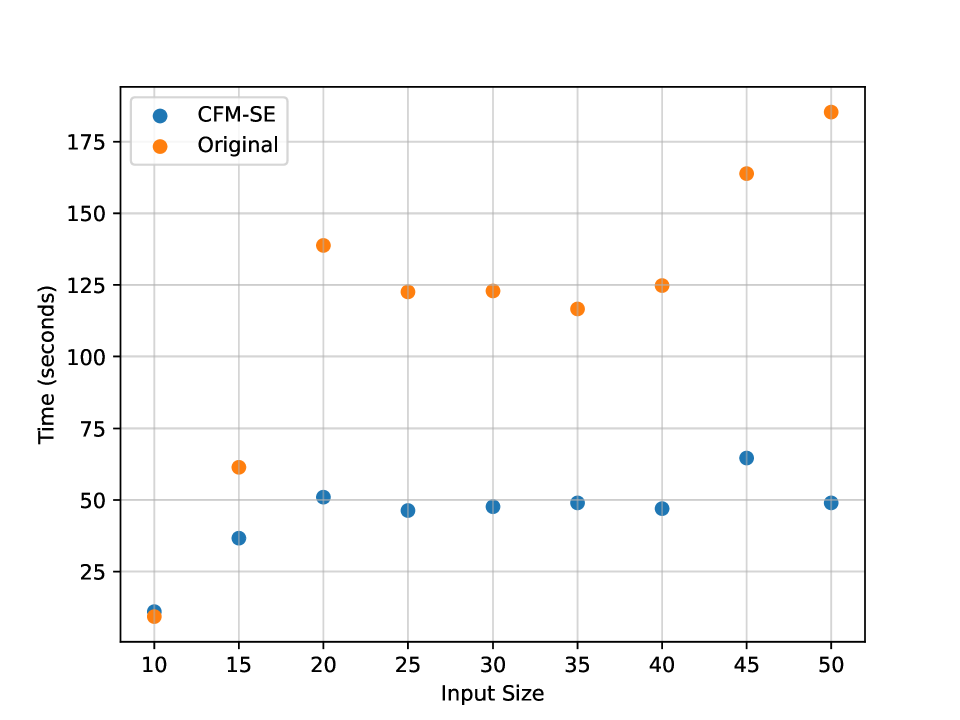}
    \caption{\code{json.h}}
    \label{fig:subfig_json}
  \end{subfigure}
  \medskip
  \\
  \hspace{\fill}%
  \begin{subfigure}[b]{0.30\linewidth}
    \includegraphics[width=\linewidth]{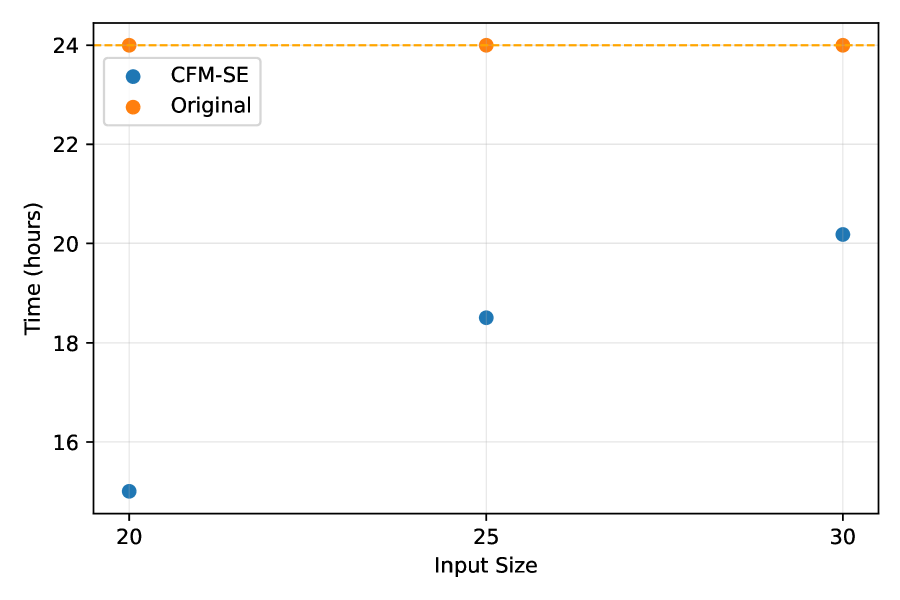}
    \caption{\code{libosip}}
    \label{fig:subfig_libosip}
  \end{subfigure}
  \begin{subfigure}[b]{0.30\linewidth}
    \includegraphics[width=\linewidth]{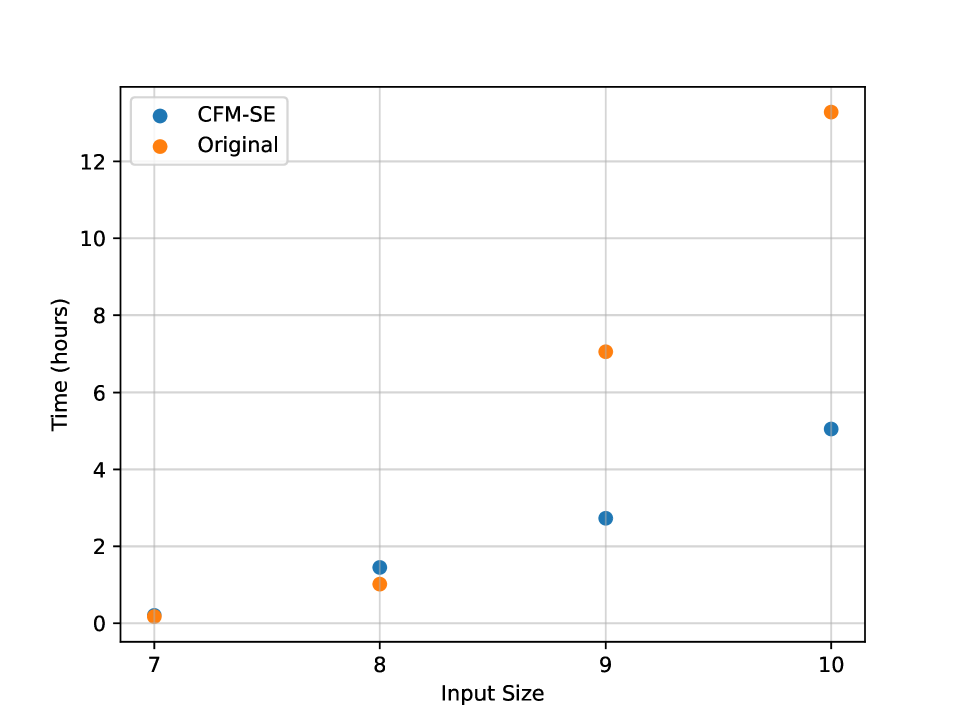}
    \caption{\code{libyaml}}
    \label{fig:subfig_libyaml}
  \end{subfigure}%
  \hspace{\fill}
  \vspace{-1em}
  \caption{Time taken (s) for KLEE to find bugs with and without \name.}
  \label{fig:overall}
  \vspace{-1em}
\end{figure}

Figure~\ref{fig:assertion_bug} and  Figure~\ref{fig:memory_bug} show examples of the original code snippet and modified one where the program has been altered to induce a crash during DSE with an assertion (in \code{json.h}) and an invalid memory access (in \code{tiny-regex-c}), respectively. 
We run these programs under the constraints of a 2-hours timeout and 32GB max-memory.  
The results reveal that the bug is found faster in \name-transformed program than in the original one.
The difference becomes more evident as we increase the input size of programs. 
For example, in Figure~\ref{fig:subfig_regex}, we can observe that for an input size of 13, the bug was found in 1370s in the \name-transformed program, whereas it took 1447s in the original program.
The difference is more pronounced in the case of \code{utf-8}; as shown in Figure~\ref{fig:subfig_utf}, it takes 1075s in the \name-transformed program while 2246s in the original one for the input size of 13.

Our experiment with \code{json.h} parser program shows an interesting trend. 
Outlined in the Figure~\ref{fig:subfig_json}, as the input size increases, the bug-finding time quickly stabilizes for the \name-transformed program. 
This experiment is based on 9 different input sizes, and the plateau is evidently reached around input size 20.  
This behavior is better explained by the \name pass, as it tried to consolidate multiple potential paths into a significantly smaller number of paths. 
Therefore, regardless of the size of the input, symbolic execution would quickly fall through to identify the bug. 
On the other hand, for the original program, KLEE still has to explore all these possible paths, thus taking a much longer time. 
Table~\ref{tab:overhead} further supports this observation, showing that \code{json} exhibits the highest merges and overhead of select instructions relative to its code size. 

\begin{figure}[t]
  \centering
  \begin{subfigure}[b]{0.30\textwidth}
    \centering
    \includegraphics[width=\linewidth]{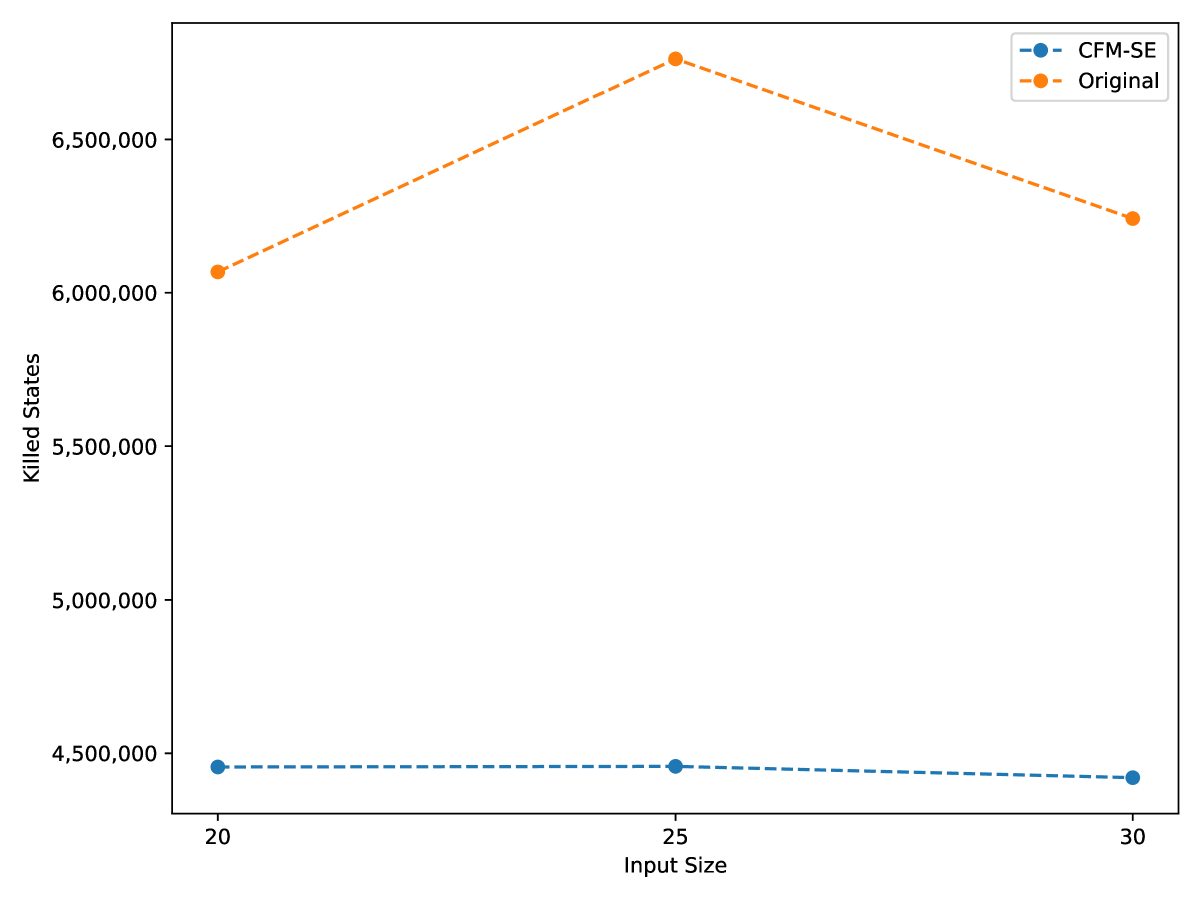}
    \caption{\texttt{libosip}}
    \label{fig:libosip_killed_states}
  \end{subfigure}
  \hspace{2em}
  \begin{subfigure}[b]{0.30\textwidth}
    \centering
    \includegraphics[width=\linewidth]{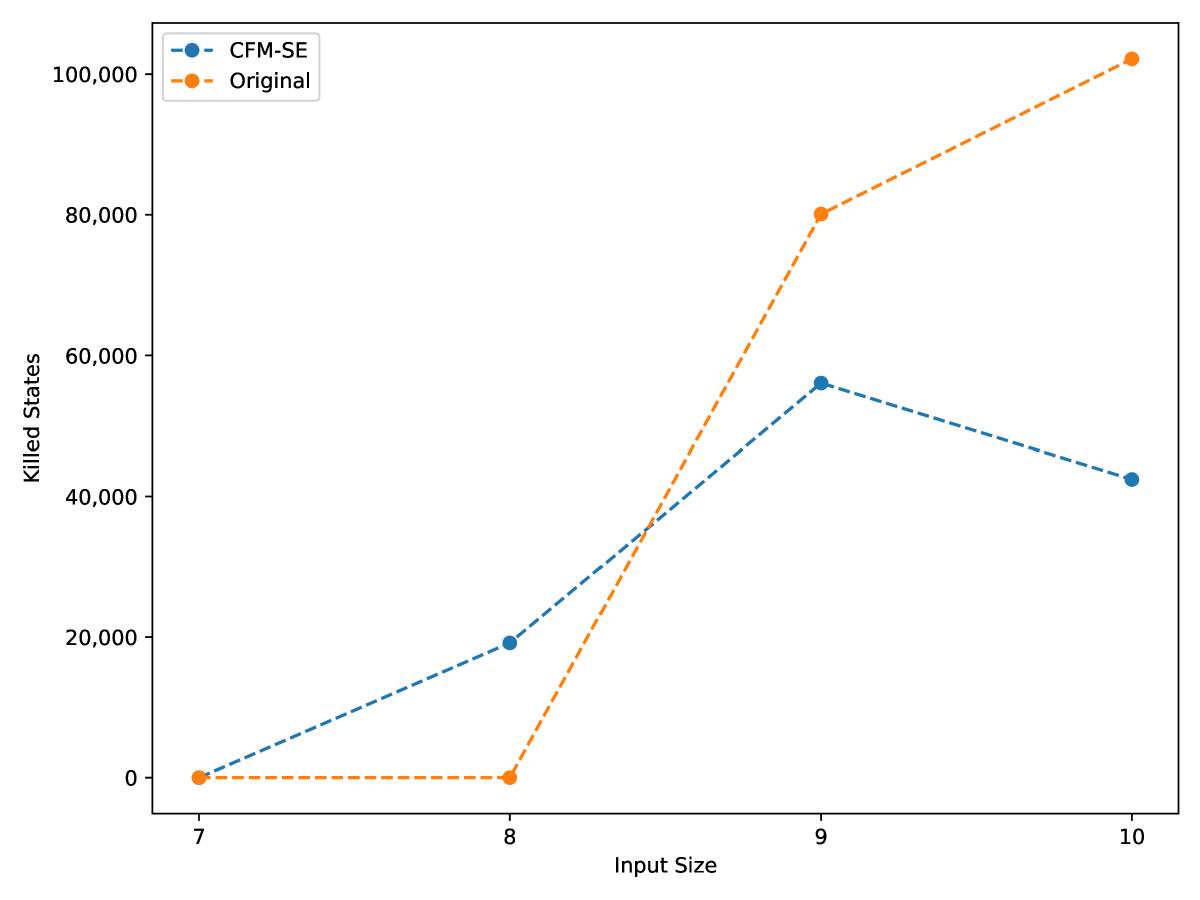}
    \caption{\texttt{libyaml}}
    \label{fig:libyaml_killed_states}
  \end{subfigure}
  \vspace{-1em}
  \caption{Killed states vs.\ input size}
  \label{fig:killed_states}
  \vspace{-1em}
\end{figure}

In addition, we also evaluated \code{libosip 4.0.0}, a moderately sized network library under resource constraints of a 24-hour timeout and a maximum memory of 75 GB. 
The target bug here is a \code{NULL} pointer passed to the \code{strcmp} function, which would result in undefined behavior as specified by the C11 standard~\cite{ISO-C11}, causing a segmentation fault. 
This bug was subsequently fixed in later versions of \code{libosip}.
For an input size of 25 and 30 \name-transformed code detects the bug in 66,631s ($\sim$18.5 hours), and 72,647s ($\sim$20.2 hours), respectively. 
However, the original code fails to find the bug within the given time budget (Figure~\ref{fig:subfig_libosip}). 
Both of these programs hit the given memory limit, but interestingly, in the original program, KLEE suffers from higher state-killing due to high memory usage (Figure~\ref{fig:libosip_killed_states}), as the bug is deeply embedded in the program.
This shows that the \name-transformed program simplifies the control-flow and reduces state-forking memory usage, reaching deep inside the program with a smaller input size.
\ie in \name-transformed program, states get simplified, leading to a cheaper workload for the solver and less memory-intensive execution for KLEE, allowing it to reach the program point with the critical failure. 

We also show that how \name helps to scale DSE when the memory is limited with \code{libyaml}. 
We observe that for a 75GB memory budget, in the original program, KLEE takes 55000s and finds the assertion faster than in \name-transformed one, which takes 76000s, but while doing so, KLEE consumes all 75GB of memory. 
Under the 20GB memory budget, as we already saw, Figure~\ref{fig:subfig_libyaml}, as input size grows, \name-transformed program outperforms the original one. 
Beyond input size 8, the time to reach the bug shows a near exponential trend for the original program. 
This can be explained by the memory pressure, where in the original program, KLEE maintains a higher number of concurrent states, leading to more states killed due to the memory cap.
Hence, results in a slower progress and a lower probability of finding a the bug embedded deep in the program. 
We observe that when increasing the memory budget shifts the inflection point (\ie input size where both original and \name-transformed versions take the same time to find the bug) to the right. 
With more RAM available, original program on KLEE becomes comparable or sometimes better than \name-transformed one for smaller input sizes.
Nevertheless, as soon as memory pressure becomes apparent, \name starts to dominate.  

We design an experiment to probabilistically demonstrate that \name has a higher chance of finding a bug as programs scale. 
In earlier experiments with \code{libyaml}, we observed a memory bottleneck with original program as the input size grew larger. 
To argue about its effect on bug detection, we instrument KLEE to measure the number of states that are killed due to the memory limitation.
Figure \ref{fig:libyaml_killed_states} shows that as the input size increases, the number of killed states rises with it, with more states killed for original program as opposed to \name-transformed one. 
As more states are killed prematurely, the probability of bug-inducing states decreases with it. 
To resume this path, DSE now has to recreate equivalent states, which consumes additional resources.
\name-transformed program with fewer killed states, preserves its ability for deeper exploration opportunities, thereby increasing the probability of bug detection and overall coverage. 
This is consistent with the bug detection result in Figure~\ref{fig:subfig_libyaml}, which shows that \name-transformed program begins to outperform the original one for bug detection at the same point in input size where original program overtakes \name-transformed one in killed states. 

\subsection{Overheads of \name Transformation}
\label{sec:eval_overhead}

\begin{table}[tb]
	\centering
	\caption{Overhead of \name transformation.}
	\vspace{-1em}
	\resizebox{\linewidth}{!}{
	\begin{tabular}{@{}lccccccc@{}}
		\toprule
		\textbf{Benchmark}                                           & \textbf{\begin{tabular}[c]{@{}l@{}}Compile-time \\ Overhead (ms)\end{tabular}} & \textbf{\begin{tabular}[c]{@{}l@{}}LoC\end{tabular}} & \textbf{\begin{tabular}[c]{@{}l@{}}\#Instructions\end{tabular}} & \textbf{\begin{tabular}[c]{@{}l@{}}\#Merges\end{tabular}} & \textbf{\begin{tabular}[c]{@{}l@{}}Select Instruction \\ Overhead (\%)\end{tabular}} &
		\textbf{\begin{tabular}[c]{@{}l@{}}Branch Nesting \end{tabular}} &
		\textbf{\begin{tabular}[c]{@{}l@{}}Control-Flow Nesting\end{tabular}}                                                                                                                                                                                                                                                                                                                                                                                     \\
		\midrule
		\code{libosip}                                               & 0.38                                                                           & 19,588                                               & 37,893
		                                                             & 194                                                                            & 0.83\%                                               & 24                                                              & 34                                                                                                                                                         \\
		\code{libtasn1}                                              & 1.58                                                                           & 20,879                                               & 34,337
		                                                             & 148                                                                            & 0.56\%                                               & 9                                                               & 13                                                                                                                                                         \\
		\code{chcon}                                                 & 0.90                                                                           & 435                                                  & 35,248
		                                                             & 149                                                                            & 0.88\%                                               & 9                                                               & 10                                                                                                                                                         \\
		\code{chown}                                                 & 0.90                                                                           & 809                                                  & 36,672                                                          & 176                                                       & 0.94\%                                                                               & 9  & 10 \\
		\code{mkdir}                                                 & 0.80                                                                           & 342                                                  & 28,903
		                                                             & 104                                                                            & 0.77\%                                               & 9                                                               & 10                                                                                                                                                         \\
		\code{mkfifo}                                                & 0.76                                                                           & 107                                                  & 27,372                                                          & 90                                                        & 0.74\%                                                                               & 9  & 10 \\
		\code{json.h}                                                & 0.02                                                                           & 2,439                                                & 5,103                                                           & 18                                                        & 1.17\%                                                                               & 5  & 5  \\
		\code{protobuf}                                              & 0.10                                                                           & 3,516                                                & 6,334
		                                                             & 40                                                                             & 0.79\%                                               & 5                                                               & 6                                                                                                                                                          \\
		\code{utf-8}                                                 & 0.02                                                                           & 1,218                                                & 3,522
		                                                             & 8                                                                              & 0.32\%                                               & 2                                                               & 3                                                                                                                                                          \\

		\code{tiny-regex-c}                                          & 0.03                                                                           & 428                                                  & 3,877
		                                                             & 13                                                                             & 0.35\%                                               & 5                                                               & 6                                                                                                                                                          \\

		\code{libyaml}                                               & 7.60                                                                           & 7,400                                                & 56,688                                                          & 260                                                       & 1.23\%                                                                               & 26 & 26 \\

		\bottomrule
	\end{tabular}
	}
    \label{tab:overhead}
	\vspace{-1.5em}
\end{table}

Compile-time overhead and run-time overhead of \name transformation is not significant.
Table~\ref{tab:overhead} summarizes the compile-time overhead, number of source lines, number of LLVM-IR instructions, number of program locations where the \name transformation applied, number of additional \code{select} instructions introduced by the \name transformation, and levels of branch nesting and control-flow nesting in the IR.
We did this only for programs from RQ2 and RQ3 due to the absence of measurable overhead in small programs.
Even though \name transformation is applied to many program locations, the compile-time overhead is less than 2ms for most of the programs.
Since in most cases, the conditional branches contain shorter and similar instruction sequences, the additional \code{select} instructions added by \name are also very small (\ie less than 2\% of the total instructions)  
The program \code{libyaml} is an exception here; It has a compile-time overhead of 7.60ms due to the highest number of locations where \name transformation applied (260) and the largest number of LLVM-IR instructions (56,688) among all the programs.
Although, theoretically, in the worst case, \name tranformation can generate an exponential (in the levels of branch nesting) number of \code{select} statements, this is not observed in practice.
We perform a static nesting depth analysis at the LLVM-IR-level; therefore, reported numbers may not directly correspond to the nesting structures of the source. 
In our analysis, branch nesting denotes the maximum depth of conditional nesting within a function, whereas control-flow nesting denotes nesting arising from any control-flow constructs, including loops, conditionals, and \texttt{switch} statements.
We observe an anomaly in \code{libosip} and \code{libyaml}. 
Although their source code structures may not appear to be deeply nested, constructs such as \texttt{else-if} are lowered into nested conditional structures in the LLVM-IR, resulting in higher measured nesting depths. 

\paragraph{\bf Ablation Study} We conduct an ablation study to compare applying CFMSE only on symbolic branches vs. applying it indiscriminately on all possible branches. 
We evaluate both variants on \code{libosip} to measure their applicability and end-to-end execution time for KLEE. 

When restricting \name to symbolic branches, it is applied on 26 branches, all of which are \code{if-then} branches, resulting in a total of 26 newly added select instructions. 
On the other hand, the indiscriminate variant is applied to 158 branches, including 98 \code{if-then} branches and 60 \code{if-then-else} branches, adding a total of 242 new select instructions. 
Applying symbolic variable analysis significantly reduces the number of inserted \code{select} instructions. 
We emphasise that these results would differ from results reported in Table~\ref{tab:overhead}. 
In this experiment, we isolate the \name transformation, making it independent of the standard optimization passes that KLEE applies in its default pipeline.

We execute both variants with an input size of 9 to ensure completion within a reasonable time. 
The symbolic-branch-only variant completes in 166.4 seconds, whereas the indiscriminate variant requires 179.13 seconds. 
The performance degradation stems from blindly merging branches that can introduce additional false-positives and make solver queries long and complex. 
This observation is supported by KLEE statistics, which show that although the symbolic-branch-only variant generates more solver queries (205 vs 198), it spends less total time in solving queries (26.44s vs. 32.67s). 
The number of queries for the indiscriminate variant is reduced because we now have fewer branches, but each query becomes significantly more complex and ends up increasing the overall solver time. 
Although the time difference is not substantial, the symbolic-branch-only variant could be further optimized by employing a more fine-grained symbolic variable analysis than the one used in this work. 
The inclusion of symbolic array accesses exacerbates the symbolic execution runtime, and we therefore identify them as prime candidates where a fine-grained analysis can make a significant difference in total execution time. 

These results suggest that while aggressive merging reduces query count, the increased complexity can outweigh its benefits. 
Although an inflection point may exist as we show for \code{merge-sort} in Figure~\ref{fig:mergesort-three}, this occurs at larger runtimes, given that the solver does not timeout, making it impractical for realistic use cases. 

\section{Discussion}
\label{cfmse:sec:discuss}

Several limitations restrict the applicability and generality of \name.
We discuss those limitations in detail here.
\paragraph{{\bf Effect of fixed-sized input.}}
The \name approach works well when the symbolic input to the program has a concrete size.
If the size of the input is symbolic it leads to expensive array constraints.
In that case, every time we access this buffer in the program, the DSE engine will make expensive solver calls.
\name is not beneficial for programs that have symbolic addresses as described in Section~\ref{subsec:transformation}.
Therefore, even if \name transformation is applicable, that does not make a discernible improvement in the performance of DSE on such programs.

\paragraph{{\bf Cost model for constraint complexity.}}
Unlike DARM~\cite{darm}, \name does not use a cost model to reason about the profitability of the branch elimination.
\name uses simple heuristics based on the symbolic variable analysis to decide whether to apply the branch elimination.
For example, if the branch contains memory accesses with symbolic addresses, \name does not apply the branch elimination.
\name does not consider how complex the constraints can become if used in a future solver query.
This requires estimating the complexity of queries a given sequence of instructions could generate.
Statically estimating the query cost has been explored by previous work related to state merging~\cite{efficient-state-merging}. 
However, estimating the query cost of compile-time transformation needs to be better explored. 
This is an interesting problem that can be explored in future work.

\paragraph{{\bf Effect of loop-carried dependences.}}
The benefits of \name are sensitive to the structure of dependences in program loops. 
If all possible paths in a loop do not matter for branches outside the loop, statically merging the branches inside the loop is highly beneficial.
This kind of program behavior can be expected when the program loops {\em does not} contain any loop-carried dependences.
\iftoggle{bv}{
We observe superior performance for benchmarks such as \code{toupper}, \code{dilation}, and \code{erosion} because of this reason in accelerating bounded verification.
}{}
In these programs, constraints generated by \name~{\em do not} grow on each iteration of the loop, and therefore the solver queries are not expensive. 
However, if the loops in the program contain loop-carried dependences, the constraints generated by \name can grow on each iteration.
\iftoggle{bv}{
This is evident in \code{bitonicsort} when used to show accelerated bounded verification. 
}{}
In these programs, \name's performance is comparable to state merging even though \name significantly reduces the number of solver calls.
In other words, due to the loop-carried dependences, the constraints generated by \name keep getting more complex when the input size increases.
This limits the scalability of \name for specific programs. 
However, DSE is not intended for large input sizes where none of the techniques are scalable.
This observation gives us good insights into designing a cost model for static evaluation of query cost.

\paragraph{{\bf Suitability of the application of DSE}}
The program transformed using \name has fewer paths than its original version.
With \name, DSE will explore fewer program paths and generate fewer test cases. 
Also, \name might not generate test cases that cover specific program paths in the original program due to static path merging.
Therefore, if the goal is to generate test cases to achieve the maximum possible coverage for a given program, \name might not be the best choice because it cannot generate test cases that cover all the paths in the original program.
This limitation is standard for any approach that tries to merge program states, including state merging~\cite{efficient-state-merging}.
However, path explosion makes achieving the maximum possible coverage for larger programs unrealistic.
Static path merging capability of \name can make DSE reach new program locations faster and achieve more coverage within a limited time.
This is a valuable property for testing real-world programs.

\section{Related Work}
\label{cfmse:sec:relatedwork}


\paragraph{Dynamic Techniques}
Many attempts have been made to mitigate the path explosion problem by guiding DSE only on interesting program paths~\cite{dse-survey}. 
Function and loop summarization produces summaries of frequently executed code sections and reuses that to avoid path explosion~\cite{anand08, xie16}.
Path equivalence and subsumption-based techniques work by avoiding redundant program paths that do not reveal new information~\cite{postconditioned-se, qi13}. 
Under-constrained symbolic execution applies to functions or code regions by isolating them from the surrounding application\cite{underconstrained-se}.
Any constraints that are applied to a tested function by external (\ie global) sources are considered \emph{under-constrained}.
State merging~\cite{efficient-state-merging} attempts to combine different program paths explored during symbolic execution to avoid path explosion.

\paragraph{Compiler Techniques}
Instead of improving the heuristics for guiding symbolic execution, the application of targeted program transformations to enhance the performance of symbolic execution is also well-studied in the literature.
Testability transformations are a type of program transformation that improves the ability of a given test generation method to generate tests for the original untransformed program. 
Prior work has shown that such transformations can improve the performance for test generation techniques~\cite{testability-transformations}
Collingbourne~\etal used branch predication to convert symbolic branches into \emph{ite} expressions, thereby reducing the number of explored program paths exponentially~\cite{collingbourne11}.
Wagner~\etal proposed \emph{-OVERIFY}, a new compiler optimization switch (\ie a collection of optimizations) that enables fast verification of programs~\cite{overify}. 
Wagner used DSE as a case study to show that selective application of compiler optimizations like constant folding, loop unswitching, and if-conversion can drastically reduce the time spent in verification.  
Cadar~\etal argued that compiler optimizations must be a first-class ingredient in a practical DSE platform~\cite{targeted-transformations}.
Sharma~\etal proposes JavaRanger that focuses on path constraints merging with dynamic method region inlining and early-return summarization to accelerate DSE of Java programs~\cite{javaranger2020}.
Perry~\etal proposed a semantics-preserving program transformation to accelerate DSE on programs with array accesses~\cite{perry17}.
Inserting extra code to improve test generation techniques has also been explored in compiler testing~\cite{donaldson21}.

\section{Conclusion}
\label{sec:conclusion}

Dynamic state merging is a well-known technique used for mitigating path explosion Dynamic Symbolic Execution (DSE), where similar program states are merged to reduce the number of explored states.
This work proposes a novel non-semantics-preserving and failure-preserving compiler transformation called \name to enable better compile-time state merging.
We develop a framework for detecting false positive failures that may be introduced by failure-preserving transformations like \name.
Our evaluation shows \name's utility in improving the scalability of DSE and achieving faster code coverage.

\newpage

\section*{Data Availability Statement}
\label{sec:data-availability-stmt}

The artifact supporting this paper is publicly available in an open-access repository on Zenodo~\cite{artifact}. The repository includes the \code{LLVM} implementation of our pass, the modified KLEE version, all evaluation benchmarks, and the scripts necessary to reproduce the results in this paper. The artifact is actively maintained at \url{https://github.com/hassanaleem/CFMSE-OOPSLA26-Artifact}. 
\begin{acks}
	We sincerely thank the anonymous reviewers for their constructive feedback and careful evaluation, which substantially improved this paper. 
	We are also grateful to Matthieu Lemerre for his insightful suggestions that enriched our work.
	This work was partially supported by NSF grants CCF-2216987, CCF-1919197, and CCF-1908504.	
\end{acks}

\bibliographystyle{ACM-Reference-Format} 
\bibliography{bibliography}


\begin{thebibliography}{55}


\ifx \showCODEN    \undefined \def \showCODEN     #1{\unskip}     \fi
\ifx \showISBNx    \undefined \def \showISBNx     #1{\unskip}     \fi
\ifx \showISBNxiii \undefined \def \showISBNxiii  #1{\unskip}     \fi
\ifx \showISSN     \undefined \def \showISSN      #1{\unskip}     \fi
\ifx \showLCCN     \undefined \def \showLCCN      #1{\unskip}     \fi
\ifx \shownote     \undefined \def \shownote      #1{#1}          \fi
\ifx \showarticletitle \undefined \def \showarticletitle #1{#1}   \fi
\ifx \showURL      \undefined \def \showURL       {\relax}        \fi
\providecommand\bibfield[2]{#2}
\providecommand\bibinfo[2]{#2}
\providecommand\natexlab[1]{#1}
\providecommand\showeprint[2][]{arXiv:#2}

\bibitem[Anand et~al\mbox{.}(2008)]%
        {anand08}
\bibfield{author}{\bibinfo{person}{Saswat Anand}, \bibinfo{person}{Patrice
  Godefroid}, {and} \bibinfo{person}{Nikolai Tillmann}.}
  \bibinfo{year}{2008}\natexlab{}.
\newblock \showarticletitle{Demand-Driven Compositional Symbolic Execution}. In
  \bibinfo{booktitle}{\emph{Proceedings of the Theory and Practice of Software,
  14th International Conference on Tools and Algorithms for the Construction
  and Analysis of Systems}} (Budapest, Hungary)
  \emph{(\bibinfo{series}{TACAS'08/ETAPS'08})}.
  \bibinfo{publisher}{Springer-Verlag}, \bibinfo{address}{Berlin, Heidelberg},
  \bibinfo{pages}{367–381}.
\newblock
\showISBNx{3540787992}


\bibitem[Baldoni et~al\mbox{.}(2018)]%
        {dse-survey}
\bibfield{author}{\bibinfo{person}{Roberto Baldoni}, \bibinfo{person}{Emilio
  Coppa}, \bibinfo{person}{Daniele~Cono D’elia}, \bibinfo{person}{Camil
  Demetrescu}, {and} \bibinfo{person}{Irene Finocchi}.}
  \bibinfo{year}{2018}\natexlab{}.
\newblock \showarticletitle{A Survey of Symbolic Execution Techniques}.
\newblock \bibinfo{journal}{\emph{ACM Comput. Surv.}} \bibinfo{volume}{51},
  \bibinfo{number}{3}, Article \bibinfo{articleno}{50} (\bibinfo{date}{may}
  \bibinfo{year}{2018}), \bibinfo{numpages}{39}~pages.
\newblock
\showISSN{0360-0300}
\href{https://doi.org/10.1145/3182657}{doi:\nolinkurl{10.1145/3182657}}


\bibitem[{Batcher}(1968)]%
        {bitonic}
\bibfield{author}{\bibinfo{person}{K.~E. {Batcher}}.}
  \bibinfo{year}{1968}\natexlab{}.
\newblock \showarticletitle{Sorting networks and their applications}. In
  \bibinfo{booktitle}{\emph{Proceedings of the April 30--May 2, 1968, spring
  joint computer conference (AFIPS '68 (Spring))}}. \bibinfo{pages}{307–314}.
\newblock
\href{https://doi.org/10.1145/1468075.1468121}{doi:\nolinkurl{10.1145/1468075.1468121}}


\bibitem[Bruttomesso et~al\mbox{.}(2010)]%
        {open-smt}
\bibfield{author}{\bibinfo{person}{Roberto Bruttomesso}, \bibinfo{person}{Edgar
  Pek}, \bibinfo{person}{Natasha Sharygina}, {and} \bibinfo{person}{Aliaksei
  Tsitovich}.} \bibinfo{year}{2010}\natexlab{}.
\newblock \showarticletitle{The OpenSMT Solver}. In
  \bibinfo{booktitle}{\emph{Tools and Algorithms for the Construction and
  Analysis of Systems}}, \bibfield{editor}{\bibinfo{person}{Javier Esparza}
  {and} \bibinfo{person}{Rupak Majumdar}} (Eds.). \bibinfo{publisher}{Springer
  Berlin Heidelberg}, \bibinfo{address}{Berlin, Heidelberg},
  \bibinfo{pages}{150--153}.
\newblock
\showISBNx{978-3-642-12002-2}


\bibitem[Cadar(2015)]%
        {targeted-transformations}
\bibfield{author}{\bibinfo{person}{Cristian Cadar}.}
  \bibinfo{year}{2015}\natexlab{}.
\newblock \showarticletitle{Targeted Program Transformations for Symbolic
  Execution}. In \bibinfo{booktitle}{\emph{Proceedings of the 2015 10th Joint
  Meeting on Foundations of Software Engineering}} (Bergamo, Italy)
  \emph{(\bibinfo{series}{ESEC/FSE 2015})}. \bibinfo{publisher}{Association for
  Computing Machinery}, \bibinfo{address}{New York, NY, USA},
  \bibinfo{pages}{906–909}.
\newblock
\showISBNx{9781450336758}
\href{https://doi.org/10.1145/2786805.2803205}{doi:\nolinkurl{10.1145/2786805.2803205}}


\bibitem[Cadar et~al\mbox{.}(2008)]%
        {klee}
\bibfield{author}{\bibinfo{person}{Cristian Cadar}, \bibinfo{person}{Daniel
  Dunbar}, {and} \bibinfo{person}{Dawson Engler}.}
  \bibinfo{year}{2008}\natexlab{}.
\newblock \showarticletitle{KLEE: Unassisted and Automatic Generation of
  High-Coverage Tests for Complex Systems Programs}. In
  \bibinfo{booktitle}{\emph{Proceedings of the 8th USENIX Conference on
  Operating Systems Design and Implementation}} (San Diego, California)
  \emph{(\bibinfo{series}{OSDI'08})}. \bibinfo{publisher}{USENIX Association},
  \bibinfo{address}{USA}, \bibinfo{pages}{209–224}.
\newblock


\bibitem[Chen et~al\mbox{.}(2003)]%
        {gen-tail-merge-sas03}
\bibfield{author}{\bibinfo{person}{Wen-Ke Chen}, \bibinfo{person}{Bengu Li},
  {and} \bibinfo{person}{Rajiv Gupta}.} \bibinfo{year}{2003}\natexlab{}.
\newblock \showarticletitle{Code Compaction of Matching Single-Entry
  Multiple-Exit Regions}. In \bibinfo{booktitle}{\emph{Proceedings of the 10th
  International Conference on Static Analysis}} (San Diego, CA, USA)
  \emph{(\bibinfo{series}{SAS'03})}. \bibinfo{publisher}{Springer-Verlag},
  \bibinfo{address}{Berlin, Heidelberg}, \bibinfo{pages}{401–417}.
\newblock
\showISBNx{3540403256}


\bibitem[Chuang et~al\mbox{.}(2003)]%
        {phi-predication}
\bibfield{author}{\bibinfo{person}{Weihaw Chuang}, \bibinfo{person}{B. Calder},
  {and} \bibinfo{person}{J. Ferrante}.} \bibinfo{year}{2003}\natexlab{}.
\newblock \showarticletitle{Phi-predication for light-weight if-conversion}. In
  \bibinfo{booktitle}{\emph{International Symposium on Code Generation and
  Optimization, 2003. CGO 2003.}} \bibinfo{pages}{179--190}.
\newblock
\href{https://doi.org/10.1109/CGO.2003.1191544}{doi:\nolinkurl{10.1109/CGO.2003.1191544}}


\bibitem[Collingbourne et~al\mbox{.}(2011)]%
        {collingbourne11}
\bibfield{author}{\bibinfo{person}{Peter Collingbourne},
  \bibinfo{person}{Cristian Cadar}, {and} \bibinfo{person}{Paul~H.J. Kelly}.}
  \bibinfo{year}{2011}\natexlab{}.
\newblock \showarticletitle{Symbolic Crosschecking of Floating-Point and SIMD
  Code}. In \bibinfo{booktitle}{\emph{Proceedings of the Sixth Conference on
  Computer Systems}} (Salzburg, Austria) \emph{(\bibinfo{series}{EuroSys
  '11})}. \bibinfo{publisher}{Association for Computing Machinery},
  \bibinfo{address}{New York, NY, USA}, \bibinfo{pages}{315–328}.
\newblock
\showISBNx{9781450306348}
\href{https://doi.org/10.1145/1966445.1966475}{doi:\nolinkurl{10.1145/1966445.1966475}}


\bibitem[Converse et~al\mbox{.}(2017)]%
        {converse-17}
\bibfield{author}{\bibinfo{person}{Hayes Converse}, \bibinfo{person}{Oswaldo
  Olivo}, {and} \bibinfo{person}{Sarfraz Khurshid}.}
  \bibinfo{year}{2017}\natexlab{}.
\newblock \showarticletitle{Non-Semantics-Preserving Transformations for
  Higher-Coverage Test Generation Using Symbolic Execution}. In
  \bibinfo{booktitle}{\emph{2017 IEEE International Conference on Software
  Testing, Verification and Validation (ICST)}}. \bibinfo{pages}{241--252}.
\newblock
\href{https://doi.org/10.1109/ICST.2017.29}{doi:\nolinkurl{10.1109/ICST.2017.29}}


\bibitem[Cormen et~al\mbox{.}(2009)]%
        {CLRS}
\bibfield{author}{\bibinfo{person}{Thomas~H. Cormen},
  \bibinfo{person}{Charles~E. Leiserson}, \bibinfo{person}{Ronald~L. Rivest},
  {and} \bibinfo{person}{Clifford Stein}.} \bibinfo{year}{2009}\natexlab{}.
\newblock \bibinfo{booktitle}{\emph{Introduction to Algorithms, Third Edition}
  (\bibinfo{edition}{3rd} ed.)}.
\newblock \bibinfo{publisher}{The MIT Press}.
\newblock
\showISBNx{0262033844}


\bibitem[{Coutinho} et~al\mbox{.}(2011)]%
        {branch-fusion}
\bibfield{author}{\bibinfo{person}{B. {Coutinho}}, \bibinfo{person}{D.
  {Sampaio}}, \bibinfo{person}{F.~M.~Q. {Pereira}}, {and} \bibinfo{person}{W.
  {Meira Jr.}}} \bibinfo{year}{2011}\natexlab{}.
\newblock \showarticletitle{Divergence Analysis and Optimizations}. In
  \bibinfo{booktitle}{\emph{2011 International Conference on Parallel
  Architectures and Compilation Techniques}}. \bibinfo{pages}{320--329}.
\newblock
\href{https://doi.org/10.1109/PACT.2011.63}{doi:\nolinkurl{10.1109/PACT.2011.63}}


\bibitem[de~Moura and Bj{\o}rner(2008)]%
        {z3}
\bibfield{author}{\bibinfo{person}{Leonardo de Moura} {and}
  \bibinfo{person}{Nikolaj Bj{\o}rner}.} \bibinfo{year}{2008}\natexlab{}.
\newblock \showarticletitle{Z3: An Efficient SMT Solver}. In
  \bibinfo{booktitle}{\emph{Tools and Algorithms for the Construction and
  Analysis of Systems}}, \bibfield{editor}{\bibinfo{person}{C.~R. Ramakrishnan}
  {and} \bibinfo{person}{Jakob Rehof}} (Eds.). \bibinfo{publisher}{Springer
  Berlin Heidelberg}, \bibinfo{address}{Berlin, Heidelberg},
  \bibinfo{pages}{337--340}.
\newblock
\showISBNx{978-3-540-78800-3}


\bibitem[Donaldson et~al\mbox{.}(2021)]%
        {donaldson21}
\bibfield{author}{\bibinfo{person}{Alastair~F. Donaldson},
  \bibinfo{person}{Paul Thomson}, \bibinfo{person}{Vasyl Teliman},
  \bibinfo{person}{Stefano Milizia}, \bibinfo{person}{Andr\'{e}~Perez Maselco},
  {and} \bibinfo{person}{Antoni Karpi\'{n}ski}.}
  \bibinfo{year}{2021}\natexlab{}.
\newblock \showarticletitle{Test-Case Reduction and Deduplication Almost for
  Free with Transformation-Based Compiler Testing}. In
  \bibinfo{booktitle}{\emph{Proceedings of the 42nd ACM SIGPLAN International
  Conference on Programming Language Design and Implementation}} (Virtual,
  Canada) \emph{(\bibinfo{series}{PLDI 2021})}. \bibinfo{publisher}{Association
  for Computing Machinery}, \bibinfo{address}{New York, NY, USA},
  \bibinfo{pages}{1017–1032}.
\newblock
\showISBNx{9781450383912}
\href{https://doi.org/10.1145/3453483.3454092}{doi:\nolinkurl{10.1145/3453483.3454092}}


\bibitem[{Free Software Foundation}(2018)]%
        {variadic-func}
\bibfield{author}{\bibinfo{person}{{Free Software Foundation}}.}
  \bibinfo{year}{2018}\natexlab{}.
\newblock \bibinfo{title}{{V}ariadic {F}unctions ({T}he {G}{N}{U} {C}
  {L}ibrary) --- gnu.org}.
\newblock
  \bibinfo{howpublished}{\url{https://www.gnu.org/software/libc/manual/html_node/Variadic-Functions.html}}.
\newblock
\newblock
\shownote{[Accessed 23-Feb-2023]}.


\bibitem[Ganesh and Dill(2007)]%
        {stp}
\bibfield{author}{\bibinfo{person}{Vijay Ganesh} {and}
  \bibinfo{person}{David~L. Dill}.} \bibinfo{year}{2007}\natexlab{}.
\newblock \showarticletitle{A Decision Procedure for Bit-Vectors and Arrays}.
  In \bibinfo{booktitle}{\emph{Computer Aided Verification}},
  \bibfield{editor}{\bibinfo{person}{Werner Damm} {and} \bibinfo{person}{Holger
  Hermanns}} (Eds.). \bibinfo{publisher}{Springer Berlin Heidelberg},
  \bibinfo{address}{Berlin, Heidelberg}, \bibinfo{pages}{519--531}.
\newblock
\showISBNx{978-3-540-73368-3}


\bibitem[{GNU Project}(2008a)]%
        {chcon}
\bibfield{author}{\bibinfo{person}{{GNU Project}}.}
  \bibinfo{year}{2008}\natexlab{a}.
\newblock \bibinfo{title}{{chcon(1) - Linux man page}}.
\newblock
\urldef\tempurl%
\url{https://linux.die.net/man/1/chcon}
\showURL{%
\tempurl}
\newblock
\shownote{[Accessed 18-Apr-2023]}.


\bibitem[{GNU Project}(2008b)]%
        {chown}
\bibfield{author}{\bibinfo{person}{{GNU Project}}.}
  \bibinfo{year}{2008}\natexlab{b}.
\newblock \bibinfo{title}{{chown(1) - Linux man page}}.
\newblock
\urldef\tempurl%
\url{https://linux.die.net/man/1/chown}
\showURL{%
\tempurl}
\newblock
\shownote{[Accessed 05-Oct-2025]}.


\bibitem[{GNU Project}(2008c)]%
        {gnu-coreutils}
\bibfield{author}{\bibinfo{person}{{GNU Project}}.}
  \bibinfo{year}{2008}\natexlab{c}.
\newblock \bibinfo{title}{{Coreutils -- GNU core utilities}}.
\newblock
\urldef\tempurl%
\url{https://www.gnu.org/software/coreutils/}
\showURL{%
\tempurl}
\newblock
\shownote{Accessed: 2023-04-18}.


\bibitem[{GNU Project}(2008d)]%
        {mkdir}
\bibfield{author}{\bibinfo{person}{{GNU Project}}.}
  \bibinfo{year}{2008}\natexlab{d}.
\newblock \bibinfo{title}{{mkdir(1) - Linux man page}}.
\newblock
\urldef\tempurl%
\url{https://linux.die.net/man/1/mkdir}
\showURL{%
\tempurl}
\newblock
\shownote{[Accessed 05-Oct-2025]}.


\bibitem[{GNU Project}(2008e)]%
        {mkfifo}
\bibfield{author}{\bibinfo{person}{{GNU Project}}.}
  \bibinfo{year}{2008}\natexlab{e}.
\newblock \bibinfo{title}{{mkfifo(1) - Linux man page}}.
\newblock
\urldef\tempurl%
\url{https://linux.die.net/man/1/mkfifo}
\showURL{%
\tempurl}
\newblock
\shownote{[Accessed 05-Oct-2025]}.


\bibitem[{GNU Project}(2023a)]%
        {gnu-libtasn1}
\bibfield{author}{\bibinfo{person}{{GNU Project}}.}
  \bibinfo{year}{2023}\natexlab{a}.
\newblock \bibinfo{title}{{GNU Libtasn1}}.
\newblock
\urldef\tempurl%
\url{https://www.gnu.org/software/libtasn1/}
\showURL{%
\tempurl}
\newblock
\shownote{Accessed: 2023-04-18}.


\bibitem[{GNU Project}(2023b)]%
        {gnu-osip}
\bibfield{author}{\bibinfo{person}{{GNU Project}}.}
  \bibinfo{year}{2023}\natexlab{b}.
\newblock \bibinfo{title}{{The GNU oSIP library}}.
\newblock
\urldef\tempurl%
\url{https://www.gnu.org/software/osip/}
\showURL{%
\tempurl}
\newblock
\shownote{Accessed: 2023-04-18}.


\bibitem[Godefroid et~al\mbox{.}(2005)]%
        {dart}
\bibfield{author}{\bibinfo{person}{Patrice Godefroid}, \bibinfo{person}{Nils
  Klarlund}, {and} \bibinfo{person}{Koushik Sen}.}
  \bibinfo{year}{2005}\natexlab{}.
\newblock \showarticletitle{DART: Directed Automated Random Testing}. In
  \bibinfo{booktitle}{\emph{Proceedings of the 2005 ACM SIGPLAN Conference on
  Programming Language Design and Implementation}} (Chicago, IL, USA)
  \emph{(\bibinfo{series}{PLDI '05})}. \bibinfo{publisher}{Association for
  Computing Machinery}, \bibinfo{address}{New York, NY, USA},
  \bibinfo{pages}{213–223}.
\newblock
\showISBNx{1595930566}
\href{https://doi.org/10.1145/1065010.1065036}{doi:\nolinkurl{10.1145/1065010.1065036}}


\bibitem[{Google}(2023)]%
        {googleprotobuf}
\bibfield{author}{\bibinfo{person}{{Google}}.}
  \bibinfo{year}{{2023}}\natexlab{}.
\newblock \bibinfo{title}{{Protocol Buffers Documentation}}.
\newblock
\urldef\tempurl%
\url{https://protobuf.dev/}
\showURL{%
\tempurl}
\newblock
\shownote{[Accessed 11-Dec-2023]}.


\bibitem[Harman et~al\mbox{.}(2004)]%
        {testability-transformations}
\bibfield{author}{\bibinfo{person}{Mark Harman}, \bibinfo{person}{Lin Hu},
  \bibinfo{person}{Rob Hierons}, \bibinfo{person}{Joachim Wegener},
  \bibinfo{person}{Harmen Sthamer}, \bibinfo{person}{Andr\'{e} Baresel}, {and}
  \bibinfo{person}{Marc Roper}.} \bibinfo{year}{2004}\natexlab{}.
\newblock \showarticletitle{Testability Transformation}.
\newblock  \bibinfo{volume}{30}, \bibinfo{number}{1} (\bibinfo{date}{jan}
  \bibinfo{year}{2004}), \bibinfo{pages}{3–16}.
\newblock
\showISSN{0098-5589}
\href{https://doi.org/10.1109/TSE.2004.1265732}{doi:\nolinkurl{10.1109/TSE.2004.1265732}}


\bibitem[{ISO/IEC JTC 1/SC 22}(2011)]%
        {ISO-C11}
\bibfield{author}{\bibinfo{person}{{ISO/IEC JTC 1/SC 22}}.}
  \bibinfo{year}{2011}\natexlab{}.
\newblock \bibinfo{booktitle}{\emph{Programming Languages --- C}}.
\newblock \bibinfo{type}{International Standard} ISO/IEC 9899:2011.
  \bibinfo{institution}{International Organization for Standardization},
  \bibinfo{address}{Geneva, Switzerland}.
\newblock
\urldef\tempurl%
\url{https://www.iso.org/standard/57853.html}
\showURL{%
\tempurl}


\bibitem[Karrenberg and Hack(2012)]%
        {llvm-div-analysis}
\bibfield{author}{\bibinfo{person}{Ralf Karrenberg} {and}
  \bibinfo{person}{Sebastian Hack}.} \bibinfo{year}{2012}\natexlab{}.
\newblock \showarticletitle{Improving Performance of OpenCL on CPUs}. In
  \bibinfo{booktitle}{\emph{Compiler Construction}},
  \bibfield{editor}{\bibinfo{person}{Michael O'Boyle}} (Ed.).
  \bibinfo{publisher}{Springer Berlin Heidelberg}, \bibinfo{address}{Berlin,
  Heidelberg}, \bibinfo{pages}{1--20}.
\newblock
\showISBNx{978-3-642-28652-0}


\bibitem[Khurshid et~al\mbox{.}(2003)]%
        {khurshid-se}
\bibfield{author}{\bibinfo{person}{Sarfraz Khurshid},
  \bibinfo{person}{Corina~S. P\u{a}s\u{a}reanu}, {and} \bibinfo{person}{Willem
  Visser}.} \bibinfo{year}{2003}\natexlab{}.
\newblock \showarticletitle{Generalized Symbolic Execution for Model Checking
  and Testing}. In \bibinfo{booktitle}{\emph{Proceedings of the 9th
  International Conference on Tools and Algorithms for the Construction and
  Analysis of Systems}} (Warsaw, Poland) \emph{(\bibinfo{series}{TACAS'03})}.
  \bibinfo{publisher}{Springer-Verlag}, \bibinfo{address}{Berlin, Heidelberg},
  \bibinfo{pages}{553–568}.
\newblock
\showISBNx{3540008985}


\bibitem[{KLEE}(2023)]%
        {coreutils-experiment}
\bibfield{author}{\bibinfo{person}{{KLEE}}.} \bibinfo{year}{{2023}}\natexlab{}.
\newblock \bibinfo{title}{{OSDI'08 Coreutils Experiments}}.
\newblock
\urldef\tempurl%
\url{https://klee.github.io/docs/coreutils-experiments/}
\showURL{%
\tempurl}
\newblock
\shownote{[Accessed 18-Apr-2023]}.


\bibitem[Korczynski(2020)]%
        {utf8-bug}
\bibfield{author}{\bibinfo{person}{David Korczynski}.}
  \bibinfo{year}{2020}\natexlab{}.
\newblock \bibinfo{booktitle}{\emph{Some minor overflow bugs}}.
\newblock
\urldef\tempurl%
\url{https://github.com/sheredom/utf8.h/issues/70}
\showURL{%
\tempurl}
\newblock
\shownote{GitHub issue \#70 in \texttt{sheredom/utf8.h}}.


\bibitem[Kuznetsov et~al\mbox{.}(2012)]%
        {efficient-state-merging}
\bibfield{author}{\bibinfo{person}{Volodymyr Kuznetsov},
  \bibinfo{person}{Johannes Kinder}, \bibinfo{person}{Stefan Bucur}, {and}
  \bibinfo{person}{George Candea}.} \bibinfo{year}{2012}\natexlab{}.
\newblock \showarticletitle{Efficient State Merging in Symbolic Execution}. In
  \bibinfo{booktitle}{\emph{Proceedings of the 33rd ACM SIGPLAN Conference on
  Programming Language Design and Implementation}} (Beijing, China)
  \emph{(\bibinfo{series}{PLDI '12})}. \bibinfo{publisher}{Association for
  Computing Machinery}, \bibinfo{address}{New York, NY, USA},
  \bibinfo{pages}{193–204}.
\newblock
\showISBNx{9781450312059}
\href{https://doi.org/10.1145/2254064.2254088}{doi:\nolinkurl{10.1145/2254064.2254088}}


\bibitem[{Lattner} and {Adve}(2004)]%
        {llvm}
\bibfield{author}{\bibinfo{person}{C. {Lattner}} {and} \bibinfo{person}{V.
  {Adve}}.} \bibinfo{year}{2004}\natexlab{}.
\newblock \showarticletitle{LLVM: a compilation framework for lifelong program
  analysis transformation}. In \bibinfo{booktitle}{\emph{International
  Symposium on Code Generation and Optimization, 2004. CGO 2004.}}
  \bibinfo{pages}{75--86}.
\newblock
\href{https://doi.org/10.1109/CGO.2004.1281665}{doi:\nolinkurl{10.1109/CGO.2004.1281665}}


\bibitem[LLVM(2023)]%
        {llvm-cost-model}
\bibfield{author}{\bibinfo{person}{LLVM}.} \bibinfo{year}{2023}\natexlab{}.
\newblock \bibinfo{title}{CostModel.cpp}.
\newblock
\urldef\tempurl%
\url{https://llvm.org/doxygen/CostModel_8cpp_source.html}
\showURL{%
\tempurl}
\newblock
\shownote{[Accessed 24-Mar-2023]}.


\bibitem[{LLVM Compiler Infrastructure}(2003)]%
        {llvm-language-ref}
\bibfield{author}{\bibinfo{person}{{LLVM Compiler Infrastructure}}.}
  \bibinfo{year}{2003}\natexlab{}.
\newblock \bibinfo{title}{LLVM Language Reference Manual}.
\newblock \bibinfo{howpublished}{\url{https://llvm.org/docs/LangRef.html}}.
\newblock
\newblock
\shownote{[Accessed 23-Feb-2023]}.


\bibitem[{LLVM Project}(2022)]%
        {llvm-simplifycfg}
\bibfield{author}{\bibinfo{person}{{LLVM Project}}.}
  \bibinfo{year}{2022}\natexlab{}.
\newblock \bibinfo{title}{{SimplifyCFG.cpp}}.
\newblock
\urldef\tempurl%
\url{https://llvm.org/doxygen/SimplifyCFG_8cpp_source.html}
\showURL{%
\tempurl}
\newblock
\shownote{[Accessed 12-Apr-2023]}.


\bibitem[Peng et~al\mbox{.}(2018)]%
        {tfuzz}
\bibfield{author}{\bibinfo{person}{Hui Peng}, \bibinfo{person}{Yan
  Shoshitaishvili}, {and} \bibinfo{person}{Mathias Payer}.}
  \bibinfo{year}{2018}\natexlab{}.
\newblock \showarticletitle{T-Fuzz: Fuzzing by Program Transformation}. In
  \bibinfo{booktitle}{\emph{2018 IEEE Symposium on Security and Privacy (SP)}}.
  \bibinfo{pages}{697--710}.
\newblock
\href{https://doi.org/10.1109/SP.2018.00056}{doi:\nolinkurl{10.1109/SP.2018.00056}}


\bibitem[Perry et~al\mbox{.}(2017)]%
        {perry17}
\bibfield{author}{\bibinfo{person}{David~M. Perry}, \bibinfo{person}{Andrea
  Mattavelli}, \bibinfo{person}{Xiangyu Zhang}, {and} \bibinfo{person}{Cristian
  Cadar}.} \bibinfo{year}{2017}\natexlab{}.
\newblock \showarticletitle{Accelerating Array Constraints in Symbolic
  Execution}. In \bibinfo{booktitle}{\emph{Proceedings of the 26th ACM SIGSOFT
  International Symposium on Software Testing and Analysis}} (Santa Barbara,
  CA, USA) \emph{(\bibinfo{series}{ISSTA 2017})}.
  \bibinfo{publisher}{Association for Computing Machinery},
  \bibinfo{address}{New York, NY, USA}, \bibinfo{pages}{68–78}.
\newblock
\showISBNx{9781450350761}
\href{https://doi.org/10.1145/3092703.3092728}{doi:\nolinkurl{10.1145/3092703.3092728}}


\bibitem[Pezoa et~al\mbox{.}(2016)]%
        {json-schema}
\bibfield{author}{\bibinfo{person}{Felipe Pezoa}, \bibinfo{person}{Juan~L.
  Reutter}, \bibinfo{person}{Fernando Suarez}, \bibinfo{person}{Mart\'{\i}n
  Ugarte}, {and} \bibinfo{person}{Domagoj Vrgo\v{c}}.}
  \bibinfo{year}{2016}\natexlab{}.
\newblock \showarticletitle{Foundations of JSON Schema}. In
  \bibinfo{booktitle}{\emph{Proceedings of the 25th International Conference on
  World Wide Web}} (Montr\'{e}al, Qu\'{e}bec, Canada)
  \emph{(\bibinfo{series}{WWW '16})}. \bibinfo{publisher}{International World
  Wide Web Conferences Steering Committee}, \bibinfo{address}{Republic and
  Canton of Geneva, CHE}, \bibinfo{pages}{263–273}.
\newblock
\showISBNx{9781450341431}
\href{https://doi.org/10.1145/2872427.2883029}{doi:\nolinkurl{10.1145/2872427.2883029}}


\bibitem[Phillips(2008)]%
        {Phillips-2008}
\bibfield{author}{\bibinfo{person}{Dwayne Phillips}.}
  \bibinfo{year}{2008}\natexlab{}.
\newblock \bibinfo{booktitle}{\emph{Image Processing in C}}.
\newblock \bibinfo{publisher}{BPB Publications}, \bibinfo{address}{New Delhi,
  India}.
\newblock
\showISBNx{9788170295150}


\bibitem[Qi et~al\mbox{.}(2013)]%
        {qi13}
\bibfield{author}{\bibinfo{person}{Dawei Qi}, \bibinfo{person}{Hoang D.~T.
  Nguyen}, {and} \bibinfo{person}{Abhik Roychoudhury}.}
  \bibinfo{year}{2013}\natexlab{}.
\newblock \showarticletitle{Path Exploration Based on Symbolic Output}.
\newblock  \bibinfo{volume}{22}, \bibinfo{number}{4}, Article
  \bibinfo{articleno}{32} (\bibinfo{date}{oct} \bibinfo{year}{2013}),
  \bibinfo{numpages}{41}~pages.
\newblock
\showISSN{1049-331X}
\href{https://doi.org/10.1145/2522920.2522925}{doi:\nolinkurl{10.1145/2522920.2522925}}


\bibitem[Ramos and Engler(2015)]%
        {underconstrained-se}
\bibfield{author}{\bibinfo{person}{David~A. Ramos} {and}
  \bibinfo{person}{Dawson Engler}.} \bibinfo{year}{2015}\natexlab{}.
\newblock \showarticletitle{Under-Constrained Symbolic Execution: Correctness
  Checking for Real Code}. In \bibinfo{booktitle}{\emph{Proceedings of the 24th
  USENIX Conference on Security Symposium}} (Washington, D.C.)
  \emph{(\bibinfo{series}{SEC'15})}. \bibinfo{publisher}{USENIX Association},
  \bibinfo{address}{USA}, \bibinfo{pages}{49–64}.
\newblock
\showISBNx{9781931971232}


\bibitem[Rocha et~al\mbox{.}(2023)]%
        {hybf}
\bibfield{author}{\bibinfo{person}{Rodrigo C.~O. Rocha},
  \bibinfo{person}{Charitha Saumya}, \bibinfo{person}{Kirshanthan
  Sundararajah}, \bibinfo{person}{Pavlos Petoumenos}, \bibinfo{person}{Milind
  Kulkarni}, {and} \bibinfo{person}{Michael F.~P. O{'}Boyle}.}
  \bibinfo{year}{2023}\natexlab{}.
\newblock \showarticletitle{HyBF: A Hybrid Branch Fusion Strategy for Code Size
  Reduction}. In \bibinfo{booktitle}{\emph{Proceedings of the 32nd ACM SIGPLAN
  International Conference on Compiler Construction}} (Montr\'{e}al, QC,
  Canada) \emph{(\bibinfo{series}{CC 2023})}. \bibinfo{publisher}{Association
  for Computing Machinery}, \bibinfo{address}{New York, NY, USA},
  \bibinfo{pages}{156–--167}.
\newblock
\showISBNx{9798400700880}
\href{https://doi.org/10.1145/3578360.3580267}{doi:\nolinkurl{10.1145/3578360.3580267}}


\bibitem[Saumya et~al\mbox{.}(2026)]%
        {artifact}
\bibfield{author}{\bibinfo{person}{Charitha Saumya}, \bibinfo{person}{Muhammad
  Hassan}, \bibinfo{person}{Rohan Gangaraju}, \bibinfo{person}{Milind
  Kulkarni}, {and} \bibinfo{person}{Kirshanthan Sundararajah}.}
  \bibinfo{year}{2026}\natexlab{}.
\newblock \bibinfo{booktitle}{\emph{Taming the Hydra: Targeted Control-Flow
  Transformations for Dynamic Symbolic Execution}}.
\newblock
\href{https://doi.org/10.5281/zenodo.19120042}{doi:\nolinkurl{10.5281/zenodo.19120042}}


\bibitem[Saumya et~al\mbox{.}(2022)]%
        {darm}
\bibfield{author}{\bibinfo{person}{Charitha Saumya},
  \bibinfo{person}{Kirshanthan Sundararajah}, {and} \bibinfo{person}{Milind
  Kulkarni}.} \bibinfo{year}{2022}\natexlab{}.
\newblock \showarticletitle{DARM: Control-Flow Melding for SIMT Thread
  Divergence Reduction}. In \bibinfo{booktitle}{\emph{2022 IEEE/ACM
  International Symposium on Code Generation and Optimization (CGO)}}.
  \bibinfo{pages}{1--13}.
\newblock
\href{https://doi.org/10.1109/CGO53902.2022.9741285}{doi:\nolinkurl{10.1109/CGO53902.2022.9741285}}


\bibitem[Sen et~al\mbox{.}(2005)]%
        {cute}
\bibfield{author}{\bibinfo{person}{Koushik Sen}, \bibinfo{person}{Darko
  Marinov}, {and} \bibinfo{person}{Gul Agha}.} \bibinfo{year}{2005}\natexlab{}.
\newblock \showarticletitle{CUTE: A Concolic Unit Testing Engine for C}. In
  \bibinfo{booktitle}{\emph{Proceedings of the 10th European Software
  Engineering Conference Held Jointly with 13th ACM SIGSOFT International
  Symposium on Foundations of Software Engineering}} (Lisbon, Portugal)
  \emph{(\bibinfo{series}{ESEC/FSE-13})}. \bibinfo{publisher}{Association for
  Computing Machinery}, \bibinfo{address}{New York, NY, USA},
  \bibinfo{pages}{263–272}.
\newblock
\showISBNx{1595930140}
\href{https://doi.org/10.1145/1081706.1081750}{doi:\nolinkurl{10.1145/1081706.1081750}}


\bibitem[Sharma et~al\mbox{.}(2020)]%
        {javaranger2020}
\bibfield{author}{\bibinfo{person}{Vaibhav Sharma}, \bibinfo{person}{Soha
  Hussein}, \bibinfo{person}{Michael~W. Whalen}, \bibinfo{person}{Stephen
  McCamant}, {and} \bibinfo{person}{Willem Visser}.}
  \bibinfo{year}{2020}\natexlab{}.
\newblock \showarticletitle{Java Ranger: statically summarizing regions for
  efficient symbolic execution of Java}. In
  \bibinfo{booktitle}{\emph{Proceedings of the 28th ACM Joint Meeting on
  European Software Engineering Conference and Symposium on the Foundations of
  Software Engineering}} (Virtual Event, USA) \emph{(\bibinfo{series}{ESEC/FSE
  2020})}. \bibinfo{publisher}{Association for Computing Machinery},
  \bibinfo{address}{New York, NY, USA}, \bibinfo{pages}{123–134}.
\newblock
\showISBNx{9781450370431}
\href{https://doi.org/10.1145/3368089.3409734}{doi:\nolinkurl{10.1145/3368089.3409734}}


\bibitem[{sheredom}(2023a)]%
        {json-h}
\bibfield{author}{\bibinfo{person}{{sheredom}}.}
  \bibinfo{year}{{2023}}\natexlab{a}.
\newblock \bibinfo{title}{{json.h : A simple single header solution to parsing
  JSON in C and C++.}}
\newblock
\urldef\tempurl%
\url{https://github.com/sheredom/json.h/tree/master}
\showURL{%
\tempurl}
\newblock
\shownote{[Accessed 11-Dec-2023]}.


\bibitem[{sheredom}(2023b)]%
        {utf-8}
\bibfield{author}{\bibinfo{person}{{sheredom}}.}
  \bibinfo{year}{{2023}}\natexlab{b}.
\newblock \bibinfo{title}{{Utf-8: A simple one header solution to supporting
  utf8 strings in C and C++}}.
\newblock
\urldef\tempurl%
\url{https://github.com/sheredom/utf8.h/tree/master}
\showURL{%
\tempurl}
\newblock
\shownote{[Accessed 11-Dec-2023]}.


\bibitem[Shiloach and Vishkin(1982)]%
        {shiloach82}
\bibfield{author}{\bibinfo{person}{Yossi Shiloach} {and} \bibinfo{person}{Uzi
  Vishkin}.} \bibinfo{year}{1982}\natexlab{}.
\newblock \showarticletitle{An O(logn) parallel connectivity algorithm}.
\newblock \bibinfo{journal}{\emph{Journal of Algorithms}} \bibinfo{volume}{3},
  \bibinfo{number}{1} (\bibinfo{year}{1982}), \bibinfo{pages}{57--67}.
\newblock
\showISSN{0196-6774}
\href{https://doi.org/10.1016/0196-6774(82)90008-6}{doi:\nolinkurl{10.1016/0196-6774(82)90008-6}}


\bibitem[Sobel(2014)]%
        {sobel1968}
\bibfield{author}{\bibinfo{person}{Irwin Sobel}.}
  \bibinfo{year}{2014}\natexlab{}.
\newblock \showarticletitle{An Isotropic 3x3 Image Gradient Operator}.
\newblock \bibinfo{journal}{\emph{Presentation at Stanford A.I. Project 1968}}
  (\bibinfo{date}{02} \bibinfo{year}{2014}).
\newblock


\bibitem[Trabish et~al\mbox{.}(2018)]%
        {chopper}
\bibfield{author}{\bibinfo{person}{David Trabish}, \bibinfo{person}{Andrea
  Mattavelli}, \bibinfo{person}{Noam Rinetzky}, {and} \bibinfo{person}{Cristian
  Cadar}.} \bibinfo{year}{2018}\natexlab{}.
\newblock \showarticletitle{Chopped Symbolic Execution}. In
  \bibinfo{booktitle}{\emph{Proceedings of the 40th International Conference on
  Software Engineering}} (Gothenburg, Sweden) \emph{(\bibinfo{series}{ICSE
  '18})}. \bibinfo{publisher}{Association for Computing Machinery},
  \bibinfo{address}{New York, NY, USA}, \bibinfo{pages}{350–360}.
\newblock
\showISBNx{9781450356381}
\href{https://doi.org/10.1145/3180155.3180251}{doi:\nolinkurl{10.1145/3180155.3180251}}


\bibitem[Wagner et~al\mbox{.}(2013)]%
        {overify}
\bibfield{author}{\bibinfo{person}{Jonas Wagner}, \bibinfo{person}{Volodymyr
  Kuznetsov}, {and} \bibinfo{person}{George Candea}.}
  \bibinfo{year}{2013}\natexlab{}.
\newblock \showarticletitle{{-OVERIFY}: Optimizing Programs for Fast
  {Verification}}. In \bibinfo{booktitle}{\emph{14th Workshop on Hot Topics in
  Operating Systems (HotOS XIV)}}. \bibinfo{publisher}{USENIX Association},
  \bibinfo{address}{Santa Ana Pueblo, NM}.
\newblock
\urldef\tempurl%
\url{https://www.usenix.org/conference/hotos13/session/wagner}
\showURL{%
\tempurl}


\bibitem[Xie et~al\mbox{.}(2016)]%
        {xie16}
\bibfield{author}{\bibinfo{person}{Xiaofei Xie}, \bibinfo{person}{Bihuan Chen},
  \bibinfo{person}{Yang Liu}, \bibinfo{person}{Wei Le}, {and}
  \bibinfo{person}{Xiaohong Li}.} \bibinfo{year}{2016}\natexlab{}.
\newblock \showarticletitle{Proteus: Computing Disjunctive Loop Summary via
  Path Dependency Analysis} \emph{(\bibinfo{series}{FSE 2016})}.
  \bibinfo{publisher}{Association for Computing Machinery},
  \bibinfo{address}{New York, NY, USA}, \bibinfo{pages}{61–72}.
\newblock
\showISBNx{9781450342186}
\href{https://doi.org/10.1145/2950290.2950340}{doi:\nolinkurl{10.1145/2950290.2950340}}


\bibitem[Yi et~al\mbox{.}(2015)]%
        {postconditioned-se}
\bibfield{author}{\bibinfo{person}{Qiuping Yi}, \bibinfo{person}{Zijiang Yang},
  \bibinfo{person}{Shengjian Guo}, \bibinfo{person}{Chao Wang},
  \bibinfo{person}{Jian Liu}, {and} \bibinfo{person}{Chen Zhao}.}
  \bibinfo{year}{2015}\natexlab{}.
\newblock \showarticletitle{Postconditioned Symbolic Execution}. In
  \bibinfo{booktitle}{\emph{2015 IEEE 8th International Conference on Software
  Testing, Verification and Validation (ICST)}}. \bibinfo{pages}{1--10}.
\newblock
\href{https://doi.org/10.1109/ICST.2015.7102601}{doi:\nolinkurl{10.1109/ICST.2015.7102601}}


\end{thebibliography}

\newpage
\appendix
\section{Artifact Appendix}
\label{sec:artifact}

\subsection{Overview}
The artifact supports the results shown in Section~\ref{cfmse:sec:eval} of the paper, including small benchmark evaluations, coverage measurements, bug detection results, and scalability experiments. 
An implementation of the \name transformation, benchmark programs, and evaluation scripts necessary to produce the results are included in the artifact.

Note that the configurations provided in this artifact are designed for a machine with more than 164~GB RAM, but can be easily adapted to systems with lower memory. 
We provide details in Section~\ref{artifact-evaluation} on the changes required in the configuration to prevent system crashes. 

\subsection{Artifact Checklist}
\begin{itemize}
  \item \textbf{LLVM version}: LLVM 14.0.0.
  \item \textbf{Compilation}: CMake with LLVM build system.
  \item \textbf{Run-time environment}: Linux operating system.
  \item \textbf{Hardware}: No specific hardware requirements. We have tested our approach on Intel(R) Xeon(R) Gold 6430 (64 cores) processor and 256 GB RAM
  \item \textbf{Metrics}: Coverage, runtime speedup, number of queries, query size, and explored path count.
  \item \textbf{Output}: Optimized LLVM IR, binaries, raw logs.
  \item \textbf{Experiment Workflow}: Download source code and benchmarks,
    compile the source code or build the Dockerfile, run the scripts to generate the results
  \item \textbf{Disk space}: Approximately 40~GB of free disk space for Docker build, or 25~GB for Dockerless builds.   
  \item \textbf{Workflow automation}: Bash and Python scripts.
\end{itemize}

\subsection{Dependencies}
This artifact is packaged as a Docker image and requires no manual dependency installation beyond Docker itself. 
All software dependencies are installed automatically upon building the container from the provided Dockerfile. 
No additional system libraries, tools, or external services are required on the host machine.

\subsection{Installation}
\label{installation}

{\bf Warning.} Building the image may take from a few minutes on a 128-core machine up to an hour on smaller machines, as LLVM and KLEE are built from source. 
By default, this build uses 4 cores to avoid out-of-memory failures. On systems with sufficient memory and CPU resources, build time can be reduced by replacing \texttt{-j4} in the Dockerfile with \texttt{-j\$(nproc)}.
 
To build the Docker image:
\begin{tcolorbox}[
   		enhanced,
    	colback=blue!5,
    	colframe=blue!60,
    	rounded corners
        ]
\begin{verbatim}
docker build -t cfmse .
\end{verbatim}
\end{tcolorbox}

To start an interactive container and mount the artifact directory:
\begin{tcolorbox}[
   		enhanced,
    	colback=blue!5,
    	colframe=blue!60,
    	rounded corners
        ]
\begin{verbatim}
docker run -it \
  -v "$(pwd):/artifact" \
  --name cfmse-session \
  --memory 220g \
  --memory-swap 240g \
  cfmse
\end{verbatim}
\end{tcolorbox}

We recommend this memory configuration for a system with 256 GB memory, used for all our experiments. For a system with lower memory, adjust the values accordingly. 

The container can be stopped and restarted later without rebuilding the image:
\begin{tcolorbox}[
   		enhanced,
    	colback=blue!5,
    	colframe=blue!60,
    	rounded corners,
        ]
\begin{verbatim}
docker start cfmse-session
docker exec -it cfmse-session bash
\end{verbatim}
\end{tcolorbox}

\subsection{Kick The Tires}
\label{kick-the-tires}
To verify that the artifact is functioning correctly, we provide a small test script. Navigate to the \texttt{KickTheTires} directory in the artifact and run the provided \texttt{runme} script. This script applies the \name transformation and run KLEE on the \texttt{ToUpper} function and produces a CSV file named \texttt{toupper.csv}.

We included a \texttt{reference-toupper.csv} file in the directory to allow comparison with a reference. The results may not be identical across systems, but should remain qualitatively consistent. 

If the generated file contains non-zero values, the artifact is working as expected. Some fields in the CSV may contain \texttt{OOT}, which is expected and indicates KLEE has hit a timeout (60s) for this particular analysis. This script takes approximately 3 minutes to complete. Once this check succeeds, you may proceed to the remaining experiments.

\subsection{Building From Scratch}
\label{artifact-building}
If you installed the artifact using the provided Dockerfile as described in the previous section, all dependencies are already installed, and the artifact is already built, so you may skip to the next section. 
However, if you wish to build the artifact from scratch, follow the instructions below.

By default, these instructions use 4 cores to avoid out-of-memory failures. On systems with sufficient memory and CPU resources, build time can be reduced by replacing \texttt{-j4} with \texttt{-j\$(nproc)}. 

\paragraph{Setup artifact directory}
Follow instructions below to download the artifact and setup the directory.
zenodo-link: \url{https://zenodo.org/records/18489026/files/CFMSE-OOPSLA26-Artifact-master.zip}
\begin{tcolorbox}[
   		enhanced,
    	colback=blue!5,
    	colframe=blue!60,
    	rounded corners,
        ]
\begin{verbatim}
# Download source Code
wget <zenodo-link>

# To unzip the file
python3 -m zipfile -e \ 
"CFMSE-OOPSLA26-Artifact-master.zip" \ 
cfmse-artifact

cd cfmse-artifact
\end{verbatim}
\end{tcolorbox}

Since this artifact is distributed as a Zenodo zip file, the executable bits may not be preserved. As a result you might run into \texttt{permission-denied} errors. Following commands fix the permission issue.
\begin{tcolorbox}[
   		enhanced,
    	colback=blue!5,
    	colframe=blue!60,
    	rounded corners,
        ]
\begin{verbatim}
find . \
    -type f \
    \( \
    -name "*.sh" \
    -o -name "configure" \
    -o -name "klee-stats" \
    \) \
    -exec chmod +x {} +

# For any remaining files
chmod +x <file>
\end{verbatim}
\end{tcolorbox}

\begin{tcolorbox}[
   		enhanced,
    	colback=blue!5,
    	colframe=blue!60,
    	rounded corners,
        title={System Dependencies}
        ]
\begin{verbatim}
sudo apt-get update
sudo apt-get install -y build-essential \
    cmake bison flex libboost-all-dev perl zlib1g-dev minisat \
    ninja-build git wget curl python3 python-is-python3 \
    libncurses5-dev libncursesw5-dev \
    libgoogle-perftools-dev libsqlite3-dev \
    gperf pkg-config libprotobuf-dev protobuf-compiler libprotoc-dev

# Create a python venv
python3 -m venv ./venv
# Activate the venv
source ./venv/bin/activate

pip3 install setuptools \
  tabulate pandas matplotlib psutil wllvm
\end{verbatim}
\end{tcolorbox}

\begin{tcolorbox}[
   		enhanced,
    	colback=blue!5,
    	colframe=blue!60,
    	rounded corners,
        title={STP Solver}
        ]
\begin{verbatim}
cd stp
mkdir -p build
cd build
cmake ..
make
cd ../..
\end{verbatim}
\end{tcolorbox}

\begin{tcolorbox}[
   		enhanced,
    	colback=blue!5,
    	colframe=blue!60,
    	rounded corners,
        title={LLVM and CFMSE}
        ]
\begin{verbatim}
cd llvm-project-cfmse
cmake -G Ninja -B build_cfmse -S llvm \
    -DLLVM_ENABLE_PROJECTS="clang" \
    -DLLVM_BUILD_EXAMPLES=ON \
    -DLLVM_TARGETS_TO_BUILD="X86" \
    -DCMAKE_BUILD_TYPE=Release \
    -DLLVM_ENABLE_ASSERTIONS=ON \
    -DCMAKE_CXX_COMPILER=g++ \
    -DLLVM_USE_LINKER=gold \
    -DCMAKE_INSTALL_PREFIX="$(pwd)/install"

ninja -C build_cfmse -j4
ninja -C build_cfmse install
cd ..
\end{verbatim}
\end{tcolorbox}
\vspace{-1.2em}
\begin{tcolorbox}[
   		enhanced,
    	colback=blue!5,
    	colframe=blue!60,
    	rounded corners,
        title={LLVM Environment Variables}
        ]
\begin{verbatim}
export EXPERIMENT_HOME="$(pwd)"
export LLVM_BUILD_DIR="${EXPERIMENT_HOME}/llvm-project-cfmse/build_cfmse"
export CLANG="${LLVM_BUILD_DIR}/bin/clang"
export OPT="${LLVM_BUILD_DIR}/bin/opt"
\end{verbatim}
\end{tcolorbox}

\begin{tcolorbox}[
   		enhanced,
    	colback=blue!5,
    	colframe=blue!60,
    	rounded corners,
        title={KLEE-uClibc}
        ]
\begin{verbatim}
cd klee-uclibc
./configure --make-llvm-lib \
    --with-cc "${LLVM_BUILD_DIR}/bin/clang" \
    --with-llvm-config "${LLVM_BUILD_DIR}/bin/llvm-config"
make -j4
cd ..
\end{verbatim}
\end{tcolorbox}

\begin{tcolorbox}[
   		enhanced,
    	colback=blue!5,
    	colframe=blue!60,
    	rounded corners,
        title={KLEE}
        ]
\begin{verbatim}
cd klee
mkdir -p build
cd build
cmake -DENABLE_SOLVER_STP=ON \
    -DENABLE_POSIX_RUNTIME=ON \
    -DENABLE_KLEE_UCLIBC=ON \
    -DKLEE_UCLIBC_PATH="${EXPERIMENT_HOME}/klee-uclibc/" \
    -DLLVM_CONFIG_BINARY="${LLVM_BUILD_DIR}/bin/llvm-config" \
    -DLLVMCC="${LLVM_BUILD_DIR}/bin/clang" \
    -DLLVMCXX="${LLVM_BUILD_DIR}/bin/clang++" \
    ..
make -j4
cd ../..
\end{verbatim}
\end{tcolorbox}

\begin{tcolorbox}[
   		enhanced,
    	colback=blue!5,
    	colframe=blue!60,
    	rounded corners,
        title={KLEE Paths}
        ]
\begin{verbatim}
export KLEE_BUILD_DIR="${EXPERIMENT_HOME}/klee/build"
export KLEE="${KLEE_BUILD_DIR}/bin/klee"
export KTEST_TOOL="${KLEE_BUILD_DIR}/bin/ktest-tool"
export KLEE_INCLUDE="${EXPERIMENT_HOME}/klee/include"
export KLEE_STATS="${KLEE_BUILD_DIR}/bin/klee-stats"
\end{verbatim}
\end{tcolorbox}

\subsection{Experiment Workflow}
The files and configurations necessary to reduce the results are organized by table or figure number. For example, to reproduce the results in Figure 7, navigate to the \texttt{fig7-coverage} folder within the artifact directory. This directory contains a runme script that can be executed to generate all the corresponding coverage plots. 

Alternatively, users may navigate to specific directories and run individual \texttt{run-cfm-driver} scripts. Configuration options for each program can be found within the \texttt{driver\_options.json} file in the program's directory.

The configuration file defines two set of options. The first includes standard KLEE flags that control symbolic execution behaviour within KLEE, such as search heuristics, runtime and memory limits, etc. The second set contains CFM-specific options, allowing users to enable CFMSE for symbolic branches, force CFMSE on all branches or restrict branch merging to loops only.

These configurations can be modified to provide flexibility to evaluate different program variants and experiments under controlled memory and runtime conditions for different machines. 

\texttt{CFM\_OPTIONS} refers to the flag introduced by our artifact within KLEE, whereas \texttt{KLEE\_OPTIONS} refers to KLEE's built-in flags. Any flags supported by KLEE may be added under \texttt{KLEE\_OPTIONS}. 
\texttt{PROG\_ARGS} specifies the command line arguments passed to the target program. Some programs in our benchmarks require parameters such as input or buffer size. When present, this usually corresponds to the size of the symbolic array in the target program. 

Table~\ref{flags} provides a list of the most relevant flags along with their descriptions.

\begin{table}[h]
	\vspace{-1em}
	\centering
	\caption{KLEE and CFMSE configuration flags}
    \label{flags}
	\vspace{-1em}
	\begin{tabular}{@{}lp{0.65\columnwidth}@{}}
		\toprule
		\textbf{Flags} & \textbf{Description} \\
		\midrule
		\texttt{max-time} &
		Sets the timeout for KLEE (within \texttt{KLEE\_OPTIONS}). \\
		
		\texttt{max-memory} &
		Specifies the memory limit for KLEE (within \texttt{KLEE\_OPTIONS}). \\
		
		\texttt{klee-cfmse} &
		Enables the CFMSE (only to symbolic branches by default). \\
		
		\texttt{klee-force-cfmse} &
		Forces CFMSE to apply to all branches, not just symbolic ones. \\
		
		\texttt{klee-cfmse\allowbreak-run\allowbreak-only\allowbreak-on\allowbreak-loops} &
		Restricts CFMSE to run only on loops when set to true. \\
		
		\texttt{klee-cfmse\allowbreak-loads\allowbreak-symbolic} &
		Treats load operations as symbolic during CFMSE. \\
		\bottomrule
	\end{tabular}
	\vspace{-1em}
\end{table}

We recommend using the following configuration: 
\begin{tcolorbox}[
   		enhanced,
    	colback=green!5,
    	colframe=green!60,
    	rounded corners,
        ]
\begin{lstlisting}[language={}]
{
  "KLEE_OPTIONS" : "--max-time=3h -max-memory=32768 -simplify-sym-indices -libc=uclibc -search=nurs:covnew --only-output-states-covering-new --write-cov",
  "CFMSE_IGNORE_JSON" : "cfmse_ignore.json",
  "CFM_OPTIONS" : "-klee-cfmse  -klee-cfmse-run-only-on-loops=false -klee-cfmse-loads-symbolic",
  "PROG_ARGS" : "  "
}
\end{lstlisting}
\end{tcolorbox}

To execute on machines with memory lower than 160GB, navigate to each benchmark program directory and modify the \texttt{driver\_options.json} based on your system specifications to avoid potential system crashes. For Table 3 modify the \texttt{makefile.in} instead of \texttt{driver\_options.json}. Ensure that the \texttt{max-memory} is set below the total memory available to your system. Additionally, configure Docker memory to avoid system crashes as described in \ref{installation}.

\subsection{List of Claims} 

\begin{itemize}
  \item \textbf{Improvement in Performance of DSE} (Section~\ref{cfmse:subsec:accel_se})  
  This claim is evaluated in the paper using Table~\ref{tab:results_without_verify}.  
  The artifact reproduces this claim using the experiment in Section~\ref{artifact:table-small-benchmarks}.

  \item \textbf{Performance Scaling of DSE} (Section~\ref{cfmse:subsec:accel_se})  
  This claim is evaluated in the paper using Figure~\ref{fig:mergesort-three}.  
  The artifact reproduces this claim using the experiment in Section~\ref{artifact-merge-sort-paths}.

  \item \textbf{Improvement in Performance of Coverage} (Section~\ref{cfmse:subsec:coverage})  
  This claim is evaluated in the paper using Figure~\ref{fig:coverage}.  
  The artifact reproduces this claim using the experiment in Section~\ref{artifact-coverage}.

  \item \textbf{Improvement in Performance of Bug Discovery} (Section~\ref{cfmse:subsec:bug})  
  This claim is evaluated in the paper using Figure~\ref{fig:overall}, which reports bug discovery performance, and Figure~\ref{fig:killed_states}, which shows the improved likelihood of finding a bug as the input to the program scales.  
  The artifact reproduced this claim using the experiment in Section~\ref{artifact-bugexperiments}.

  \item \textbf{Minimal Overhead} (Section~\ref{sec:eval_overhead})  
  This claim is evaluated in the paper using Table~\ref{tab:overhead}.  
  The artifact reproduced this claim using the experiment in Section~\ref{artifact:overhead}.
\end{itemize}

Results may differ on machines with limited RAM. Although our approach is designed to operate under lower memory constraints, extremely limited resources may cause the OS to terminate KLEE. Such terminations can affect coverage or bug detection results. In some cases, the bug may not be discovered because the analysis does not run long enough to reach the relevant execution path.

Because KLEE's memory usage is highly dependent on the explored paths, it is hard to predict how many benchmarks will be affected by memory-constrained systems. When the memory conditions used in the experiment are met, all claims are expected to hold.

\subsection{Evaluation and Expected Results}
\label{artifact-evaluation}

\subsubsection{Table~\ref{tab:results_without_verify}}
\label{artifact:table-small-benchmarks}
\par
\noindent

To generate the table, run the following commands. 
\begin{tcolorbox}[
   		enhanced,
    	colback=blue!5,
    	colframe=blue!60,
    	rounded corners,
        ]
\begin{verbatim}
cd table3-small-benchmarks
./runme.sh

# To run an individal benchmark
./run-bench-<benchmark>.sh
\end{verbatim}
\end{tcolorbox}

\paragraph{Runtime.}
The total runtime depends on the benchmark being executed. \texttt{runme.sh} executes all these benchmarks and takes an approximately 132 hours. Approximate execution times (per benchmark) are as follows:

\begin{itemize}
    \item \texttt{toupper}: 10 hours
    \item \texttt{bitonic sort}: 10 hours
    \item \texttt{connected components}: 10 hours
    \item \texttt{prim}: 10 hours
    \item \texttt{merge sort}: 40 hours
    \item \texttt{transitive closure}: 6 hours
    \item \texttt{detect edges}: 15 hours
    \item \texttt{floyd warshall}: 15 hours
    \item \texttt{erosion}: 6 hours
    \item \texttt{dijkstra}: 10 hours
    \item \texttt{total time}: 132 hours
\end{itemize}

\paragraph{Timeout and repetitions.}
Each benchmark is executed with a timeout of 1 hour and repeated for 5 runs. To reduce the number of repetitions, navigate to the corresponding script
\path{run-bench-<benchmark>.sh} and modify the third argument of the \texttt{run\_bench} invocation. For example:
\begin{tcolorbox}[
   		enhanced,
    	colback=blue!5,
    	colframe=blue!60,
    	rounded corners,
        ]
\begin{verbatim}
run_bench "bitonic_sort" "4 8 16" "5" "klee_cfm"
\end{verbatim}
\end{tcolorbox}

Replace \texttt{5} with the desired number of runs.

\paragraph{Adjusting the timeout.}
To reduce the timeout value, edit \texttt{makefile.in} and change the \texttt{max-time} flag to the desired duration. Note that modifying the timeout or the number of runs may affect the total number of generated queries.

For the CFMSE transformation, only the \texttt{merge sort} benchmark reaches the timeout limit. All other benchmarks complete within one minute.

\subsubsection{Figure~\ref{fig:mergesort-three}}
\label{artifact-merge-sort-paths}
\par
\noindent

To generate plots, run the following commands. 

\begin{tcolorbox}[
   		enhanced,
    	colback=blue!5,
    	colframe=blue!60,
    	rounded corners,
        ]
\begin{verbatim}
cd fig6-completed-paths-merge-sort
./runme.sh
\end{verbatim}
\end{tcolorbox}

This takes approximately 30 hours to run. 

\subsubsection{Figure~\ref{fig:coverage}}
\par
\noindent
\label{artifact-coverage}

To generate plots, run the following commands.

\begin{tcolorbox}[
   		enhanced,
    	colback=blue!5,
    	colframe=blue!60,
    	rounded corners,
        ]
\begin{verbatim}
cd fig7-coverage
./runme.sh

# To run individual experiments
cd <benchmark>
./run-cfm-driver.sh
\end{verbatim}
\end{tcolorbox}

\paragraph{Runtime.}
The total runtime depends on the benchmark being executed. \texttt{runme.sh} executes all these benchmarks and takes approximately 16 hours. Approximate execution times (per benchmark) are as follows:
\begin{itemize}
    \item \texttt{libosip}: 3 hours
    \item \texttt{libtasn}: 1 minute
    \item \texttt{chcon}: 1 hour
    \item \texttt{chown}: 1 hour
    \item \texttt{mkdir}: 1 hour
    \item \texttt{mkfifo}: 1 hour
    \item \texttt{json.h}: 3 hours
    \item \texttt{protobuf}: 1 hour
    \item \texttt{utf8-valid}: 1 hour
    \item \texttt{utf8-nonvaid}: 30 minutes
    \item \texttt{total}: 13 hours
\end{itemize}

We provide logs generated by rerunning the experiments on the same server configuration used for the paper. These logs can be found in the \texttt{preexisting-logs} directory within the \texttt{fig7-coverage}. To generate plots from these files, follow the steps below: 

\begin{tcolorbox}[
   		enhanced,
    	colback=blue!5,
    	colframe=blue!60,
    	rounded corners,
        ]
\begin{verbatim}
cd fig7-coverage
cd preexisting-logs
cd <benchmark>
python3 ${KLEE_BUILD_DIR}/../scripts/coverage_graph.py \
    ./coverage-cfmse.csv \
    ./coverage-nocfmse.csv
\end{verbatim}
\end{tcolorbox}

\subsubsection{Figure~\ref{fig:overall} and Figure~\ref{fig:killed_states}}
\par
\noindent
\label{artifact-bugexperiments}

To generate plots, run the following commands.

\begin{tcolorbox}[
   		enhanced,
    	colback=blue!5,
    	colframe=blue!60,
    	rounded corners,
        ]
\begin{verbatim}
cd fig10-11-bug-detection-and-killed-states
./runme.sh

# To run individual experiments
cd <benchmark>
./run.sh
\end{verbatim}
\end{tcolorbox}
\paragraph{Runtime.}
The total runtime depends on the benchmark being executed. \texttt{runme.sh} executes all these benchmarks and takes an approximately 96 hours. Approximate execution times (per benchmark) are as follows:
\begin{itemize}
    \item \texttt{tiny-regex-c}: 1 hour
    \item \texttt{utf8}: 1 hour
    \item \texttt{json.h}: 1 hour
    \item \texttt{libosip}: 72 hours
    \item \texttt{libyaml}: 21 hours
    \item \texttt{total}: 96 hours
\end{itemize}

Plots for Figure~\ref{fig:killed_states} depend on the results of \texttt{libyaml} and \texttt{libosip}. 
The framework will auto-generate them.

We provide the logs used for the paper. Each benchmark subdirectory contains a \texttt{logs-paper} directory with these logs. To generate plots directly from these log files, run the following commands below:

\begin{tcolorbox}[
   		enhanced,
    	colback=blue!5,
    	colframe=blue!60,
    	rounded corners,
        ]
\begin{verbatim}
cd fig10-11-bug-detection-and-killed-states
cd <benchmark>
python3 ../read_cfm_output.py \ 
    logs-paper <benchmark> seconds
\end{verbatim}
\end{tcolorbox}

\newpage
\subsubsection{Table~\ref{tab:overhead}}
\label{artifact:overhead}
\par
\noindent

Table~\ref{tab:overhead} relies on the coverage and bug experiments being run and can not be run standalone. To generate the table in Table~\ref{tab:overhead} first run all the experiments in Section~\ref{artifact-coverage} and Section~\ref{artifact-bugexperiments},  then run the following commands.

\begin{tcolorbox}[
   		enhanced,
    	colback=blue!5,
    	colframe=blue!60,
    	rounded corners,
        ]
\begin{verbatim}
cd table4-overhead
./runme.sh
\end{verbatim}
\end{tcolorbox}

\subsubsection{Execution Logs}
\par
\noindent

After successfully executing the \texttt{runme} script, the framework will generate the corresponding plots or tables. These outputs should closely match the results presented in the paper. Minor variations may occur due to differences in hardware architecture and nondeterminism in KLEE's execution. 

In addition to the plots and tables, the framework also generates a detailed execution log, as well as a transformed intermediate representation (IR) file. 
The log file reports summary statistics related to the CFMSE transformation, including the number of CFMSE applications and the number of select conversions achieved by the CFMSE transformation.  
Listing~\ref{lst:cfmse-log} shows a snippet of the execution log for \texttt{tiny-regex-c} (Figure~\ref{fig:subfig_regex}). 
The complete log file contains a complete execution trace of KLEE and the CFM driver, reporting all detected false-positive bugs as well as the true-positive bug. 

\begin{tcolorbox}[
   		enhanced,
    	colback=green!5,
    	colframe=green!60,
    	rounded corners,
        ]
\begin{lstlisting}[caption={Excerpt from execution log of tiny-regex-c}, label={lst:cfmse-log}]
INFO[cfmse]: Running CFMSE ...
INFO[cfmse]: Select Instructions before CFMSE : 3
INFO[cfmse]: Cond Branch Instructions before CFMSE : 183
INFO[cfmse]: Uncond Branch Instructions before CFMSE : 239
INFO[cfmse]: Total Instructions before CFMSE : 3877
INFO[cfmse]: Number of successful applications of CFMSE : 13
INFO[cfmse]: Number of if-then merges : 12
INFO[cfmse]: Number of if-then-else merges : 1
INFO[cfmse]: Number of select instructions added : 13
INFO[cfmse]: Select Instructions after CFMSE : 16
INFO[cfmse]: Cond Branch Instructions after CFMSE : 170
INFO[cfmse]: Uncond Branch Instructions after CFMSE : 238
INFO[cfmse]: Total Instructions after CFMSE : 3707
CFMSE took 1.000000e-02 ms.
\end{lstlisting}
\end{tcolorbox}

\subsection{Customization and Reusability}

{\bf Licenses.} This artifact includes modified versions of LLVM and KLEE, which remain licensed under their original licenses. 
All artifact-specific scripts, drivers, and experimental infrastructure are released under the MIT License.

\noindent {\bf Public Availability.} The artifact archived on Zenodo: \url{https://doi.org/10.5281/zenodo.18489026} and actively maintained on GitHub: 
\url{https://github.com/hassanaleem/CFMSE-OOPSLA26-Artifact}.

We now discuss how the proposed transformation can be extended to programs beyond those considered in this paper. 
Since our approach is implemented as an LLVM transformation, it can be applied to any program that can be compiled into LLVM bitcode.

We present this process in two parts. First, we describe how to apply the transformation as a standalone LLVM pass. 
Second, we explain how to use the transformed program within the KLEE-based test driver.

\paragraph{Standalone LLVM Transformation.}
Let us consider a program \texttt{tutorial.c}. To compile this program into LLVM bitcode, we use the following command:

\begin{tcolorbox}[
   		enhanced,
    	colback=blue!5,
    	colframe=blue!60,
    	rounded corners,
        ]
\begin{verbatim}
$CLANG -c -emit-llvm -Xclang -disable-O0-optnone \
-O0 tutorial.c -o tutorial.bc
\end{verbatim}
\end{tcolorbox}

Once the LLVM bitcode is generated, we apply the CFMSE transformation using \texttt{opt}:

\begin{tcolorbox}[
   		enhanced,
    	colback=blue!5,
    	colframe=blue!60,
    	rounded corners,
        ]
\begin{verbatim}
$OPT -f -cfmse tutorial.bc -o tutorial-cfmse.bc
\end{verbatim}
\end{tcolorbox}

By default, this transformation applies only to symbolic branches in the program. However, we can force the transformation to  be applied to all branches by using the \texttt{-force-cfmse} flag:

\begin{tcolorbox}[
   		enhanced,
    	colback=blue!5,
    	colframe=blue!60,
    	rounded corners,
        ]
\begin{verbatim}
$OPT -f -cfmse -force-cfmse tutorial.bc -o tutorial-cfmse.bc
\end{verbatim}
\end{tcolorbox}

Here, \texttt{\$CLANG} and \texttt{\$OPT} refer to the paths of the \texttt{clang} and \texttt{opt} binaries, respectively, as defined in our Dockerfile.

\paragraph{Using the Transformation with KLEE}
To use the transformed program with the test driver, we first compile the target program into LLVM bitcode, as described above. We then define a configuration file, \texttt{driver\_options.json}, which specifies the options passed to KLEE and the CFMSE extension.

A sample configuration file is shown below and can be used without modification:

\begin{tcolorbox}[
   		enhanced,
    	colback=green!5,
    	colframe=green!60,
    	rounded corners,
        ]
\begin{lstlisting}[language={}]
{
  "KLEE_OPTIONS" : "--max-time=86400s -max-memory=51200",
  "CFMSE_IGNORE_JSON" : "cfmse_ignore.json",
  "CFM_OPTIONS" : "-klee-cfmse -klee-cfmse-run-only-on-loops=false",
  "PROG_ARGS" : ""
}
\end{lstlisting}
\end{tcolorbox}

Additional KLEE flags may be introduced by modifying the \texttt{KLEE\_OPTIONS} field.

Finally, we run the test driver using the following command:

\begin{tcolorbox}[
   		enhanced,
    	colback=blue!5,
    	colframe=blue!60,
    	rounded corners,
        ]
\begin{verbatim}
python3 ${KLEE_BUILD_DIR}/../scripts/cfm_driver/driver.py \ 
-e -i ./test-driver.bc -k driver_options.json -r $main_dir/$benchName
\end{verbatim}
\end{tcolorbox}

In this command, the user provides the input LLVM bitcode file, the JSON configuration file containing the driver options, and the output directory where the results are stored.
The driver executes KLEE on the provided program and terminates as soon as the first true positive bug is discovered.
Since \name is implemented as LLVM transformation, it resides within the LLVM framework. Thus, for this artifact, we modify both the LLVM and KLEE source code. Our LLVM changes can be found inside \texttt{llvm/lib/Transforms/CFMSE}. Since we now need to call this transformation from within KLEE, we modify its source by adding a call to our transformation in the \texttt{KModule.cpp} file present in \texttt{klee/lib/Module}.

\end{document}